\documentclass[a4paper,fleqn,usenatbib]{mnras}

\usepackage{newtxtext,newtxmath}

\usepackage[T1]{fontenc}
\usepackage{ae,aecompl}

\usepackage{graphicx}	
\usepackage{amsmath}	
\usepackage{amssymb}	

\newcommand{\txn}{\textnormal}
\newcommand{\us}{\,} 
\newcommand{\diff}{\txn{d}}
\newcommand{\code}{\nolinkurl} 
\newcommand{\specwizard}{\textsc{specwizard}}
\newcommand{\eagle}{EAGLE}

\newcommand{\Ang}{\ensuremath{\us \txn{\AA}}}
\newcommand{\mA}{\ensuremath{\us \txn{m\AA}}}
\newcommand{\cMpc}{\ensuremath{\us \txn{cMpc}}} 
\newcommand{\ckpc}{\ensuremath{\us \txn{ckpc}}} 
\newcommand{\pkpc}{\ensuremath{\us \txn{pkpc}}} 
\newcommand{\pcmsq}{\ensuremath{\us \txn{cm}^{-2}}}
\newcommand{\kmps}{\ensuremath{\us \txn{km}\, \txn{s}^{-1}}}
\newcommand{\pcc}{\ensuremath{\us \txn{cm}^{-3}}}
\newcommand{\Msun}{\ensuremath{\us \txn{M}_{\sun}}}
\newcommand{\Lstar}{\ensuremath{\us \txn{L}_{*}}}
\newcommand{\Mstellar}{\ensuremath{\us \txn{M}_{\star}}}
\newcommand{\Mvir}{\ensuremath{\us \txn{M}_{\txn{200c}}}}
\newcommand{\Rvir}{\ensuremath{\us \txn{R}_{\txn{200c}}}}
\newcommand{\Tvir}{\ensuremath{\us \txn{T}_{\txn{200c}}}}

\newcommand{\dex}{\ensuremath{\us \txn{dex}}}
\newcommand{\K}{\ensuremath{\us \txn{K}}}

\title[X-ray/FUV absorption lines in the EAGLE CGM]{The warm-hot circumgalactic medium around EAGLE-simulation galaxies and its detection prospects with X-ray and UV line absorption}

\author[N. A. Wijers et al.]{
Nastasha A. Wijers,$^{1}$\thanks{E-mail: wijers@strw.leidenuniv.nl}
Joop Schaye,$^{1}$
Benjamin D.\ Oppenheimer$^{2,3}$
\\
$^{1}$Leiden Observatory, Leiden University, Niels Bohrweg 2, NL-2333~CA Leiden, The Netherlands\\
$^{2}$CASA, Department of Astrophysical and Planetary Sciences, University of Colorado, 389 UCB, Boulder, CO 80309, USA\\
$^{3}$Harvard-Smithsonian Center for Astrophysics, 60 Garden St, Cambridge, MA 02138, USA
}

\date{Accepted XXX. Received YYY; in original form ZZZ}

\pubyear{2019}

\begin{document}
\label{firstpage}
\pagerange{\pageref{firstpage}--\pageref{lastpage}}
\maketitle

\begin{abstract}
We use the EAGLE (Evolution and Assembly of GaLaxies and their Environments) cosmological simulation to study the distribution of baryons, and far-ultraviolet
(\ion{O}{vi}), extreme-ultraviolet (\ion{Ne}{viii}) and X-ray (\ion{O}{vii}, \ion{O}{viii}, \ion{Ne}{ix}, and \ion{Fe}{xvii}) line absorbers, around galaxies and haloes of mass $\Mvir = 10^{11}$--$10^{14.5} \Msun$ at redshift $0.1$. 
EAGLE predicts that the circumgalactic medium (CGM) contains more metals than the interstellar medium across halo masses. The ions we study here trace the warm-hot, volume-filling phase of the CGM, but are biased towards temperatures corresponding to the collisional ionization peak for each ion, and towards high metallicities. Gas well within the virial radius is mostly collisionally ionized, but around and beyond this radius, and for \ion{O}{vi}, photoionization becomes significant. When presenting observables we work with column densities, but quantify their relation with equivalent widths by analysing virtual spectra. Virial-temperature collisional
ionization equilibrium ion fractions are good predictors of column density trends with halo mass, but underestimate the diversity of ions in haloes. Halo gas dominates the highest column density absorption for X-ray lines, but lower density gas contributes to strong UV absorption lines from \ion{O}{vi} and \ion{Ne}{viii}. Of the \ion{O}{vii} (\ion{O}{viii}) absorbers detectable in an Athena X-IFU blind survey, we find that 41 (56)~per~cent arise from haloes with $\Mvir = 10^{12.0 \txn{--}13.5} \Msun$. We predict that the X-IFU will detect \ion{O}{vii} (\ion{O}{viii}) in 77 (46) per cent of the sightlines passing $\Mstellar = 10^{10.5 \txn{--}11.0} \Msun$ galaxies within $100 \pkpc$ (59 (82)~per~cent for $\Mstellar > 10^{11.0} \Msun$). Hence, the X-IFU will probe covering fractions comparable to those detected with the Cosmic Origins Spectrograph for \ion{O}{vi}.
\end{abstract}





\section{Introduction}

It is well established that galaxies are surrounded by haloes of diffuse gas: the circumgalactic medium (CGM). Observationally, this gas has been studied mainly through rest-frame ultraviolet (UV) absorption by ions tracing cool ($\sim 10^{4} \K$) or warm-hot ($\sim 10^{5.5} \K$) gas \citep[e.g.,][for a review]{tumlinson_peeples_werk_2017_cgmreview}.
It has been found that the higher ions (mainly \ion{O}{vi}) trace a different gas phase than the lower ions (e.g., \ion{H}{i}), and that the CGM is therefore multiphase. \citet{werk_proschaska_etal_2014} find that these phases and the central galaxy may add up to the cosmic baryon fraction around $\Lstar$ galaxies, but the budget is highly uncertain, mainly due to uncertainties about the ionization conditions of the warm phase. 

Theoretically, we expect hot, gaseous haloes to develop around $\sim \Lstar$ and more massive galaxies \citep[$\log_{10} \Mvir \Msun^{-1} \gtrsim 11.5$--$12.0$; e.g.,][]{dekel_birnboim_2006, keres_katz_etal_2009, van-de-voort_schaye_etal_2011, correa_schaye_etal_2018}. The hot gas phase ($\gtrsim 10^{6} \K$) mainly emits and absorbs light in X-rays. For example, high-energy ions with X-ray lines dominate the haloes of simulated $\Lstar$ galaxies \citep[{e.g.}][]{oppenheimer_etal_2016, Nelson_etal_2018_hiO}. In observations, it is, however, still uncertain how much mass is in this hot phase of the CGM.

Similarly, there are theoretical uncertainties regarding the hot CGM. For example, we can compare the {\eagle} \citep[Evolution and Assembly of GaLaxies and their Environments;][]{eagle_paper} and IllustrisTNG \citep{pillepich_springel_etal_2018} cosmological simulations. They are both calibrated to produce realistic galaxies. However, they find very different (total) gas fractions in haloes with $\Mvir \lesssim 10^{12.5} \Msun$ \citep{davies_crain_etal_2019_tngcomp}, implying that the basic central galaxy properties used for these calibrations do not constrain those of the CGM sufficiently. This means that, while difficult, observations of the CGM hot phase are needed to constrain the models. The main differences here are driven by whether the feedback from star formation and black hole growth, which (self-)regulates the stellar and black hole properties in the central galaxy, ejects gas only from the central galaxy into the CGM (a galactic fountain), or ejects it from the CGM altogether, into the intergalactic medium \citep[IGM;][]{davies_crain_etal_2019_tngcomp, mitchell_schaye_etal_2020}.

There are different ways to try to find this hot gas.
The Sunyaev--Zel'dovich (SZ) effect traces the line-of-sight free-electron pressure, and therefore hot, ionized gas. So far, it has been used to study clusters, and connecting filaments in stacked observations, as reviewed by \citet{mroczkowski_nagai_etal_2018}. Future instruments \citep[e.g., CMB-S4,][]{cmb-s4_science_2016} might be able to probe smaller angular scales with the SZ-effect, and thereby smaller/lower mass systems.

Dispersion measures from fast radio bursts (FRBs) measure the total free-electron column density along the line of sight, but are insensitive to the redshift of the absorption. They therefore probe ionized gas in general, but the origin of the electrons can be difficult to determine \citep[e.g.,][]{prochaska_zheng_2019}. \citet{ravi_2019} found, using an analytical halo model, that it might be possible to constrain the ionized gas content of the CGM and IGM using FRBs. This does require host galaxies for FRBs to be found in order to determine their redshift, uncertainties about absorption local to FRB environments to be reduced, and galaxy positions along the FRB sightline to be measured from (follow-up) surveys. 

Another way to look for this hot phase is through X-ray emission. Unlike absorption or the SZ-effect, this scales with the density squared, and is therefore best suited for studying dense gas. However, if observed, it can give a more detailed image of a system than absorption along a single sightline. 
Emission around giant spirals, such as the very massive ($\Mstellar = 3 \times 10^{11} \Msun$) isolated spiral galaxy NGC~1961, has been detected \citep{anderson_churazov_etal_2016}.
Around lower mass spirals, such hot haloes have proven difficult to find: \citet{bogdan_vogelsberger_etal_2015} stacked {\it Chandra} observations of eight  $\Mstellar = 0.7 $--$2\times 10^{11} \Msun$ spirals and found only upper limits on the X-ray surface brightness beyond the central galaxies. \citet{anderson_bregman_etal_2013} stacked {\it ROSAT} images of a much larger set of galaxies (2165), and constrained the hot gas mass in the inner CGM.

In this work, we will focus on metal-line absorption. 
\ion{O}{vi} absorption has been studied extensively using its FUV doublet at $ 1032, 1038 \Ang$ at low redshift. It has been the focus of a number of observing programmes with the \emph{Hubble Space Telescope's} \emph{Cosmic Origins Spectrograph} (HST-COS) \citep[{e.g.,}][]{tumlinson_thom_etal_2011, johnson_chen_mulchaey_2015, johnson_chen_etal_2016}. A complication with \ion{O}{vi} is that the implications of the observations depend on whether the gas is photoionized or collisionally ionized. This is often uncertain from observational data \citep[{e.g.,}][]{carswell_schaye_kim_2002, trip_sembach_etal_2008, werk_proschaska_etal_2014, werk_prochaska_etal_2016}, and simulations find that both are present in the CGM  \citep[{e.g.,}][]{tepper-garcia_richter_etal_2011, rahmati_etal_2016, oppenheimer_etal_2016, oppenheimer_2018_fossilAGN_cos, roca-fabrega_dekel_etal_2018}. The uncertainty in the ionization mechanism leads to uncertainties in which gas phase is traced, and how much mass is in it. 


The hot phase of the CGM, predicted by analytical arguments (the virial temperatures of haloes) and hydrodynamical simulations is difficult to probe in the FUV, since the hotter temperatures expected for $\sim L_{*}$ galaxies' CGM imply higher energy ions. One option,  proposed by \citet{tepper-garcia_richter_etal_2013} and used by \citet{burchett_tripp_etal_2018}, is to use HST-COS to probe the CGM with \ion{Ne}{viii} ($770, 780 \Ang$) at higher redshifts ($z > 0.5$). These lines in the extreme ultraviolet (EUV) cannot be observed at lower redshifts, so for nearby systems a different approach is needed. 

Many of the lines that might probe the CGM hot phase have their strongest absorption lines in the X-ray regime \citep[e.g.,][]{perna_loeb_1998, hellsten_gnedin_miralda-escude_1998, chen_weinberg_etal_2003, cen_fang_2006, branchini_ursino_etal_2009}.
Some extragalactic \ion{O}{vii}, \ion{O}{viii}, and \ion{Ne}{ix} X-ray-line absorption has been found with current instruments, but with difficulty.
\citet{kovacs_orsolya_etal_2019} found \ion{O}{vii} absorption by stacking X-ray observations centred on \ion{H}{i} absorption systems near massive galaxies, though they targeted large-scale structure filaments rather than the CGM, while \citet{ahoranta_nevalainen_etal_2020} found \ion{O}{viii} and \ion{Ne}{ix} at the redshift of an \ion{O}{vi} absorber. \citet{bonamente_nevalainen_etal_2016} found likely \ion{O}{viii} absorption at the redshift of a broad Lyman~$\alpha$ absorber.
These tentative detections demonstrate that more certain, and possibly blind, extragalactic detections of these lines might be possible with more sensitive instruments. 

The hot CGM of our own Milky Way galaxy can be observed more readily. Absorption from \ion{O}{vii} has been found by e.g., \citet{bregman_lloyd-davies_edward_2007} and \citet[also \ion{O}{viii}]{gupta_mathur_etal_2012}, and \citet{hodges-kluck_miller_bregman_2016} studied the velocities of \ion{O}{vii} absorbers. \citet{gatuzz_churazov_etal_2018} studied \ion{Ne}{ix} absorption alongside \ion{O}{vii} and \ion{O}{viii}, focussing on the hot CGM and the ISM. The Milky Way CGM has also been probed with soft X-ray emission \citep[{e.g.,}][]{kuntz_snowden_2000, miller_bregman_2015, das_mathur_etal_2019}, and studied using combinations of emission and absorption \citep[{e.g.,}][]{bregman_lloyd-davies_edward_2007, gupta_mathur_etal_2014, miller_bregman_2015, gupta_mathur_etal_2017, das_mathur_etal_2019}. 

Previous theoretical studies of CGM X-ray absorption include analytical modelling, which tends to focus on the Milky Way.  For example, \citet{voit_2019}, used a precipitation-limited model to predict absorption by \ion{O}{vi}--\ion{}{viii}, \ion{N}{v}, and \ion{Ne}{viii}, and \citet{stern_fielding_etal_2019} compared predictions of their cooling flow model to \ion{O}{vii} and \ion{O}{viii} absorption around the Milky Way. \citet{faerman_sternberg_mckee_2017} constructed a phenomenological CGM model, based on \ion{O}{vi}--\ion{}{viii} absorption and \ion{O}{vii} and \ion{O}{viii} emission in the Milky Way.
\citet{Nelson_etal_2018_hiO} studied \ion{O}{vii} and \ion{O}{viii} in IllustrisTNG, but focused on a wider range of halo masses: two orders of magnitude in halo mass around $\Lstar$. 

In \citet{wijers_schaye_etal_2019} we used the {\eagle} hydrodynamical simulation to predict the cosmic distribution of \ion{O}{vii} and \ion{O}{viii} for blind observational surveys. 
We found that absorbers with column densities $\mathrm{N}_{\mathrm{O\,VII}, \mathrm{O\,VIII}} \gtrsim 10^{16} \pcmsq $ typically have gas overdensities $\gtrsim 10^2$, and that absorbers with overdensities $\sim 10$ may be difficult to detect at all in planned surveys. Therefore, we expect that a large fraction of the X-ray absorbers detectable with the planned Athena X-IFU \citep{athena_ifu_2016} survey, and proposed missions such as Arcus \citep{brenneman_smith_etal_2016, smith_abraham_etal_2016_arcus}, are associated with the CGM of galaxies.
Until such missions are launched, progress can be made with deep follow-up of FUV absorption lines with current X-ray instruments. 
The simulations can also help interpret the small number of absorbers found with current instruments \citep[e.g.,][]{nicastro_etal_2018, kovacs_orsolya_etal_2019, ahoranta_nevalainen_etal_2020}; e.g.\ \citet{johnson_mulchaey_etal_2019} used galaxy information to re-interpret the lines found by \citet{nicastro_etal_2018}.

In this work, we will consider \ion{O}{vi} ($ 1032, 1038 \Ang$ FUV doublet), \ion{Ne}{viii} ($770, 780 \Ang$ EUV doublet), \ion{O}{vii} (He-$\alpha$ resonance line at $21.60 \Ang$), \ion{O}{viii} ($18.9671, 18.9725 \Ang$ doublet), \ion{Ne}{ix} ($13.45 \Ang$), and \ion{Fe}{xvii} ($15.01, 15.26 \Ang$). In collisional ionization equilibrium (CIE), the limiting ionization case for high-density gas, these ions probe gas at temperatures $T \sim 10^{5.5}$--$10^{7} \us \mathrm{K}$, covering the virial temperatures of $\sim \Lstar$ haloes to smaller galaxy clusters (see Fig.~\ref{fig:Tvir} and Table~\ref{tab:ions}), as well as the `missing baryons' temperature range in the warm-hot IGM \citep[e.g.,][]{cen_ostriker_1999}. 
We include \ion{O}{vi} because this highly ionized UV ion has proved useful in {\it HST-COS} studies, and \ion{Ne}{viii} has been used to probe a hotter gas phase, albeit at higher redshifts. \ion{O}{vii}, \ion{O}{viii}, and \ion{Ne}{ix} are strong soft X-ray lines, probing our target gas temperature range, and have proven to be detectable in X-ray absorption. \ion{Fe}{xvii} is expected to be a relatively strong line at higher energies \citep{hellsten_gnedin_miralda-escude_1998}, probing the hottest temperatures in the missing baryons range (close to $10^{7} \K$), and is therefore also included. 



We will predict UV and X-ray column densities in the CGM of {\eagle} galaxies at $z=0.1$, and explore the physical properties of the gas the various ions probe. We also investigate which haloes we are most likely to detect with the Athena X-IFU.
In \S\ref{sec:methods}, we discuss the {\eagle} simulations and the methods we use for post-processing them. In \S\ref{sec:results}, we will discuss our results. We start with a general overview of the ions and their absorption in \S\ref{sec:ionprop}, then discuss the baryon, metal, and ion contents of {\eagle} haloes in \S\ref{sec:halofracs}. 
Then, we discuss  what fraction of absorption systems of different column densities are due to the CGM (\S\ref{sec:cddfsplit}) and how those column densities translate into equivalent widths (EWs), which are more directly observable. We then switch to a galaxy-centric perspective and show absorption profiles for galaxies of different masses (\S\ref{sec:radprof}), and what the underlying spherical gas and ion distributions are (\S\ref{sec:3dprof}). In \S\ref{sec:obs}, we use those absorption profiles and the relations we found between column density and EW to predict what can be observed. In \S\ref{sec:discussion} we discuss our results in the light of previous work, and we summarize our results in \S\ref{sec:conclusions}. 

Throughout this paper, we will use $\Lstar$ for the characteristic luminosity in the present-day galaxy luminosity function ($\sim 10^{12} \Msun$ haloes), and $\Mstellar$ for the stellar masses of galaxies. Except for centimetres, which are always a physical unit, we will prefix length units with `c' if they are comoving and `p' if they are proper/physical sizes.


\section{Methods}
\label{sec:methods}

In this section, we will discuss the cosmological simulations we use to make our predictions (\S\ref{sec:eagle}), the galaxy and halo information we use (\S\ref{sec:halocat}), and how we define the CGM (\S\ref{sec:cgmdef}). We explain how we predict column densities (\S\ref{sec:ions} and \S\ref{sec:cdensmeth}), EWs (\S\ref{sec:ewmeth}), and absorption profiles (\S\ref{sec:radprofmeth}) from these simulations.

\subsection{{\eagle}}
\label{sec:eagle}

The {\eagle} \citep[`Evolution and Assembly of GaLaxies and their Environments';][]{eagle_paper, eagle_calibration, mcalpine_helly_etal_2016} simulations are cosmological, hydrodynamical simulations. Gravitional forces are calculated with the \textsc{Gadget-3} \textsc{TreePM} scheme \citep{springel_2005} and hydrodynamics is implemented using a smoothed particle hydrodynamics (SPH) method known as \textsc{Anarchy} (\citeauthor{eagle_paper} \citeyear{eagle_paper}, appendix~A; \citeauthor{anarchy_effect} \citeyear{anarchy_effect}).  {\eagle} uses a Lambda cold dark matter cosmogony with the \citet{planck_2013} cosmological parameters: $(\Omega_m,\Omega_\Lambda,\Omega_b, h, \sigma_8, n_s, Y) = (0.307, 0.693, 0.04825, 0.6777, 0.8288, 0.9611, 0.248)$, which we also adopt in this work.

Here, we use the $100^3 \cMpc^{3}$ {\eagle} simulation, though we made some comparisons to both smaller volume and higher resolution simulations to check convergence. 
It has a dark matter particle mass of $9.70 \times 10^6 \Msun$, an initial gas particle mass of $1.81 \times 10^6 \Msun$, and a Plummer-equivalent gravitational softening length of $0.70 \pkpc$ at the low redshifts we study here. 

The resolved effects of a number of unresolved processes (`subgrid physics') are modelled in order to study galaxy formation. This includes star formation, black hole growth, and the feedback those cause, as well as radiative cooling and heating of the gas, including metal-line cooling \citep{wiersma_schaye_smith_2009}. To prevent artificial fragmentation of cool, dense gas, a pressure floor is implemented at ISM densities.  

In {\eagle}, stars form in dense gas, with a pressure-dependent star formation rate designed to reproduce the Kennicutt--Schmidt relation. They return metals to surrounding gas based on the yield tables of \citet{wiersma_etal_2009_insim} and provide feedback from supernova explosions by stochastically heating gas to $10^{7.5} \K$, with a probability set by the expected energy produced by supernovae from those stars \citep{dalla-vecchia_schaye_2012}. Black holes are seeded in low-mass haloes and grow by accreting nearby gas \citep{rosas-guevara_bower_etal_2015}. They provide feedback by stochastic heating as well \citep{booth_schaye_2009}, but to $10^{8.5} \K$.
This stochastic heating is used to prevent overcooling due to the limited resolution: if the expected energy injection from single supernova explosions is injected into surrounding dense $\sim 10^{6} \Msun$ gas particles at each time-step, the temperature change is small, cooling times remain short, and the energy is radiated away before it can do any work. This means self-regulation of star formation in galaxies fails, and galaxies become too massive.  The star formation and stellar and black hole feedback are calibrated to reproduce the $z=0.1$ galaxy luminosity function, the black hole mass-stellar mass relation, and reasonable galaxy sizes \citep{eagle_calibration}.

\subsection{Galaxies and haloes in the EAGLE simulation}
\label{sec:halocat}

We use galaxy and halo information from {\eagle} in two ways. First, we look at the properties of gas around haloes. We obtain absorption profiles (column densities as a function of impact parameter), as well as spherically averaged gas properties as a function of (3D) distance to the central galaxy.
Secondly, we investigate what fraction of absorption in a random line of sight with a particular column density is, on average, due to haloes (of different masses), to help interpret what might be found in a blind survey for line absorption.

We use the {\eagle} galaxy and halo catalogues, which were publicly released as documented by \citet{mcalpine_helly_etal_2016}. The haloes are identified using the Friends-of-Friends (FoF) method \citep{davis_efstathiou_1985}, which connects dark matter particles that are close together (within $0.2$ times the mean interparticle separation, in this case), forming haloes defined roughly by a constant outer density. Other simulation particles (gas, stars, and black holes) are linked to an FoF halo if their closest dark matter particle is. Within these haloes, galaxies are then identified as subhaloes recovered by \textsc{subfind} \citep{springel_white_etal_2001, dolag_borgani_etal_2009}, which identifies self-bound overdense regions within the FoF haloes. The central galaxy is the subhalo containing the particle with the lowest gravitational potential. 

Though \textsc{subfind} and the FoF halo finder are used to identify structures, we do not characterize haloes using their masses directly.
Instead, we use {\Mvir}, for halo masses, which is calculated by growing a sphere around the FoF halo potential minimum (central galaxy) until the enclosed density is the target $200 \rho_{\mathrm{c}}$, where $\rho_{\mathrm{c}} = 3 H(z)^{2} \left(8 \pi G \right)^{-1}$ is the critical density, and $H(z)$ is the Hubble factor at redshift $z$. For stellar masses, we use the stellar mass enclosed in a sphere with a $30 \pkpc$ radius around each galaxy's lowest gravitational potential particle. 
We use centres of mass for the positions of galaxies, and the centre of mass of the central galaxy for the halo position. 
 
Since the temperature of the gas is important in determining its ionization state, we also want an estimate of the temperature of gas in haloes of different masses. For this, we use the virial temperature  
\begin{equation}
\label{eq:Tvir}
\Tvir = \frac{\mu \mathrm{m}_{\mathrm{H}}}{3 k}  G  \Mvir^{2/3} (200 \rho_{\mathrm{c}})^{1/3},
\end{equation}
where $\mathrm{m}_{\mathrm{H}}$ is the hydrogen mass, $G$ is Newton's constant, and $k$ is the Boltzmann constant.
We use a mean molecular weight $\mu = 0.59$, which is appropriate for primordial gas, with both hydrogen and helium fully ionized. 

We will look into the properties of haloes mostly as a function of $\Mvir$. For this, we use halo mass bins $0.5 \dex$ wide, starting at $10^{11} \Msun$. Table~\ref{tab:galsample} shows the sample size this yields for different halo masses. There is a halo with a mass $> 10^{14.5} \Msun$, but we mostly choose not to include a separate bin for this single $10^{14.53}  \Msun$ halo, and group all haloes with $\Mvir > 10^{14} \Msun$ together instead.
The second column shows the total number of haloes in the $100^3 \cMpc^3$ volume we use, and the third column shows the number of haloes that are not `cut in pieces' by the box slicing method we use to obtain column densities (\S\ref{sec:cdensmeth}). The sample size in the second column is used when calculating absorption as a function of impact parameter. However, to reduce calculation times, we use a subsample of 1000 randomly chosen haloes when we calculate total baryon and ion masses in the CGM, and gas properties as a function of (3D) radius. This is shown in the fourth column. 	 
  
\begin{table}
	\caption{The halo sample size from {\eagle} \code{L0100N1504} at $z=0.1$, with the total number of haloes (equal to the number used for the 2D radial profiles), the number outside $\Rvir$ of any $6.25 \cMpc$ slice edge, and the number used for 3D radial profiles.}
	\label{tab:galsample}
	\centering
	\begin{tabular}{r@{.}l@{--}r@{.}l r r r}
		\hline
		\multicolumn{4}{c}{$\Mvir$} &
		\multicolumn{1}{c}{Total}   &
		\multicolumn{1}{c}{Off edges} &
		\multicolumn{1}{c}{3D profiles} \\
		\multicolumn{4}{c}{$\log_{10} \Msun$} &
		\multicolumn{1}{c}{} &
		\multicolumn{1}{c}{} &
		\multicolumn{1}{c}{} \\
		\hline
		11&0	& 11&5 & 6295	& 6044	& 1000 \\
		11&5	& 12&0 & 2287	& 2159 	& 1000 \\
		12&0	& 12&5 & 870	& 792 	& 870 \\
		12&5	& 13&0 & 323	& 288 	& 323 \\
		13&0	& 13&5 & 119	& 103 	& 119 \\
		13&5	& 14&0 & 26		& 20 	& 26 \\
		\multicolumn{4}{l}{$\geq 14.0$} & 9 & 8 & 9\\
		\hline
	\end{tabular}
\end{table}

\subsection{CGM definitions}
\label{sec:cgmdef}

Roughly speaking, the CGM is the gas surrounding a central galaxy, in a region similar to that of the dark-matter halo containing the galaxy. This definition is not very precise, because there is no clear physical boundary between the CGM and IGM or between the CGM and ISM. We will make use of a few different definitions. Here, we discuss how to identify individual SPH particles as part of the CGM. In \S\ref{sec:radprofmeth}, we discuss two methods for identifying (line-of-sight-integrated) absorption due to haloes. We mention the used definition in each figure caption, but summarize the definitions here.

The simplest approach we take is to ignore any explicit halo membership and just consider all gas as a function of distance to halo centres. We use this method for column densities and covering fractions as a function of impact parameter (though we do limit what is included along the line of sight; see \S\ref{sec:cdensmeth}), and for the temperature, density, and metallicity profiles we calculate. This is what we use in Fig.~\ref{fig:cddfs}, the solid, black lines in Fig.~\ref{fig:cddfsplits_abs}, the solid lines in Fig.~\ref{fig:radprof_Rvir}, Figs.~\ref{fig:radprof_obs}--\ref{fig:radprof_fcov_obs}, and~\ref{fig:radprof_zvar}, and the black lines in Fig.~\ref{fig:cddfsplit_zev}.

The first CGM definition we use is based on the FoF groups we discussed in \S\ref{sec:halocat}. Here, we define the CGM as all gas in the FoF group defining a halo, as well as any other gas within the $\Rvir$ sphere of that halo. We use this definition when we want to identify all gas within a set of haloes (the haloes in different mass bins), because for each {\eagle} gas particle, a halo identifier following this definition is stored \citep{eagle-team_2017}. We use this in Fig.~\ref{fig:simple_iondist}, and in the halo-projection method discussed in \S\ref{sec:radprofmeth}, used in the brown and rainbow-coloured lines in Figs.~\ref{fig:cddfsplits_abs} and~\ref{fig:cddfsplit_zev} and the dashed lines in Fig.~\ref{fig:radprof_Rvir}. This method is also one of the options explored in Fig.~\ref{fig:cddfsplits_techdep} (see also \S\ref{sec:radprofmeth} and Appendix~\ref{app:techsplit}). 

In \S\ref{sec:halofracs}, we also describe the composition of haloes using other CGM definitions. For Figs.~\ref{fig:baryinhalo} and~\ref{fig:ionfrac}, we define all gas within $\Rvir$ of the halo centre as part of the halo. When we split the gas mass into CGM and ISM in Fig.~\ref{fig:baryinhalo}, we define the ISM to be all star-forming gas and the CGM to be all other gas inside the halo. In Fig.~\ref{fig:ionfrac}, we explore the ion content of the halo. Here, we roughly excise the central galaxy by excluding gas within $0.1 \Rvir$ of the halo centre. However, we explore some variations of these definitions.

\subsection{The ions considered in this work}
\label{sec:ions}

We consider six different ions in this work: \ion{O}{vi}, \ion{O}{vii}, \ion{O}{viii}, \ion{Ne}{viii}, \ion{Ne}{ix}, and \ion{Fe}{xvii}. We list the atomic data we use for the absorption lines of these ions in Table~\ref{tab:lines}. To calculate the fraction of each element in an ionization state of interest, we use tables giving these fractions as a function of temperature, density, and redshift.
These are the tables of \citet{bertone_schaye_etal_2010, bertone_schaye_etal_2010_uv}.  The density- and redshift-dependence comes from the assumed uniform, but redshift-dependent \citep{HM01} UV/X-ray background. The tables were generated using using \textsc{Cloudy} \citep{cloudy}, version~c07.02.00. This is consistent with the radiative cooling and heating used in the {\eagle} simulations \citep{wiersma_schaye_smith_2009}. 

Unfortunately, this main set of tables we use does not include all the ionization states of oxygen, and we want to examine the overall partition of oxygen ions in haloes. Therefore, we also use a second set of tables, though only for the oxygen ions in Fig.~\ref{fig:ionfrac}. This second set of tables was made under the same assumptions as our main set: the uniform but time-dependent UV/X-ray background \citep{HM01} used for the {\eagle} cooling tables, assuming optically thin gas in ionization equilibrium. However, they were generated using a newer \textsc{Cloudy} version: 13 \citep{cloudy_2013}. We checked by comparing the tables and a smaller {\eagle} simulation that the differences between these tables are small for \ion{O}{vi}--\ion{}{viii}. In a part of a smaller {\eagle} volume, and in the column density regimes of interest, the \ion{O}{vi} column densities differed by $\lesssim 0.1 \dex$. The \ion{O}{vii} and \ion{O}{viii} column densities differed even less. The tables differ most clearly in the photoionized regime, where the column densities are small.

\subsection{Column densities from the simulated data}
\label{sec:cdensmeth}

Using these ion fractions, we calculate column densities in the same way as in \citet{wijers_schaye_etal_2019}. In short, we use the ion fraction tables we described in \S\ref{sec:ions}, which we linearly interpolate in redshift, log density, and log temperature to get each SPH particle's ion fraction. We multiply this by the tracked element abundance and mass of each SPH particle to calculate the number of ions in each particle. 

We then make a two-dimensional column density map from this ion distribution. Given an axis to project along and a region of the simulation volume to project, we calculate the number of ions in long, thin columns parallel to the projection axis. We then divide by the area of the columns perpendicular to the projection axis to get the column density in each pixel of a two-dimensional map. In order to divide the ions in each SPH particle over the columns, we need to assume a spatial ion distribution for each particle. For this, we use the same C2-kernel used for the hydrodynamics in the {\eagle} simulations \citep{wendland_1995}, although we only input the two-dimensional distance to each pixel centre.

A simple statistic that can be obtained from these maps is the column density distribution function (CDDF). This is a probability density function for absorption system column density, normalized to the comoving volume probed along a line of sight. The CDDF is defined by
\begin{equation}
f(\mathrm{N}, z) = \frac{\partial^2 n}{\partial \log_{10} \mathrm{N} \, \partial X},
\end{equation}
where N is the column density, $n$ is the number of absorbers, $z$ is the redshift, and $X$ is the absorption length given by
\begin{equation}
\diff X = (1 + z)^2 \,  (H(0) \,/\, H(z)) \, \diff z, 
\end{equation}
where $H(z)$ is the Hubble parameter. 

In practice, we make column density maps along the z-axis of the simulation box, which is a random direction for haloes. We use $32000^2$~pixels of size $3.125^2 \ckpc^2$ for the column density maps, and 16~slices along the line of sight, which means the slices are $6.25 \cMpc$ thick. 

\citet{wijers_schaye_etal_2019} found that this produces converged results for \ion{O}{vii} and \ion{O}{viii} CDDFs up to column densities $ N \approx 10^{16.5} \pcmsq$. Here we mean converged with respect to pixel size, simulation size, and simulation resolution.
By default, we set the temperature of star-forming gas to be $10^4 \K$, since the equation of state for this high-density gas does not reflect the temperatures we expect from the ISM.
However, this has negligible impacts on the column densities of \ion{O}{vii} and \ion{O}{viii}. Note that all our results do neglect a hot ISM phase, which is not modelled in {\eagle}, but may affect column densities in observations for very small impact parameters.

\citet{rahmati_etal_2016} used {\eagle} to study UV ion CDDFs and tested convergences for \ion{O}{vi} and \ion{Ne}{viii}. They used the same slice thickness at low redshift, but a lower map resolution: $10000^2$~pixels. At that resolution, they find \ion{O}{vi} CDDFs are converged to $\mathrm{N} \approx 10^{15} \pcmsq$, and \ion{Ne}{viii} to $\mathrm{N} \approx 10^{14.5} \pcmsq$. The volume and resolution of the simulation do affect CDDFs down to lower column densities. For \ion{O}{vi}, resolution has effects down to $\mathrm{N} \approx 10^{14} \pcmsq$.  

We checked the convergence of \ion{Ne}{ix} and \ion{Fe}{xvii} CDDFs with slice thickness, pixel size, box size, and box resolution in the same way as \citet{wijers_schaye_etal_2019}. We found that \ion{Ne}{ix} column densities are converged up to  $\mathrm{N} \approx 10^{16} \pcmsq$, with $\lesssim 20$~per~cent changes in the CDDF at $\mathrm{N} \gtrsim 10^{12} \pcmsq$ due to factor of~2 changes in slice thickness. 
For \ion{Fe}{xvii}, CDDFs are converged to  $\mathrm{N} \approx 10^{15.4} \pcmsq$, with mostly smaller dependences on slice thickness than the other X-ray ions. (We will later see that this ion tends to be more concentrated within haloes, so on smaller scales, than the others we investigate.) 
The trends of effect size with column density, and the relative effect sizes of changing pixel size, slice thickness, simulation volume, and simulation resolution on the CDDFs, are similar to those for \ion{O}{vii} and \ion{O}{viii}.  We note that the resolution test for \ion{Fe}{xvii} may not be reliable, since at larger column densities, this ion is largely found in high-mass haloes which are very rare or entirely absent in the smaller volume ($25^3 \cMpc^3$) used for this test.

\subsection{EWs from the simulated data}
\label{sec:ewmeth}

\begin{table}
	\caption{Atomic data for the absorption lines we study.  For each ion, we record the wavelengths $\lambda$, oscillator strengths $\mathrm{f}_{\mathrm{osc}}$, and transition probabilities A we use to calculate the EWs in Fig.~\ref{fig:N-EW}. For resolved doublets, we only use the stronger line. The last column indicates the source of the line data: M03 for \citet{morton_2003}, V96 for \citet{verner_verner_ferland_1996}, and K18 for \citet{kaastra_2018_pc}.}
	\label{tab:lines}
	\centering
	\begin{tabular}{l   r@{.}l r@{.}l r@{$\times$}l c}
		\hline
		Ion  & \multicolumn{2}{c}{$\lambda$} &
		\multicolumn{2}{c}{$\mathrm{f}_{\mathrm{osc}}$} &  \multicolumn{2}{c}{A} &Source \\
		& \multicolumn{2}{c}{($\txn{\AA}$)} &
		\multicolumn{2}{c}{} & 
		\multicolumn{2}{c}{($\mathrm{s}^{-1}$)} &
		\\
		\hline
		\ion{O}{vi} 	& 1031&9261	& 0&1325 & 4.17&$10^{8}$	&M03 \\
		\ion{Ne}{viii}	& 770&409 	& 0&103  & 5.79&$10^{8}$	& V96 \\
		\ion{O}{vii}		& 21&6019 	& 0&696  & 3.32&$10^{12}$	&V96/K18 \\	
		\ion{O}{viii}	& 18&9671	& 0&277  & 2.57&$10^{12}$	&V96 \\
					& 18&9725	& 0&139  & 2.58&$10^{12}$	&V96 \\
		\ion{Ne}{ix}	& 13&4471   	& 0&724  & 8.90&$10^{12}$	&V96/K18 \\
		\ion{Fe}{xvii}	& 15&0140 	& 2&72    & 2.70&$10^{13}$	& K18 \\
		\hline
	\end{tabular}
\end{table}

In observations, column densities are not directly observable. Instead, they must be inferred from absorption spectra. The EW can be calculated from the spectrum more directly, and for X-ray absorption, determines whether a line is observable. (Linewidths can play a role, but for the Athena X-IFU, those will be below the spectral resolution of the instrument in all cases, as we will later show.)

We compute the EWs in mostly the same way as \citet{wijers_schaye_etal_2019}, using \textsc{specwizard} \citep[e.g.,][\S3.1]{tepper-garcia_richter_etal_2011}. Briefly, in \citet{wijers_schaye_etal_2019}, we extracted absorption spectra along $100 \cMpc$ sightlines through the full {\eagle} simulation box, then calculated the EW for the whole sightline, and compared that to the total column density calculated in the same code. 

In \textsc{specwizard}, sightlines are divided into pixels (one-dimensional), and ion densities, ion-density-weighted peculiar velocities and ion-density-weighted temperatures are calculated in those pixels. The spectrum is then calculated by adding up the optical depth contributions from the position-space pixels in each spectral pixel. The optical depth profile used for each position-space pixel is Gaussian, with the centre determined by the pixel position and peculiar velocity, the width by the temperature (thermal line broadening only), and the normalization by the column density. Since, in reality, spectral lines are better described as Voigt profiles, a convolution of a Gaussian with a Cauchy--Lorentz profile, we convolve the (Gaussian-line) spectra from \textsc{specwizard} with the appropriate Cauchy--Lorentz profile for each spectral line, using the transition probabilities from Table~\ref{tab:lines}.  

Comparing EWs calculated over the full sightlines with and without the additional line broadening (eq.~\ref{eq:cl}), we find that for \ion{O}{vi} and \ion{Ne}{viii}, the differences are $< 0.01 \dex$ everywhere. For the X-ray ions, the vast majority of sightlines show differences $< 0.1 \dex$, with larger differences occurring in $\lesssim 10$ sightlines at the highest column densities. The differences are largest for \ion{Fe}{xvii}. 

In this work, we do not measure column densities and EWs along full $100 \cMpc$ sightlines. Instead, we use velocity windows around the line-of-sight velocity where the optical depth is largest. We calculate EWs in these velocity ranges by integrating the synthetic spectra over that velocity range. For the column densities in those windows, we use the fact that the total optical depth is proportional to the column density. Therefore, the fraction of the total column density in each velocity window is the same as the fraction of the total (integrated) optical depth contained within the window.

Note that we do not necessarily use all absorption systems in the sightline.
This may bias our results, but so does using full sightline values. Identifying and fitting individual absorbers and absorption systems is beyond the scope of this paper. In Appendix~\ref{app:bpar}, we show that our results are insensitive to the precise choice of velocity window.

For the UV ions, we mimic velocity windows used to define absorption systems by observers: $\pm 300 \kmps$ (rest frame). This matches how \citet{burchett_tripp_etal_2018} defined absorption systems in their CASBaH study of \ion{Ne}{viii}. For \ion{O}{vi}, \citet{johnson_chen_mulchaey_2015} searched $\Delta v = \pm 300 \kmps$ regions around galaxy redshifts for the eCGM survey. \citet{tumlinson_thom_etal_2011} searched a larger region of $\Delta v = \pm 600 \kmps$ in the COS-Haloes survey, but found that the absorbers were strongly clustered within  $\Delta v = \pm 200 \kmps$.

For the X-ray lines, we want to use velocity windows resolvable by the Athena X-IFU: the full width at half-maximum resolution (FWHM) should be $2.5 \us \mathrm{eV}$ \citep{Athena_2018_07}. This corresponds to different velocity windows for the different lines (at different energies) we consider: 
$\approx 1200 \kmps$ for \ion{O}{vii}, 
$\approx 1000 \kmps$ for \ion{O}{viii}, 
$\approx 800 \kmps$ for \ion{Fe}{xvii}, and  
$\approx 700 \kmps$ for \ion{Ne}{ix}  
at $z=0.1$. Based on the dependence of the best-fitting $b$-parameters on the velocity ranges, we choose to use a half-width $\Delta v = \pm 800 \kmps$ for the X-ray ions. We discuss this choice in Appendix~\ref{app:bpar}. 

We started with the sample of spectra for the sightlines used in \citet{wijers_schaye_etal_2019} for $z=0.1$. This sample was a combination of three subsamples, selected to have high column density in \ion{O}{vi}, \ion{O}{vii} or \ion{O}{viii}. Subsamples were selected uniformly in log column density for $N \geq 10^{13} \pcmsq$ in each ion, iterating the selection until the desired total sample size of 16384 sightlines was reached. For this work, we added a sample of the same size, but with subsamples selected by \ion{Ne}{viii}, \ion{Ne}{ix}, and \ion{Fe}{xvii} column density. Some sightlines in the two samples overlapped, giving us a total sample of 31706~sightlines. For each ion, we only use the sightlines selected for that ion specifically. These subsamples contain $\approx 5600$ sightlines each. 
 

Table~\ref{tab:lines} lists the wavelengths, oscillator strengths, and transition probabilities we use for the ions. If an ion absorption line is actually a close doublet (expected to be unresolved), we calculate the EWs from the total spectrum of the doublet lines. This is only the case for \ion{O}{viii} 
\citep[e.g.\ fig.~4 of ][]{wijers_schaye_etal_2019}. 
For \ion{Fe}{xvii}, the $15.26, 15.02 \Ang$ doublet has a rest-frame velocity difference of $4.75 \times 10^{3} \, \mathrm{km}\,\mathrm{s}^{-1}$. This is well above the linewidths we find, so the lines will not generally be intrinsically blended, and
should be resolvable by the Chandra LETG\footnote{\url{http://cxc.harvard.edu/cdo/about_chandra/}} and the XMM-Newton RGS \citep[][fig.~11]{den-herder_brinkman_etal_2001}. 
The Athena X-IFU will have a higher resolution \citep{Athena_2018_07}. 
We only use the stronger component for the \ion{O}{vi} $1031.9, 1037.6 \Ang$ and \ion{Ne}{viii} $770.4, 780.3 \Ang$ doublets, which are easily resolved with current FUV spectrographs. 

We note that for \ion{Fe}{xvii}, the atomic data for the line are under debate, with theoretical calculations and experiments finding different values \citep[e.g.,][]{gu_chen_etal_2007, de-plaa_zhuravleva_etal_2012, bernitt_brown_etal_2012, wu_gao_2019, gu_raassen_etal_2019}. Indeed the  \citet{kaastra_2018_pc} wavelength and oscillator strength that we  use for this ion do not agree with the \citet{verner_verner_ferland_1996} values. The wavelengths only differ by $0.001 \Ang$ (a relative difference of 0.007~per~cent), but the oscillator strengths and transition probabilities differ by 8~per~cent.


We will use these spectra to infer the relation between the more directly observable EWs, and the more physically interesting column densities we use throughout the paper. 
We parametrize this relation using the relation between column density and EW for a single absorber (so-called `curves of growth'), using linewidths $b$. These relations are for a single Voigt profile (or doublet of Voigt profiles). They consist of a Gaussian absorption line convolved with a Cauchy--Lorentz profile. The line is described by a continuum-normalized spectrum  $\exp(- \tau(\Delta v))$, where $\Delta v$ is the velocity offset from the line centre and $\tau$ is the optical depth. The Gaussian part of the optical depth profiles is described by
\begin{equation}
\label{eq:bpar}
\tau(\Delta v) \propto \mathrm{N} \, b^{-1} \exp\left(-(\Delta v \, b^{-1})^{2}\right),
\end{equation}
where N is the column density of the ion. The constant of proportionality is governed by the atomic physics of the transition in question. For such a line, $\mathrm{FWHM} = 1.67 b$.
However, the line is additionally broadened by the Cauchy--Lorentz component 
\begin{equation}
\label{eq:cl}
f(\nu) = \frac{1}{4 \pi^2} \frac{A}{(\Delta \nu)^2 + (A / 4 \pi)^2},
\end{equation}
where $\Delta \nu$ is the frequency offset and $A$ is the transition probability. When we fit $b$ parameters, we model the Voigt profile of the lines (convolution of eqs.~\ref{eq:bpar} and~\ref{eq:cl}), and $b$ refers to the width of the Gaussian component (eq.~\ref{eq:bpar}) alone.

We will fit these $b$ parameters to the column densities and EWs measured along the different sightlines for the different ions, by minimizing
\begin{equation}
\label{eq:logfit}
\sum_{i} \left(\log_{10}\mathrm{EW}_{i} - \log_{10}\mathrm{EW}(N_i, b) \right)^2,
\end{equation} 
where the sum is over the sightlines, $N$ is the column density, and $\mathrm{EW}(N, b)$ is obtained by integrating the spectrum produced by the Voigt profile in eqs.~\ref{eq:bpar} and~\ref{eq:cl}. 
Fitting the EWs themselves instead of the log~EWs makes little difference: only a few {\kmps}. 
Using the velocity windows instead of the full sightlines only makes a substantial difference for \ion{O}{viii}. 
We discuss the dependence of the best-fitting $b$ values on the velocity range used in Appendix~\ref{app:bpar}.   

Note that the indicative $b$-parameters we find here from the curve of growth should not be directly compared with observed values: in UV observations, linewidths are often measured by fitting Voigt profiles to individual absorption components, instead of inferred from theoretically known column densities and EWs of whole absorption systems as we do here.

\subsection{Absorption profiles}
\label{sec:radprofmeth}

We extract absorption profiles around galaxies from the two-dimensional maps described in \S\ref{sec:cdensmeth}. We extract profiles from both full maps and from maps created using only gas in haloes in particular mass ranges (i.e., gas in the FoF groups or $\Rvir$ regions of these haloes; see \S\ref{sec:cgmdef}). 
Given the positions of the galaxies, we obtain radial profiles by extracting column densities and distances from pixel centres to galaxy centres, then binning column densities by distance. 

We use only two-dimensional distances (impact parameters) here, but only use the column density map for the Z-coordinate range that includes the galaxy centre. 
We compared this method to two variations for obtaining radial profiles (not shown): adding up column densities from the two slices closest to the halo centre, and using only galaxies at least $\Rvir$ away from slice edges for radial profiles. We found that this made little difference for the median column densities: profiles excluding haloes close to slice edges were indistinguishable from those using all haloes, in part because the excluded haloes were only a small part of the sample (Table~\ref{tab:galsample}). The exceptions were the most massive haloes ($\Mvir > 10^{13.5} \Msun$), where larger haloes and small sample sizes mean the effect on the sample is larger. Even there, differences remained $\lesssim 0.2 \dex$. Using two slices instead of one made a significant difference only where both predicted median column densities were well below observable limits we consider, and well below the highest halo column densities we find.


To obtain the contributions of different haloes to the CDDF, we use two approaches. In the first, which we call the \emph{halo-projection method}, we make CDDFs by counting ions in long, thin, columns as for the total CDDFs, but we only use particles that are part of a halo's FoF group, or inside its $\Rvir$ sphere. 
Alternatively, we make maps describing which pixels in the full column density maps belong to which haloes, if any, by checking if a pixel is within $\Rvir$ of a halo (in projected distance $r_{\perp}$): the \emph{pixel-attribution method}.
To do this, we make 2D maps of the same regions, and at the same resolution, as the column density maps. These are simple True/False maps, and we make them for every set of haloes we consider. 
However, the map does not include any pixel that is closer, in units of {\Rvir}, to a halo from a different mass-defined set.
We compare these methods for splitting up the CDDFs in Appendix~\ref{app:techsplit}. Typically, the results are similar for larger column densities, but the halo-projection CDDFs contain more small column density values, coming largely from sightlines probing only short paths through the edges of the haloes.  


The advantage of using the pixel-attribution method is that it is more comparable to observations, where large-scale structure around haloes will also be present. (Note, however, that we neglect peculiar velocities.) For the CDDFs, it also allows us to attribute specific pixels in the maps to a halo or the IGM, meaning we can truly split up the CDDF into different contributions. A downside is that some haloes will be close to an edge of the projected slice, meaning that absorption due to a halo in one slice will be missed, while that of another is underestimated. However, the fraction of such haloes is small (Table~\ref{tab:galsample}). 
On the other hand, absorption may also be attributed to haloes that just happen to be close (in projection) to the absorber. This is mainly an issue for lower mass haloes. We also explore this effect Appendix~\ref{app:techsplit}.

\section{Results}
\label{sec:results}

First, we investigate some of the simplest data on our ions: what temperatures and densities they exist at (\S\ref{sec:ionprop}). We then discuss the contents of haloes (\S\ref{sec:halofracs}). Then we discuss the contributions of different haloes to the ion CDDFs, and the relation between column densities and EWs (\S\ref{sec:cddfsplit}), absorption around haloes as a function of impact parameter (\S\ref{sec:radprof}), and the 3D ion distribution around galaxies (\S\ref{sec:3dprof}). For predictions that should be comparable to observations, we refer the reader to \S\ref{sec:obs}. These results are for $z=0.1$. In Appendix~\ref{app:zev}, we compare some results to those for $z=0.5$.

\subsection{Ion properties}
\label{sec:ionprop}

\begin{figure}
	\includegraphics[width=\columnwidth]{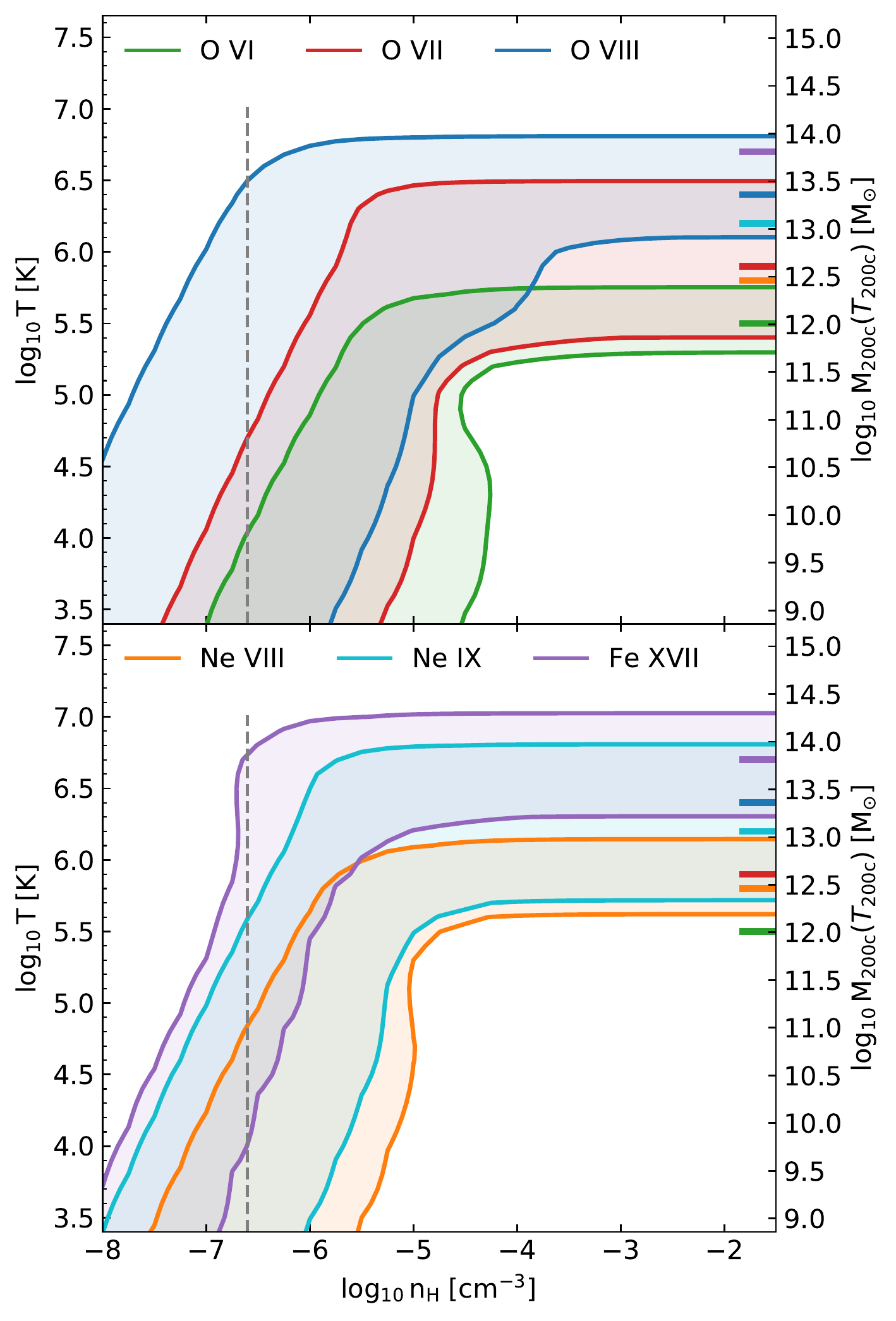}
	\caption{The temperatures and densities where different ions occur at $z=0.1$, assuming a \citet{HM01} UV/X-ray background as the only photoionizing source. The contours for each of the indicated ions are at 10~per~cent of the maximum ion fraction in CIE. The vertical, dashed line indicates the cosmic average baryon density. The right axis indicates the halo masses with virial temperatures (eq.~\ref{eq:Tvir}) matching the temperatures on the y-axis, and the coloured ticks indicate where each ion's fraction peaks in CIE.}
	\label{fig:Tvir}
\end{figure}

First, we will examine at which densities and temperatures the ions we investigate exist in meaningful quantities, which can be used to make a simple estimate of which ions are most prominent in which haloes. Table~\ref{tab:ions} and Fig.~\ref{fig:Tvir} show the energies and temperatures associated with each ion. Fig.~\ref{fig:Tvir} visualizes the \citet{bertone_schaye_etal_2010, bertone_schaye_etal_2010_uv} ionization tables we use throughout the paper.

\begin{table}
	\caption{Data for the ions we study. $\mathrm{E}_{\mathrm{ion}}$ is the energy needed to remove the least bound electron from each ion, and $\mathrm{T}_{\mathrm{CIE}}$ is the preferred CIE temperature of the ions. The CIE ranges are the upper and lower temperatures at which the ion fraction is 10~per~cent of the CIE maximum. Ionization energies are from \citet{crc_handbook}.}
	\label{tab:ions}
	\centering
	\begin{tabular}{l  r@{.}l r@{.}l@{--}r@{.}l }
		\hline
		Ion 	& \multicolumn{2}{c}{$\mathrm{E}_{\mathrm{ion}}$}	& \multicolumn{4}{c}{$\mathrm{T}_{\mathrm{CIE}}$} \\
		& \multicolumn{2}{c}{(eV)}	& \multicolumn{4}{c}{$\log_{10} \, \mathrm{K}$} \\
		\hline
		\ion{O}{vi} 	& 138&12	& 5&3	& 5&8 	 \\
		\ion{Ne}{viii}	& 239&10	& 5&6	& 6&1 	 \\
		\ion{O}{vii}		& 739&29	& 5&4	& 6&5 	\\	
		\ion{O}{viii}	& 871&41	& 6&1	& 6&8 	\\
		\ion{Ne}{ix}	& 1195&83	& 5&7	& 6&8 \\
		\ion{Fe}{xvii}	& \multicolumn{2}{l}{1266} 	& 6&3	& 7&0  \\
		\hline
	\end{tabular}
\end{table}

The shaded regions for each ion in Fig.~\ref{fig:Tvir} show the temperatures and densities where the ion fraction is at least $0.1$ times the maximum fraction in CIE. The temperature range this corresponds to in CIE is given in Table~\ref{tab:ions}. 

Fig.~\ref{fig:Tvir} shows two regimes for each ion. The first is the high-density regime where ionization by the UV/X-ray background is negligible compared to ionization by electron-ion collisions. Since recombinations and ionizations both increase as $\mathrm{n}_{\mathrm{H}}^2$ in this regime, ion fractions are only dependent on temperature here. Since we assume ionization equilibrium, this is the CIE regime. 
The second is the low-density regime where ionization by the UV/X-ray background dominates, and the density of the gas becomes important. This is the photoionization equilibrium (PIE) regime. The transition between these regimes occurs at $\mathrm{n}_{\mathrm{H}} \sim 10^{-5} \pcc$. 

The long, coloured tick marks on the right axis indicate the temperature where each ion's fraction is largest in CIE, and the right axis shows the halo mass with $\Tvir$ (eq.~\ref{eq:Tvir}) corresponding to the temperature on the left axis. Since the densities in the CGM are typically $\mathrm{n}_{\mathrm{H}} \gtrsim 10^{-5} \pcc$ (see \S\ref{sec:3dprof}), comparing the halo masses on the right axis to the temperatures where the ion fractions are high in CIE gives a reasonable estimate of which haloes contain the highest masses of the different ions, and have the highest column densities of those ions (as shown later in Figs.~\ref{fig:simple_iondist} and~\ref{fig:radprof_Rvir}).

\subsection{The baryonic content of haloes}
\label{sec:halofracs}

\begin{figure}
	\includegraphics[width=\columnwidth]{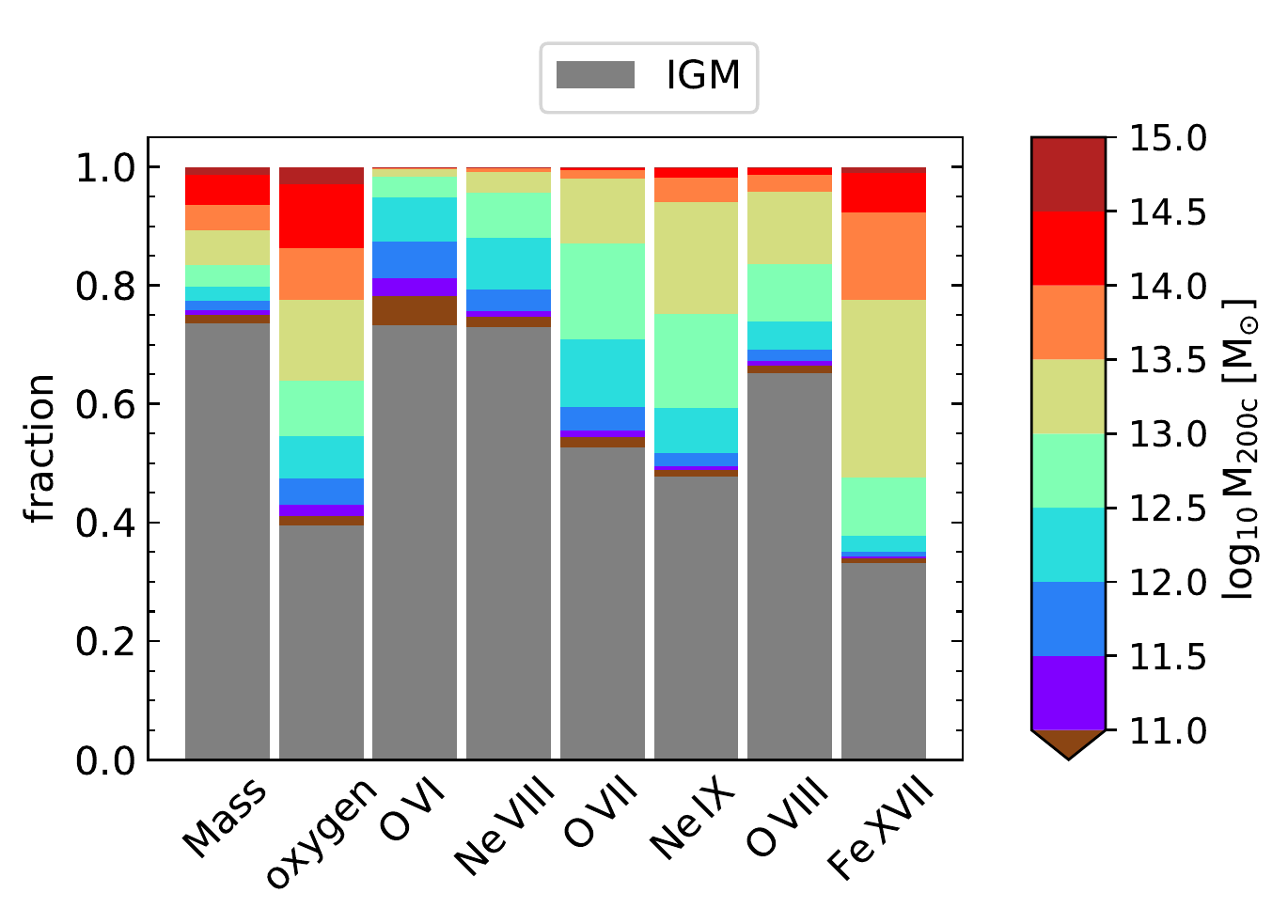}
\caption{The fraction of total gas mass and the (gas-phase) elements and ions we investigate contributed by haloes of different masses at $z=0.1$ in the {\eagle} simulation. Colours indicate halo masses according to the colour bar, with grey indicating gas that does not belong to any halo. Gas is considered part of a halo if it is part of its FoF group or $\Rvir$ sphere. Neon and iron (not shown) are distributed similarly to oxygen.}	
\label{fig:simple_iondist}
\end{figure}

Next, we look into how the ions relate to haloes in {\eagle}. Fig.~\ref{fig:simple_iondist} shows the contributions of haloes of different masses to the total mass and ion budget in the simulated $100^3 \cMpc^{3}$. An SPH particle is considered part of a halo if it is within the halo's FoF group or $\Rvir$ region. We include the $14.5$--$15$ bin for consistent spacing, but this bin contains only a single halo with $\Mvir = 10^{14.53} \Msun$, so in rest of the paper, we will group all nine haloes with masses $\Mvir \geq 10^{14} \Msun$ into one halo mass bin.



Fig.~\ref{fig:simple_iondist} shows the ions inside haloes are mostly found at halo masses where $\Tvir \sim \mathrm{T}_{\mathrm{CIE}}$. The differences between ions, and between the ion, metal, and mass distributions show that these trends are not simply a result of the ions tracing mass or metals. The importance of haloes with $\Tvir \sim \mathrm{T}_{\mathrm{CIE}}$ can be explained by a few factors. First, the temperature of the warm/hot gas in haloes is roughly $\Tvir$. Secondly, in haloes, the ions are mostly found in whatever gas there is at $\sim \mathrm{T}_{\mathrm{CIE}}$. This is because, third, the density of the warm/hot phase is mostly high enough that the gas is collisionally ionized. (In lower mass haloes, with $\Mvir \lesssim 10^{12} \Msun$, and/or gas at $\sim \Rvir$, photoionization does become relevant.) This means that haloes with $\Tvir \sim \mathrm{T}_{\mathrm{CIE}}$ contain larger amounts of ion-bearing gas than haloes at higher or lower temperatures (masses). We will demonstrate these properties of the halo gas in Fig.~\ref{fig:3dprof}.


\begin{figure*}
\includegraphics[width=0.495\textwidth]{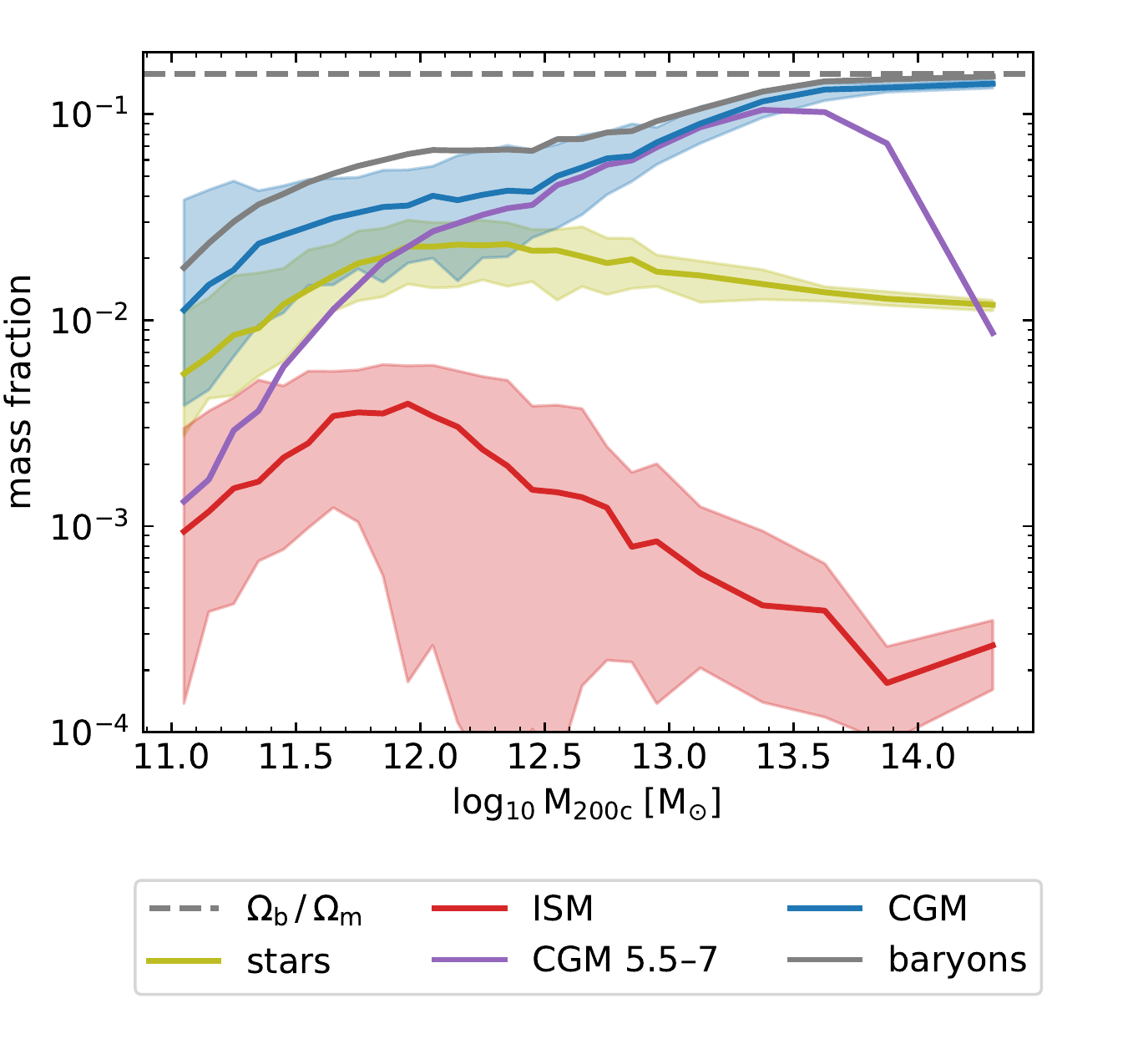}
\includegraphics[width=0.495\textwidth]{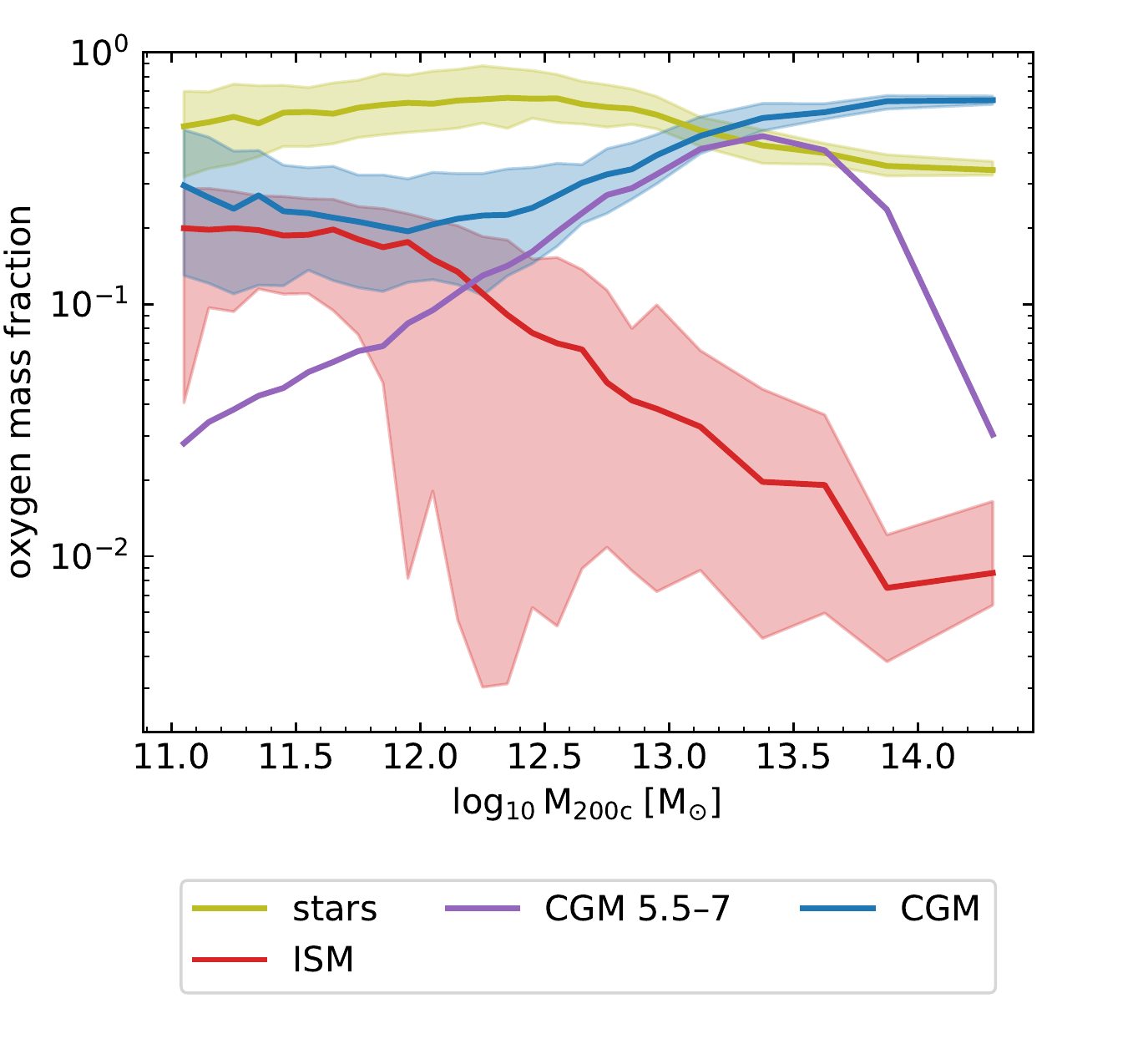}
\caption{\emph{Left-hand panel}: the fraction of halo mass (i.e., $< \Rvir$) in stars, ISM, and CGM as a function of halo mass in the {\eagle} simulation at $z=0.1$. The grey line shows the total baryon mass fraction, and the purple line shows CGM gas with temperatures in the $10^{5.5}$--$10^{7} \K$ range. The dashed, horizontal line indicates the cosmic baryon fraction. 
\emph{Right-hand panel} (note the different y-axis range): The fraction of halo oxygen mass (oxygen ejected by stars, currently within $\Rvir$) in stars, ISM, and CGM in haloes of different masses. The halo oxygen budget (total and in stars) does not include metals produced in stars that have never been ejected, or any oxygen captured by black holes. 
The solid lines show medians and shaded regions show the 80~per~cent halo-to-halo scatter in each halo mass bin; the shading is omitted for legibility for the total baryons and $10^{5.5}$--$10^{7} \K$ CGM. Here, ISM is defined as all star-forming gas and CGM as the other gas. We use $0.1 \dex$ halo mass bins for $\Mvir < 10^{13} \Msun$ haloes, then $0.25 \dex$ bins, and one bin for the haloes above $10^{14} \Msun$. The CGM is typically the largest baryon mass component in haloes, and typically contains more metals than the ISM at all halo masses we study.}
\label{fig:baryinhalo}
\end{figure*}


Besides all gas, we also want to investigate the gas in the CGM specifically. We show the mass fraction in different baryonic components as a function of halo mass in the left-hand panel of Fig.~\ref{fig:baryinhalo}. Here, we consider everything within $\Rvir$ of the central galaxy to be part of the halo. The black hole contribution is too small to appear on the plot.
The total baryon fraction increases with halo mass, and is substantially smaller than the cosmic fraction for $\Mvir < 10^{13} \Msun$. The trend at lower halo masses ($\mathrm{M}_{\mathrm{500c}} < 10^{13} \us \txn{M}_{\sun}$) is not currently constrained by observations. 
The {\eagle} baryon fractions are somewhat too high for $\Mvir > 10^{14} \Msun$ \citep{barnes_kay_etal_2017}. The observations do support the trend of rising baryon fractions with halo mass at high masses.

The CGM mass fraction increases with halo mass, while the stellar and ISM fractions peak at $\Mvir \sim 10^{12} \Msun$, with the ISM fraction declining particularly steeply towards higher masses. This is likely a result of star formation quenching starting in $\sim \Lstar$ galaxies. The `missing baryons' CGM at $10^{5.5}$--$10^{7} \K$ dominates for halo masses $\Mvir \sim 10^{12}$--$10^{13.5} \Msun$, which is what we would expect according to $\Tvir$. 
The  $\Mvir \sim 10^{12}$--$10^{13.5} \Msun$ haloes where this gas dominates are indeed the ones that dominate the ion budgets in Fig.~\ref{fig:simple_iondist}, except for \ion{O}{vi}, which probes cooler gas, and \ion{Fe}{xvii}, which probes gas in this temperature range, but where the dominant haloes include some higher mass ones, in agreement with $\Tvir$ (Fig.~\ref{fig:Tvir}).


The right panel of Fig.~\ref{fig:baryinhalo} similarly shows the fraction of oxygen 
in different baryon components for haloes of different masses. Oxygen produced in stars, but never ejected is not counted. A smaller fraction of the oxygen that was swallowed by black holes is not tracked in {\eagle}. The fraction in stars therefore reflects the metallicity of the gas the stars were born with. The fractions for neon are nearly identical to those for oxygen, while the curves for iron have the same shape, but with a somewhat smaller mass fraction in stars and more in CGM and ISM.

We see that at lower halo masses, most of the metals in haloes reside in stars, while for $\Mvir  \gtrsim 10^{13} \Msun$, more metals are found in the CGM. The changes with halo mass seem to be in line with the overall mass changes in ISM and CGM as halo mass increases (Fig.~\ref{fig:baryinhalo}), though the stars and ISM contain higher metal fractions than mass fractions, reflecting their higher metallicities. 
Interestingly, there are more metals in the CGM than in the ISM for all halo masses, though the difference is small for $\Mvir < 10^{12} \Msun$. This is similar to what \citet{oppenheimer_etal_2016} found for a smaller set of haloes with {\eagle}-based halo zoom simulations. They considered all the oxygen produced in galaxies within $\Rvir$, in 20 zoom simulations  of $\Mvir = 10^{11}$--$10^{13} \Msun$ haloes, and found that a substantial fraction of that oxygen ($\sim 30$--$70$~per~cent) is outside $\Rvir$ at $z=0.2$. That oxygen is not included in the census in Fig.~\ref{fig:baryinhalo}. 

The mass and oxygen fractions in the CGM and ISM do depend somewhat on the definition of the ISM. (The CGM is all gas within $\Rvir$ that is not ISM in all our definitions.) In Fig.~\ref{fig:baryinhalo}, we define the ISM as all gas with a non-zero star formation rate. Since the minimum density for star formation in {\eagle} is lower for higher metallicity, higher metallicity gas is more likely to be counted as part of the ISM. If we define the ISM as gas with $n_{\mathrm{H}} > 10^{-1} \pcc$ instead, the mass fractions change. Per halo, the ISM mass changes by a median of $\approx -30$--$-50$~per~cent for $\Mvir \lesssim 10^{12} \Msun$, $\approx 0$~per~cent at $\sim 10^{13} \Msun$, and up to $+30$~per~cent at higher masses. The central $80$~per~cent range is large, including differences comparable to the total ISM mass using the star formation definition in both directions. The scatter in differences is largest at low masses. The median trend with halo mass makes sense given the higher central metallicities (meaning lower minimum $n_{\mathrm{H}}$ for star formation) we find in lower mass haloes (Fig.~\ref{fig:3dprof}). If we count gas that is star-forming or meets the $n_{\mathrm{H}}$ threshold as ISM, the ISM mass can only increase relative to the star-forming definition. Median differences are $\lesssim 3$~per~cent at $\Mvir \lesssim 10^{12} \Msun$, but increase to $\approx 30$--$60$~per~cent at $\Mvir \gtrsim 10^{13} \Msun$. Since the CGM contains more mass overall, differences in the CGM mass using the two alternative ISM definitions are typically $\lesssim 11$~per~cent (central 80~per~cent of differences).

The ISM definitions also affect how oxygen is split between the ISM and CGM. Using the $n_{\mathrm{H}} > 10^{-1} \pcc$ definition results in lower ISM oxygen fractions, with median per-halo differences $\approx -20$--$-55$~per~cent, and a central 80~per~cent range of differences mostly between $\approx -10$ and $-90$~per~cent. CGM fractions are consistently higher, with median per-halo differences of up to $\approx 40$~per~cent at $\Mvir  < 10^{12.5} \Msun$, deceasing to close to zero between $\Mvir = 10^{12}$ and $10^{13} \Msun$. Using the combination ISM definition ($n_{\mathrm{H}} > 10^{-1} \pcc$ or star-forming) does not change the oxygen masses by much, since dense, but non-star-forming gas has a low metallicity. The central 80~per~cent of per halo differences is $< 1$~per~cent at all $\Mvir$.





\begin{figure*}
\includegraphics[width=0.495\textwidth]{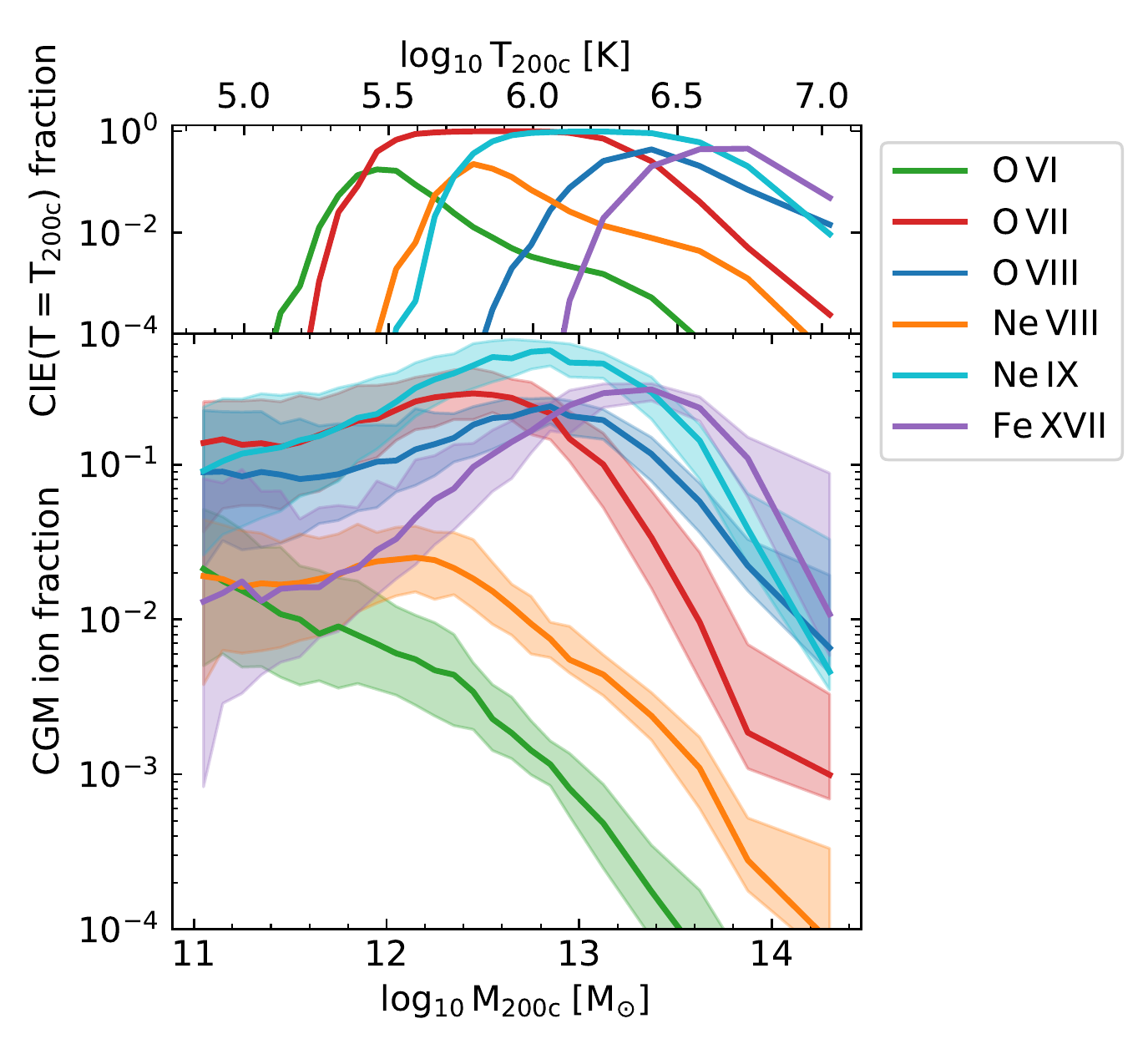}
\includegraphics[width=0.495\textwidth]{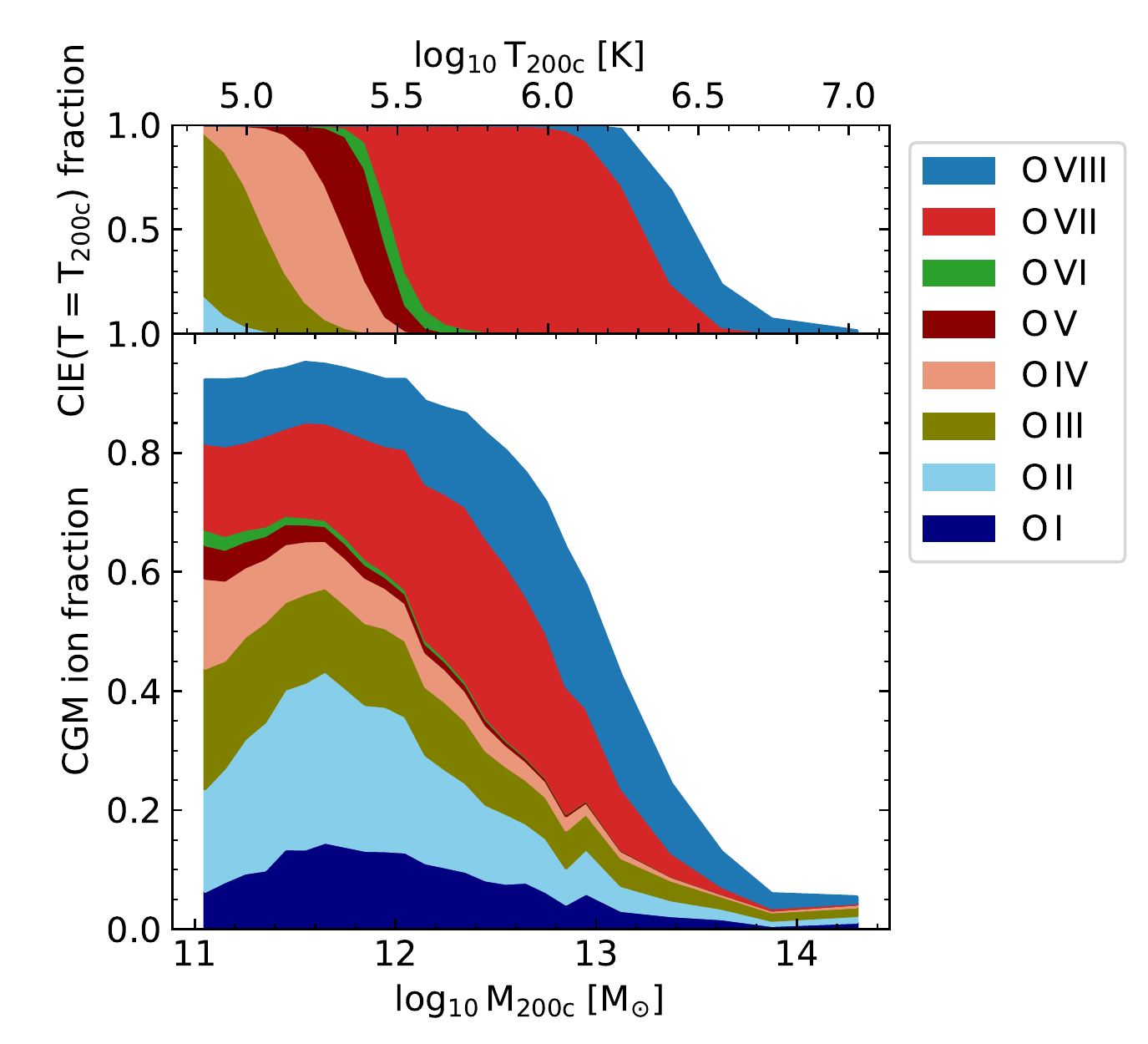}
\caption{Total fraction of each element in the ionization states indicated in the legends in the $z=0.1$ {\eagle} CGM, as a function of halo mass ($0.1$--$1 \Rvir$; lower panels). The top panels show the CIE ion fractions as a function of temperature, assuming the temperatures match $\Tvir$ for the different halo masses. The left-hand panels show these fractions for the six ions we focus on in this work. The solid lines show median ion fractions in different mass bins, while the shaded regions in the same colour show the scatter (percentiles 10--90). The virial temperature and CIE ionization curves predict the qualitative trends of ionization fraction as a function of halo mass, but strongly underestimate the ion fractions in low-mass haloes. The right panels show the average fraction of oxygen in the indicated ionization states in the CGM. The ions \ion{O}{i}--\ion{}{viii} are shown from bottom to top in order.}
\label{fig:ionfrac}
\end{figure*}
 

Finally, since we are primarily interested in ions in this work, we look into the ionization states of the metals in haloes of difference mass.\footnote{Here, we use \textsc{Cloudy} version~13 ionization tables for the oxygen ions, instead of the version~7.02 tables used in the rest of the paper. }  
The bottom panels of Fig.~\ref{fig:ionfrac} show the fraction of ions in the CGM (all gas at $0.1$--$1 \Rvir$) as a function of halo mass, compared to the CIE ion fractions at the halo virial temperatures in the top panels. In the left-hand panels, we show median ion fractions with the $10^{\mathrm{th}}$--$90^{\mathrm{th}}$ percentile range, for the ions we focus on in this work. In the right-hand panels, we show the average fractions of all the ionization states of oxygen. Note that the ionization table we use does not include the effects of self-shielding (or local radiation sources), so the lowest ionization state, \ion{O}{i}, could be underestimated.  

For the ions we focus on in this work, including the gas within $0.1 \Rvir$ has a negligible effect, since there is very little highly ionized gas there (Fig.~\ref{fig:absloc}). Including gas out to $2 \Rvir$ does make a difference. If that gas is included, these ion fractions rise, especially at the low- and high-mass ends, and the peaks of the ionization curves shift to slightly higher masses. The larger overall ion fractions are likely due to the increased amount of gas photoionized to the higher states we examine here at larger distances. The slight shifts are likely due to the lower gas temperatures in the same haloes at larger distances (Fig.~\ref{fig:3dprof}). 

For the lower ionization states in the bottom right-hand panel, whether or not we include gas at radii $< 0.1 \Rvir$ has more of an effect: including this gas increases the \ion{O}{i} and \ion{O}{ii} content by large amounts; the fraction of the total increases by $\approx 0.2$--$0.4$ for $\Mvir \sim 10^{11}$--$10^{12} \Msun$, with the effect decreasing toward higher halo masses. The difference will be due to the fact that the central galaxy contains plenty of cold gas, but very little of the more highly ionized species. (\citet{wijers_schaye_etal_2019} verified that the \ion{O}{vii} and \ion{O}{viii} CDDFs are negligibly impacted by whether or not star-forming gas is accounted for.) Including gas at larger radii (out to $2 \Rvir$) increases the fraction of oxygen in the \ion{O}{vi}--\ion{}{viii} states, at the cost of gas in lower states, but also at the cost of \ion{O}{ix} at $\Mvir \gtrsim 10^{12} \Msun$.

For the high ions in the left-hand panels, we confirm by comparing the top and bottom panels that the CIE ionization peak and halo virial temperatures are good predictors of the qualitative trends of halo ion content as a function of halo mass, but the $\txn{CIE}(\mathrm{T}=\Tvir)$ curves strongly underestimate the ion fractions at low mass, where photoionization dominates.

The CIE curves peak at slightly larger halo masses than {\eagle} haloes show. This might be because the temperature inside $\Rvir$ is typically higher than $\Tvir$. We will show this using the mass- and volume-weighted temperature profiles in Fig.~\ref{fig:3dprof}. Alternatively, or additionally, photoionization may be responsible, by lowering the typical temperature at which the ions are preferentially found. Fig.~\ref{fig:3dprof} shows this would mostly be important at lower halo masses ($\Mvir \lesssim 10^{12} \Msun$) or at radii approaching $\Rvir$.

For \ion{O}{vi}, we do not find a peak at all in the halo mass range we examine. This is due to photoionization becoming important at and below the halo masses where CIE would produce an \ion{O}{vi} peak, in the same regime where other halo ion fractions flatten out. 




As in the left-hand panels, the CIE curve for a single temperature predicts much more extreme ion fractions than we see in the Eagle haloes. In particular, Fig~\ref{fig:ionfrac} shows that lower mass haloes contain many high ions, and that the lowest ionization states peak at much higher masses than  $\txn{CIE}(\mathrm{T}=\Tvir)$ predicts, suggesting the presence of significant amounts of gas with $\mathrm{T} \ll \Tvir$. 

On the other hand, the higher high-ion fractions than suggested by the CIE curves indicate the presence of $T \gg \Tvir$ gas in sub-$\Lstar$ haloes. This is likely a result  of gas heating by stellar (and at higher masses, AGN) feedback. Temperature distributions indicate this is not only a result of the direct heating of particles due to feedback in {\eagle}, but that sub-$\Lstar$ haloes have smooth mass- and volume-weighted temperature distributions that can extend to $\sim 10^{6} \K$ or somewhat higher at $\sim \Rvir$. Besides this hotter gas, photoionized gas close to $\Rvir$ also plays a part: at these radii in $\Mvir \lesssim 10^{12} \Msun$ haloes, gas densities can reach $\mathrm{n}_{\mathrm{H}} \sim 10^{-5} \pcc$ (Fig.~\ref{fig:3dprof}), where photoionization becomes important. The importance of photoionization for the CGM ion content was previously pointed out by \citet{faerman_sternberg_mckee_2020} in their isentropic model of the CGM of an $\Lstar$ galaxy.

\subsection{Column density distributions and EWs}
\label{sec:cddfsplit}

\begin{figure}
	\includegraphics[width=\columnwidth]{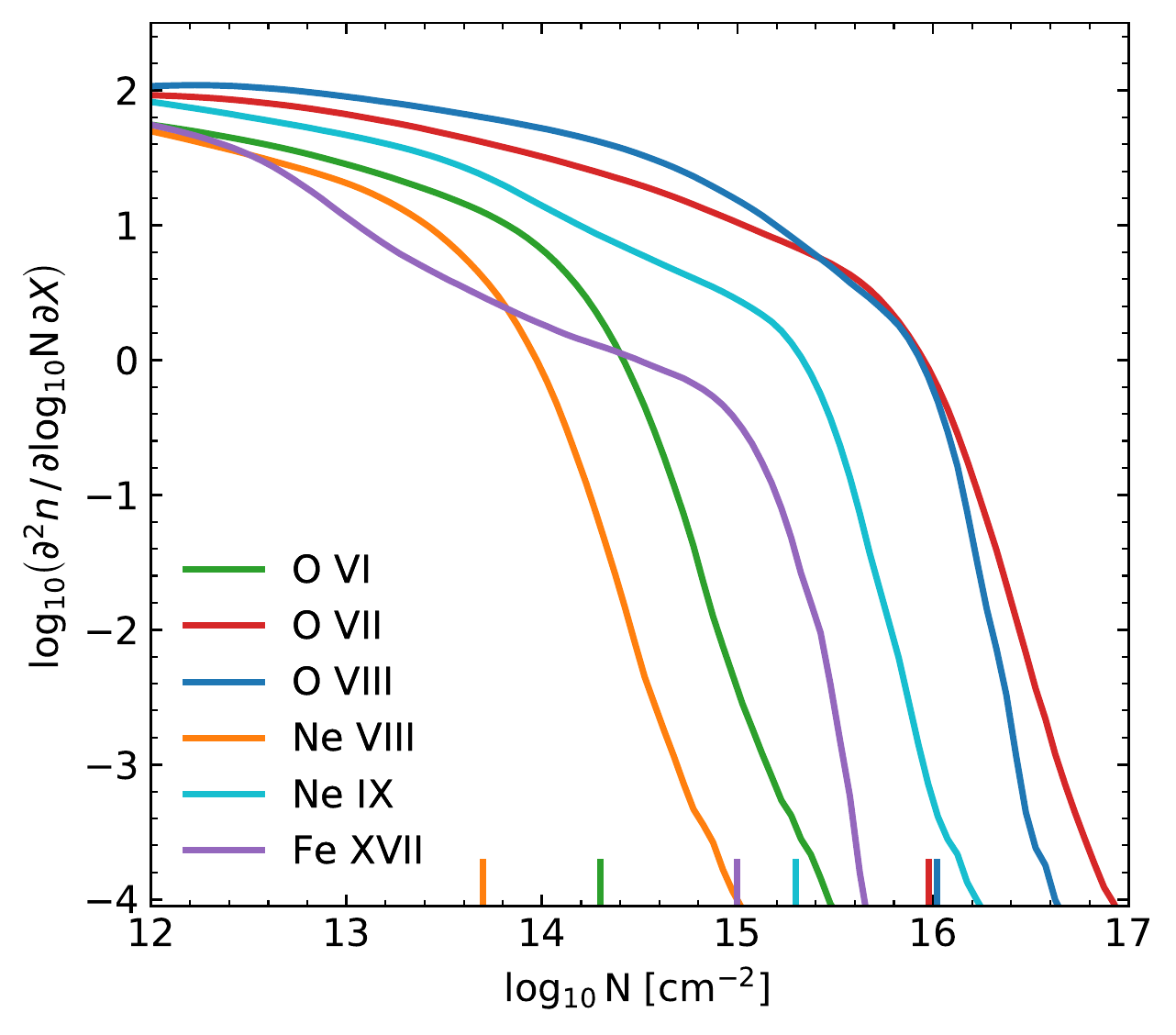}
	\caption{The CDDFs for the ions we consider in this paper, for the {\eagle} simulation at $z=0.1$. Coloured ticks on the x-axis roughly indicate the positions of breaks in the CDDFs (determined visually), which serve as reference points in further figures. These are at the same position for \ion{O}{vii} and \ion{O}{viii}, but the ticks are slightly offset for legibility.}
	\label{fig:cddfs}
\end{figure}

Before we look into metal-line absorption around haloes, we consider metal-line absorption at random locations. We consider how their column densities relate to the more directly observable EWs of absorption systems, and how haloes contribute to the absorbers we expect to find in a blind survey.

In Fig.~\ref{fig:cddfs}, we show the column density distributions for the six ions we focus on. The ions all show distributions with roughly two regimes, with a shallow and steep slope at low and high column densities, respectively. The coloured ticks on the x-axis indicate the `knees' which mark the transition between these regimes, determined visually. The ticks are for reference in other figures. 


\begin{figure*}
	\includegraphics[width=\textwidth]{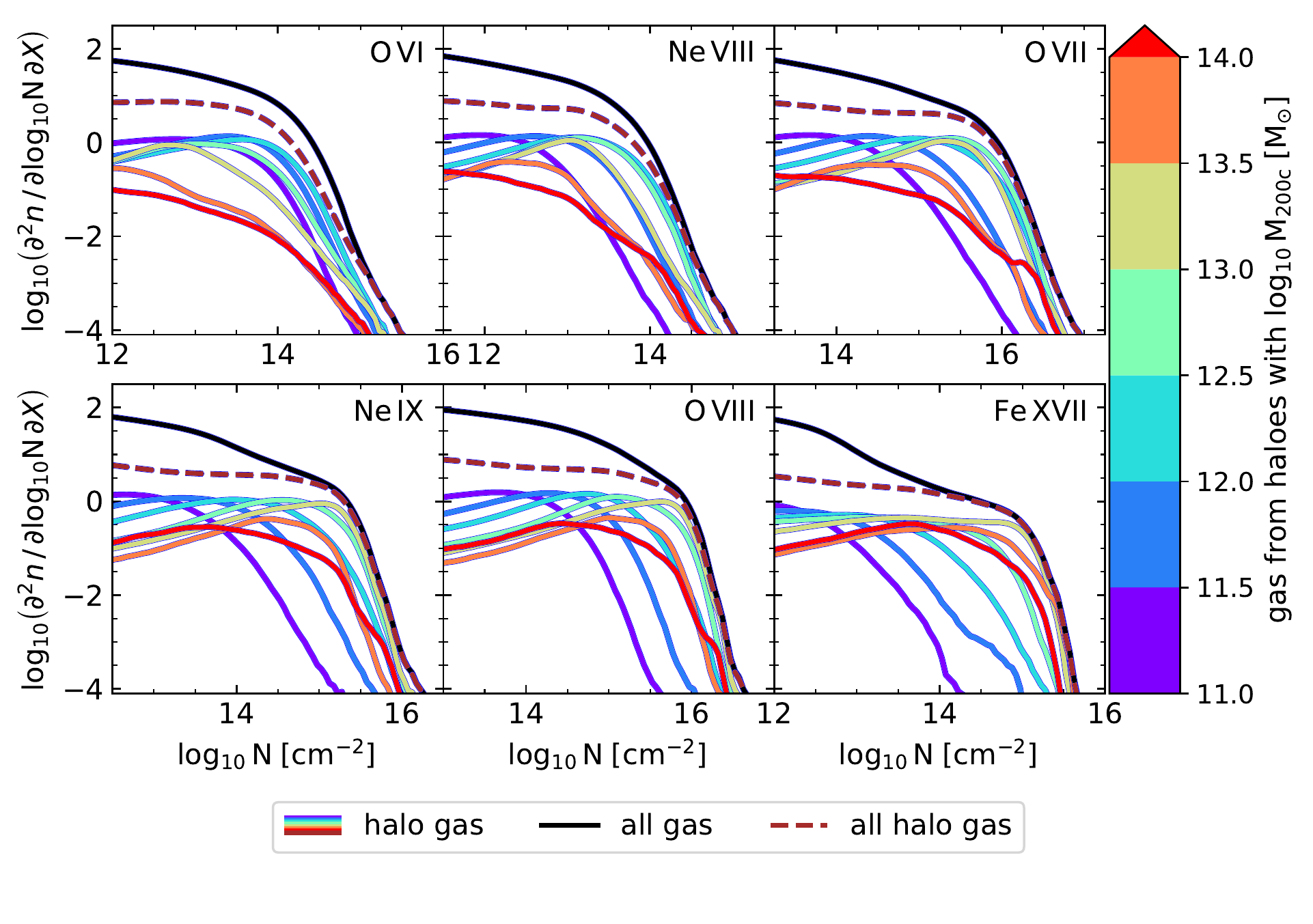}
	\caption{The contribution of absorption by haloes of different masses to the column density distributions of the ions indicated in the panels at $z=0.1$ in the {\eagle} simulations. The black line indicates the distribution of all absorption systems, while the brown, dashed line indicates the contribution of all haloes (including those with $\Mvir < 10^{11} \Msun$). The colour bar indicates the mass range for which each solid, coloured line represents the contribution to the CDDF. Contributions are determined by computing CDDFs from column density maps made with only gas in each halo mass range (in a FoF group or $\Rvir$ sphere): the halo-projection method in \S\ref{sec:radprofmeth}.}
	\label{fig:cddfsplits_abs}
\end{figure*}

In Fig.~\ref{fig:cddfsplits_abs}, we explore how haloes contribute to this absorption along randomly chosen sightlines. It shows the contributions of different halo masses to the CDDFs of our six ions. 
The CDDFs for each halo mass bin are generated from the simulations in the same way as the total CDDFs, but using only SPH particles belonging to a halo of that mass (the halo-projection method from \S\ref{sec:radprofmeth}). An SPH particle belongs to a halo if it is in the halo's FoF group, or within $\Rvir$ of the halo centre.

From Fig.~\ref{fig:cddfsplits_abs}, we see that for the X-ray ions, most absorption at column densities higher than the knee of the CDDF is due to haloes. This confirms the suspicion of \citet{wijers_schaye_etal_2019} that this was the case for \ion{O}{vii} and \ion{O}{viii}, based on the typical gas overdensity of absorption systems at these column densities. However, for the FUV/EUV ions \ion{O}{vi} and \ion{Ne}{viii}, there is a substantial contribution from gas outside haloes at these relatively high column densities. 

For all these ions, we also note the following trend. The absorption at higher column densities tends to be dominated by more massive haloes until a turn-around is reached.
These turn-around masses are consistent with the temperatures preferred by the ions, suggesting they are being driven by the increase in virial temperature with halo mass (compare to Fig.~\ref{fig:Tvir}). We have verified that trends with halo mass are not driven simply by the covering fraction of haloes of different masses.    

\begin{figure*}
	\includegraphics[width=\textwidth]{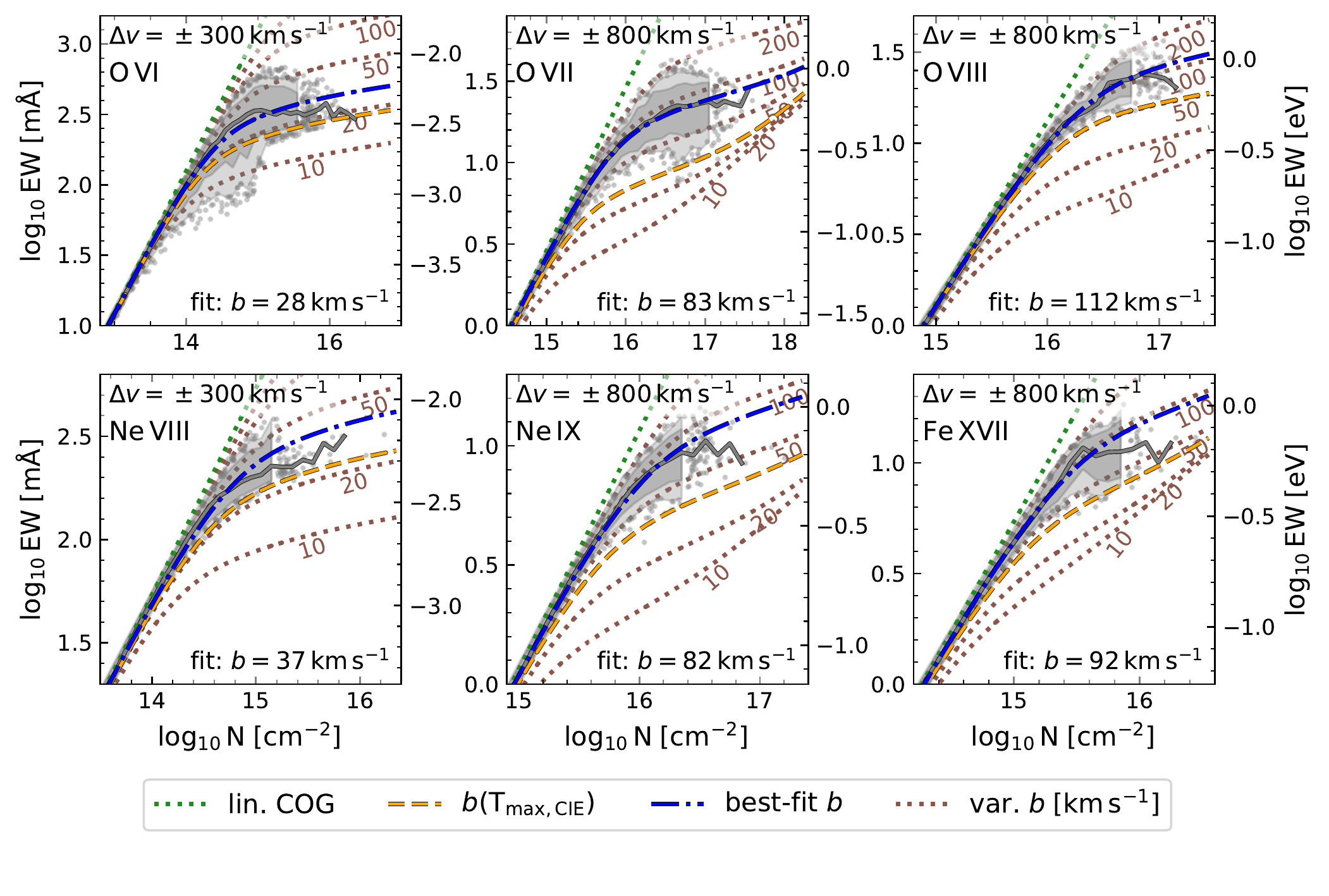}
	\caption{Rest-frame EWs for the ions we investigate as a function of ion column density at $z=0.1$ in the {\eagle} simulation. The left axes show EWs in $\log_{10} \mA$, the right axes show $\log_{10}$~eV.
	The solid, grey line shows the median EW in bins of $0.1 \dex$ in column density, while the shading shows the central 80~per~cent (darker grey) and central 96~per~cent (lighter grey) of the EWs in the bins. For EWs outside these ranges, and column density bins with fewer then 50~sightlines, we show each sightline as a single grey point. 
	 We also show best-fitting values (using eq.~\ref{eq:logfit}) for the Gaussian line broadening $b$ (eq.~\ref{eq:bpar}) in blue dot-dashed lines. The best-fitting values are indicated in the bottom right of each panel. The relation for unsaturated absorption is shown with a dotted green line. The orange, dashed line shows the thermal broadening for ions at the temperature where their ion fraction is at a maximum in CIE (equation~\ref{eq:tmcie}, Fig.~\ref{fig:Tvir}). The various dotted brown lines show the column density-EW relation for Voigt profiles with different Gaussian line broadening values (i.e., $b$ parameters): 10, 20, 50, 100, and 200~$\kmps$, from bottom to top in the panels. The spectra and column density-EW relations are for absorption lines at a single rest-fame wavelength, except for \ion{O}{viii}, where we model doublet absorption.}
	\label{fig:N-EW}
\end{figure*}

To get a sense of what column densities might be detectable with different instruments (\S\ref{sec:obs}), we look into what rest-frame EWs these column densities typically correspond to. Though we will work with column densities in the rest of this paper, the fits we find can be used to (roughly) convert between the two.
Fig.~\ref{fig:N-EW} shows typical EW as a function of column density. 

We parametrize the column density-EW relation using the width of the Gaussian part of the Voigt profile $b$, as described in \S\ref{sec:ewmeth}. We list the best-fitting parameters in Table~\ref{tab:bfit}, and show the relation for these parameters in Fig.~\ref{fig:N-EW}.  The shadings in Fig.~\ref{fig:N-EW} give an indication of how broad the $b$-parameter distribution is (10$^{\txn{th}}$ and 90$^{\txn{th}}$,  2$^{\txn{nd}}$ and 98$^{\txn{th}}$ percentiles). We will use these best-fitting $b$-parameters in \S\ref{sec:obs} to estimate the minimum column densities observable with the Athena X-IFU, Arcus, and the Lynx XGS. We explore the dependence of the best-fitting values on the velocity windows in which we measure column densities and EWs in Appendix~\ref{app:bpar}.

\begin{table}
	\caption{Best-fitting $b$ parameters to the column density-EW relation for the different ions, derived from {\eagle} mock spectra at $z=0.1$. Rest-frame EWs are calculated for each ion (first column) using the absorption lines in Table~\ref{tab:lines}. The second column indicates the half-width of the velocity windows used to calculate the EWs. Column~3 shows the best-fitting $b$ parameters (eq.~\ref{eq:logfit}), using the velocity windows in column~2. }
	\label{tab:bfit}
	\centering
	\begin{tabular}{l r r}
		\hline
		ion & $\Delta v$ & $b(\Delta  v)$ \\
		& {\kmps} & {\kmps} \\
		\hline
		\ion{O}{vi}    &  300 &  28  \\
		\ion{Ne}{viii} &  300 &  37  \\
		\ion{O}{vii}   &  800 &  83  \\
		\ion{Ne}{ix}   &  800 &  82  \\
		\ion{O}{viii}  &  800 & 112  \\
		\ion{Fe}{xvii} &  800 &  92  \\
		\hline
	\end{tabular}
\end{table}

Generally, the thermal line broadening expected at the temperature where the ion fraction peaks in CIE, 
\begin{equation}
\label{eq:tmcie}
b(T_{\max, \txn{CIE}}) = \sqrt{2 k T_{\max, \txn{CIE}} \, m_{\txn{ion}}^{-1}},
\end{equation} 
gives a good lower limit\footnote{For a given column density, non-thermal broadening or multiple absorption components spread out the ions in velocity space, meaning the absorption is less saturated. Therefore, a single line or doublet with only thermal broadening should give a lower limit to the EW of an absorption system at fixed column density.} to the EWs (dashed orange lines). Here, 
$m_{\txn{ion}}$ is the ion mass.
For \ion{O}{vii}, \ion{Ne}{ix}, \ion{Fe}{xvii}, and particularly \ion{O}{vi}, lower values do occur.  For \ion{O}{vii}, \ion{Ne}{ix}, and \ion{Fe}{xvii}, this is still consistent with the lower end of the CIE temperature range in Table~\ref{tab:ions}: $b = 16$, $20$, and $24 \kmps$, respectively. For \ion{O}{vii}, this was previously described by \citet{wijers_schaye_etal_2019}. For \ion{O }{vi}, the lower CIE end gives  $b = 14 \kmps$, which does not cover this range. Such low $b$ values are rare for this ion, but their occurrence suggests at least some high-column-density \ion{O}{vi} is photoionized. 

In Fig.~\ref{fig:N-EW}, we can also see the importance of Lorentz broadening for the EWs of the different absorption lines. The single-component absorber curves (all lines except the grey ones) show an upturn where the `wings' of the Voigt profile become important. This becomes relevant for narrow, high-column-density absorbers  for the X-ray lines, especially \ion{O}{viii} and \ion{Fe}{xvii}. For the UV lines, the effect of the Lorentz broadening is negligible, since the extra broadening is smaller relative to their wavelengths compared to the X-ray lines (Table~\ref{tab:lines}).

\subsection{Column density profiles}
\label{sec:radprof}

\begin{figure*}
	\includegraphics[width=\textwidth]{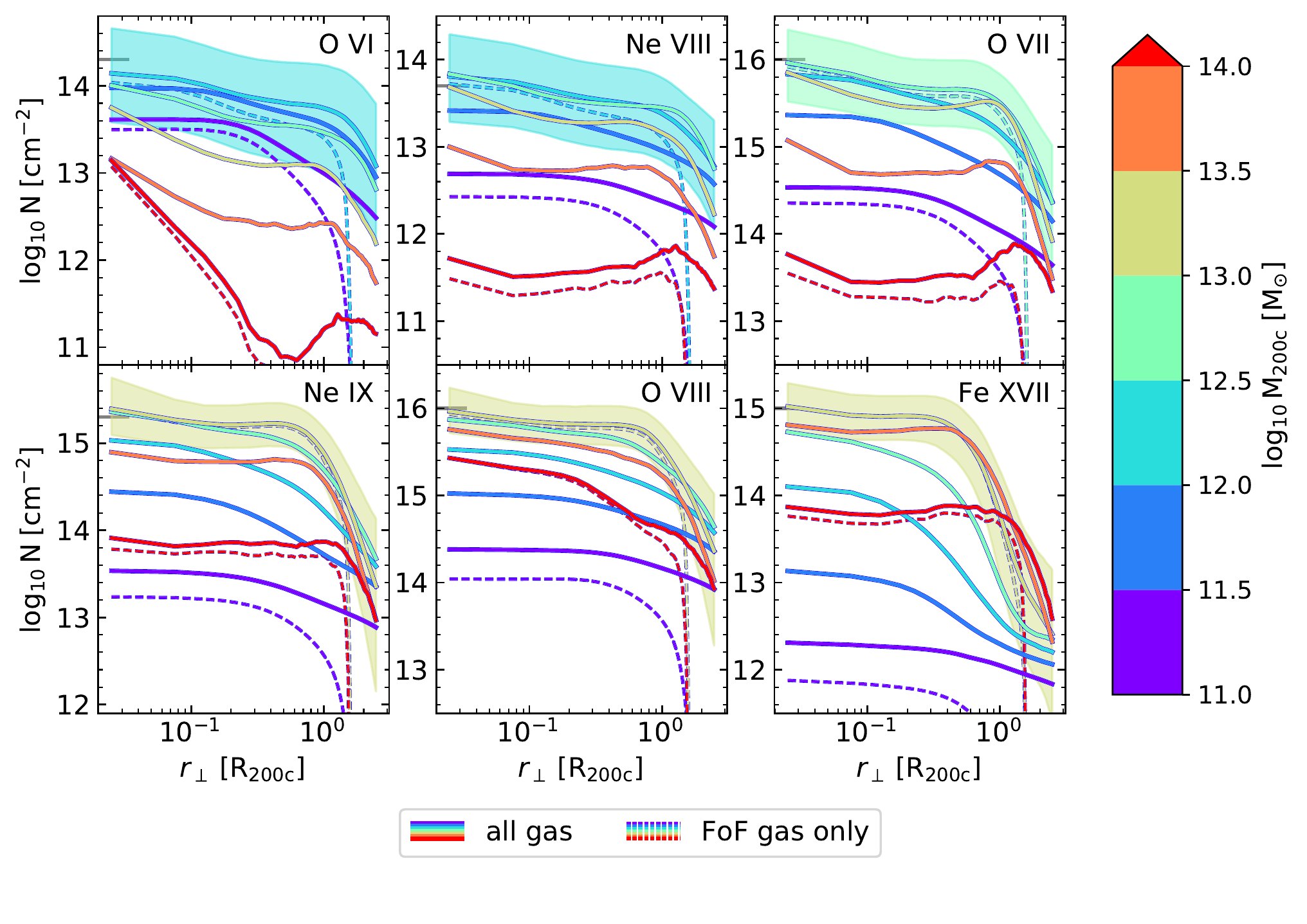}
	\caption{Radial column density profiles for the different ions around central galaxies in haloes of different masses at $z=0.1$ in the {\eagle} simulation. The median column density at different impact parameters is shown, with shaded regions showing the $10^{\txn{th}}$--$90^{\txn{th}}$~percentile range for a halo mass bin with a high central peak column density. Column densities are measured in the single $6.25 \cMpc$ slice of the box that contains each central galaxy's centre of mass. The horizontal, grey ticks on the y-axis indicate roughly where breaks in the CDDFs for the different ions occur (see Fig.~\ref{fig:cddfs}). Solid lines include all gas from a given slice, while dashed lines show absorption around haloes in each slice coming only from gas within haloes (FoF group or within $\Rvir$) in the same $\Mvir$ range. For legibility, we only show the halo contributions for three halo mass bins: the mass yielding the maximum median column density for each ion, and the highest and lowest masses.}
	\label{fig:radprof_Rvir}
\end{figure*}

The CDDFs we examined are predictions for finding absorption lines at random in the spectra of background sources. However, it is also common to look for absorption close to galaxies specifically, especially in stacking studies. Therefore, we consider what we might find if we looked for absorption around haloes of different masses. For this, we use the radial profiles computed as described in \S\ref{sec:radprofmeth}. 
The column density radial profiles are shown in Fig.~\ref{fig:radprof_Rvir}. The solid lines show absorption by all gas in the same $6.25 \cMpc$ slice as the halo centres, while the dashed lines show absorption only by gas in a halo (FoF group or otherwise inside $\Rvir$) with $\Mvir$ in the matched halo mass range. In principle, this means that single-halo profiles might include absorption by gas in different haloes of similar mass, but the fact that the dashed lines for all ions drop off sharply at the same $r_{\perp} \approx 1.5 \Rvir$ indicates that this effect is negligible, at least for the median profiles.

We see a clear pattern: the median column density increases with halo mass until it reaches a peak, which corresponds to the halo mass where the relative contribution to the CDDF (at higher column densities) peaks in Fig.~\ref{fig:cddfsplits_abs}. This again supports the idea that the column densities of these haloes are largely driven by the halo virial temperature. 

We also note more qualitative trends. 
Column densities at large distances ($\gtrsim 2 \Rvir$) increase considerably less with halo mass than central column densities do. At halo masses beyond the peak, the median column density declines and the profile flattens within $\Rvir$, even having a deficit of absorption somewhere in the range $\sim 0.1$--$1 \Rvir$ compared to $\sim \Rvir$ for the lower energy ions in the largest halo mass bins. We will examine the causes of these trends in \S\ref{sec:3dprof}, using (3D) radial profiles of the halo gas properties.

The fraction of absorption caused by gas in the haloes (dashed curves) also shows a clear trend: the halo contributions are largest in halo centres, and for haloes at the mass where the median column density peaks. Halo contributions drop as typical column densities decrease, towards both higher and lower halo masses. 

Comparing to the column densities where breaks in the CDDFs occur (long horizontal grey ticks on the left), we see the absorption in the high column density tails of the overall distribution, at column densities above those indicated, comes from absorbers that are stronger than typical for haloes of any mass. Therefore, the low occurrence of stronger absorbers does not simply reflect the low volume density of haloes in the ion's preferred mass range, it is also due to the fact that they are relatively high column density absorbers for such haloes. Note that the scatter here includes both interhalo and intrahalo scatter, so it is possible that such absorbers are more common in a subset of haloes at some halo mass. 



In observations, halo masses can be uncertain, especially around low-mass galaxies. Therefore, we also show show radial profiles in bins of central galaxy stellar mass, 
as a function of projected distance to the galaxy centre of mass. Unlike before, we obtain the median and scatter in column density in bins of physical impact parameter. We only consider central galaxies here. The profiles are shown in Fig.~\ref{fig:radprof_obs}.


We use bins spaced by $0.5 \dex$ in stellar mass. However, we do not use a separate bin for $\Mstellar = 10^{11.5}$--$10^{11.7} \Msun$, since this bin would only contain six galaxies. Instead, we group all $\Mstellar > 10^{11} \Msun$ galaxies into one bin.
We use at most 1000~galaxies (randomly selected) for the profiles for each $\Mstellar$ bin, which is relevant for galaxies with $\Mstellar < 10^{10.5} \Msun$. 

\begin{figure}
	\includegraphics[width=\columnwidth]{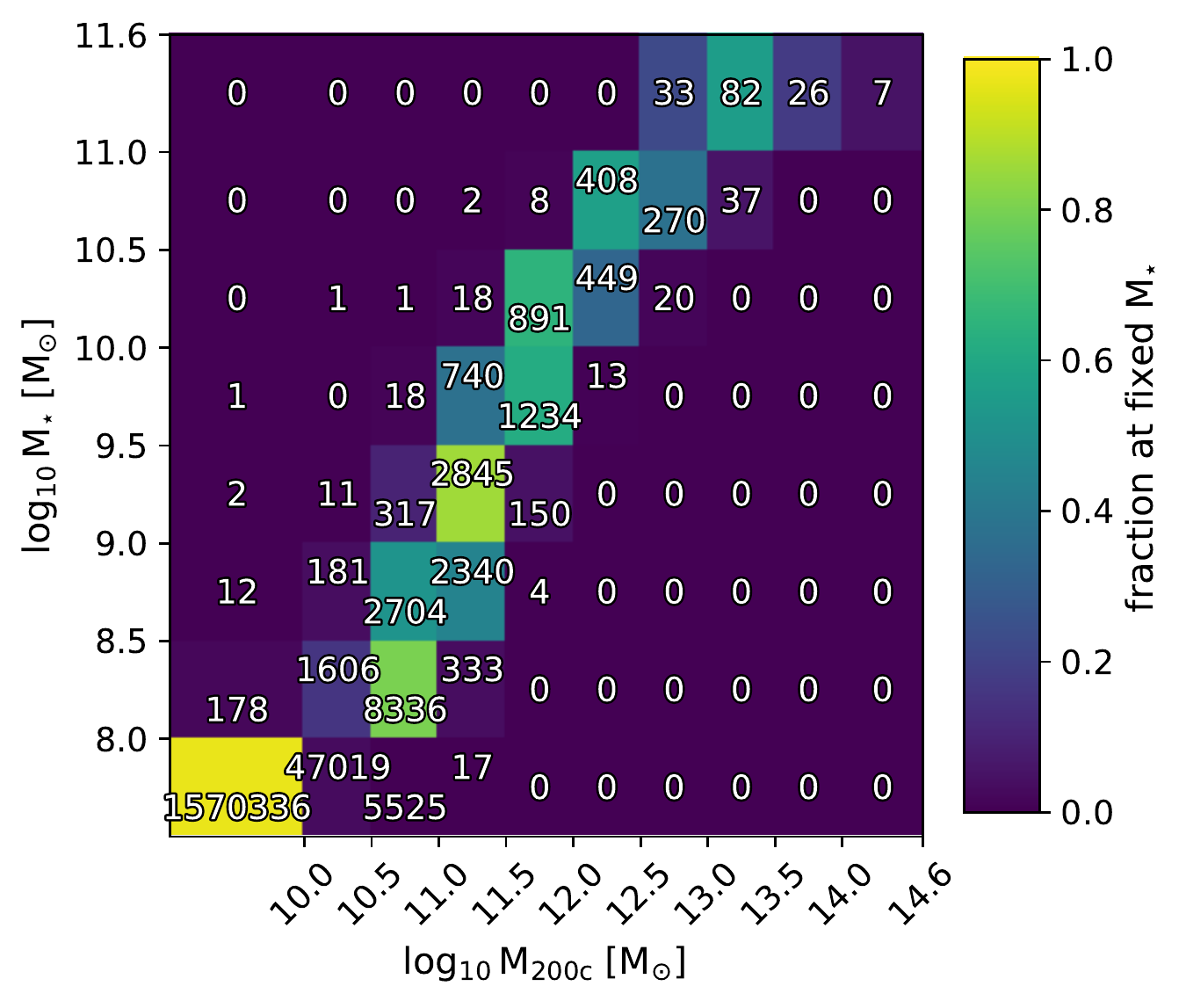}
	\caption{The {\eagle} $\Mstellar$-$\Mvir$ relation for central galaxies at $z=0.1$ shown as a confusion matrix, demonstrating how our $\Mstellar$ and $\Mvir$ bins compare. The number of central galaxies in each $\Mstellar, \Mvir$ bin is shown. The lowest mass bins have no lower limit, and include galaxies and haloes that are unresolved in the simulation. The colours show what fraction of galaxies in each $\Mstellar$ bin are in haloes in different $\Mvir$ bins.}
	\label{fig:confmatrix}
\end{figure}

\begin{figure*}
	\includegraphics[width=\textwidth]{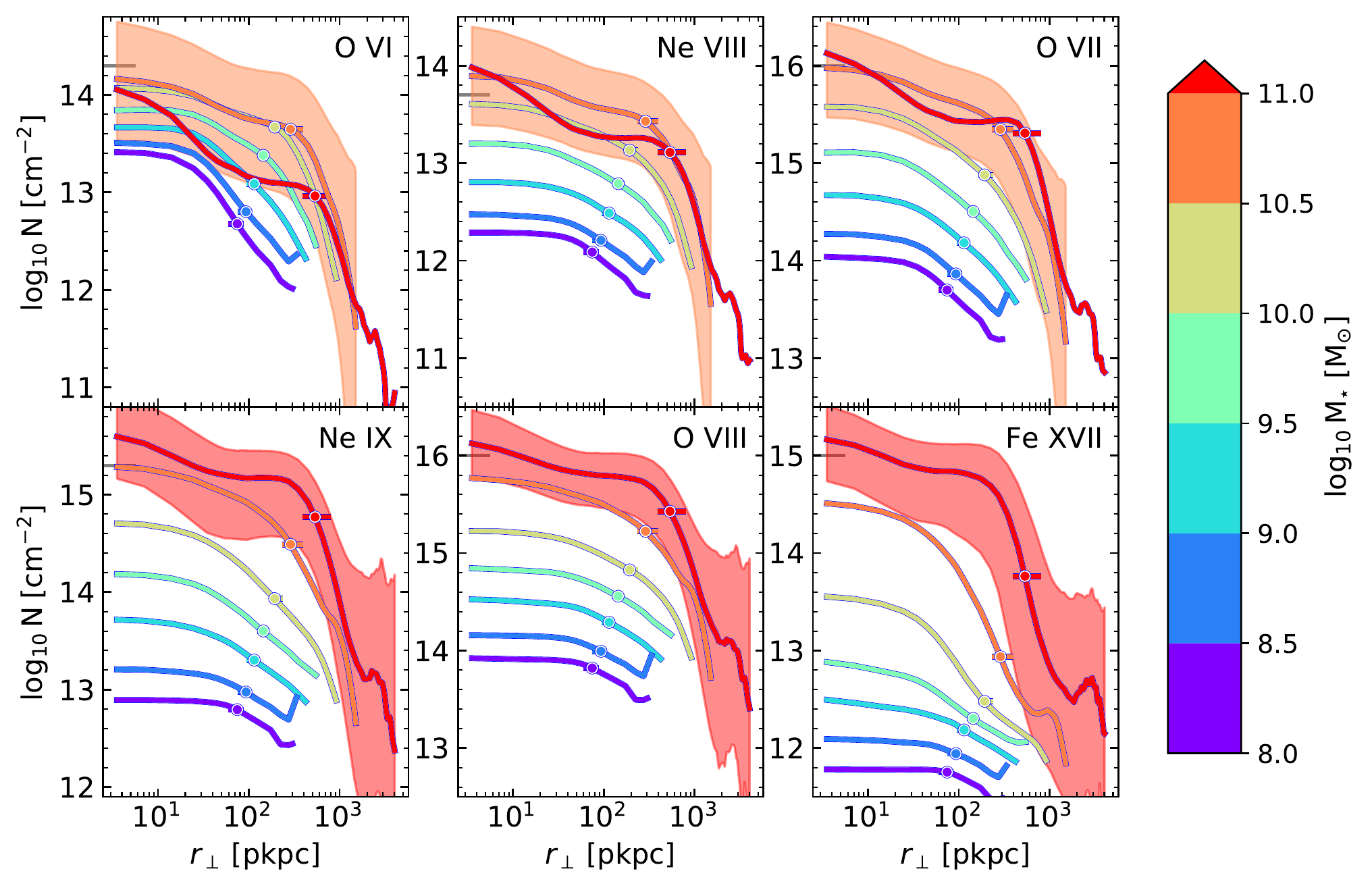}
	\caption{Median column density as a function of impact parameter (physical distance) and stellar mass of the central galaxy at $z=0.1$ in the {\eagle} simulation. Different panels are for different ions. Long ticks on the y-axes (left-hand panel) indicate approximately where the breaks in the CDDFs occur (Fig.~\ref{fig:cddfs}). The profiles extend out to $3 \Rvir$ of the 99$^{\txn{th}}$ percentile of the $\Mvir$ distribution of each $\Mstellar$ bin (see Fig.~\ref{fig:confmatrix}). Points on each curve mark the median virial radius in each $\Mstellar$ bin, and horizontal lines show the central 80~per~cent of virial radii in those bins. Shaded regions show the $10^{\txn{th}}$--$90^{\txn{th}}$~percentile range for a stellar mass bin with a high central column density.}
	\label{fig:radprof_obs}
\end{figure*}

Fig.~\ref{fig:confmatrix} shows the stellar-mass-halo-mass relation for {\eagle} central galaxies as a  `confusion matrix'. It shows how the stellar mass bins we use in this section map onto the halo mass bins used in the rest of the paper. According to \citet{eagle_paper}, the galaxy stellar mass function is converged with resolution down to stellar masses $\Mstellar \approx 2 \times 10^{8} \Msun$, though for other properties, such as star formation rates, the lower limit is $\sim 10^{9} \Msun$ or somewhat more massive. In the lowest halo mass bin we considered ($\Mvir = 10^{11}$--$10^{11.5} \Msun$), we do find a substantial contribution from $\Mstellar < 10^{9} \Msun$ galaxies, but most central galaxies in this halo mass bin have $\Mstellar > 2 \times 10^{8} \Msun$.
Fig.~\ref{fig:confmatrix} also shows that the highest three halo mass bins will have little impact outside the largest stellar mass bin, and the very largest halo mass bin contains too few galaxies to contribute significantly for any stellar mass. 

Fig.~\ref{fig:radprof_obs} shows the same main trends of column density with $\Mstellar$ in physical distance units as Fig.~\ref{fig:radprof_Rvir} showed for normalized distance and halo mass. However, for \ion{O}{viii}, \ion{Ne}{ix}, and \ion{Fe}{xviii}, the fact that the highest stellar-mass bin contains mostly $\Mvir < 10^{13.5} \Msun$ haloes means we do not see a decrease in column density towards the highest stellar masses. The overall correspondence implies that, with sufficiently sensitive instruments and large enough sample sizes, the column density trends with halo mass should be observable.

Note that the innermost parts of these profiles ($\mathrm{r}_{\perp} \ll 10 \pkpc$) might be less reliable, where they probe the central galaxy or gas close to it. 
\citet{wijers_schaye_etal_2019} found that including or excluding star-forming gas altogether has very little effect on the CDDFs of \ion{O}{vii} and \ion{O}{viii}. Indeed, in making the column density maps, we assumed all this gas had a temperature of $10^{4} \K$, too cool for these ions at these high densities (Fig.~\ref{fig:Tvir}). However, in reality, a hot phase in the ISM may contain such ions. 
On the other hand, in {\eagle}, there is hot, low-density gas in halo centres in (\S\ref{sec:3dprof}), which may have been directly heated by star formation or AGN feedback, and might cause absorption that is sensitive to the adopted subgrid heating temperatures associated with these processes.

\begin{figure*}
	\includegraphics[width=\textwidth]{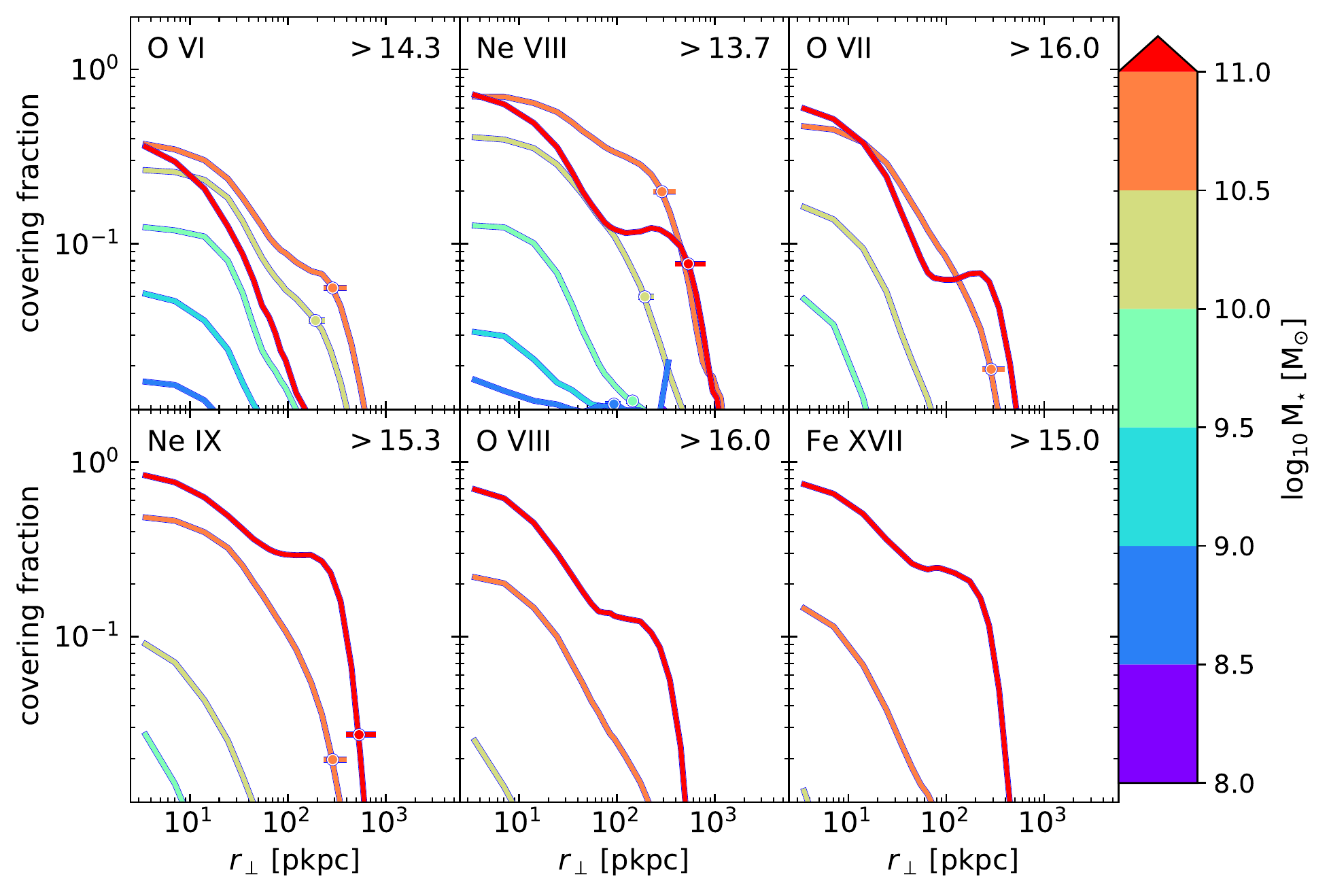}
	\caption{Covering fractions around central galaxies of different stellar masses in {\eagle} at $z=0.1$ as a function of physical impact parameter. The covering fraction is the fraction of sightlines with column densities larger than a threshold value at each impact parameter. The threshold column densities, in $\log_{10} \pcmsq$ units, are shown in the panels. These covering fractions are for column densities equal to the respective CDDF breaks. Points on each curve mark the median virial radius in each $\Mstellar$ bin, and horizontal lines show the central 80~per~cent of virial radii in those bins. (Some are outside the range of the plot.)}
	\label{fig:radprof_fcov_break}
\end{figure*}

Now that we have examined median column densities, we consider the extreme end of the distribution: how much absorption we find at very high column densities as a function of impact parameter. By very high column densities, we mean those above the CDDF breaks in Fig.~\ref{fig:cddfs}. The values are listed in Table~\ref{tab:fcov}, and shown in Fig.~\ref{fig:radprof_fcov_break}. 
For a number of ions, the absorption lines we analyse here (Table~\ref{tab:lines}) (start to) become saturated at these column densities (Fig.~\ref{fig:N-EW}). For unresolved X-ray lines, these covering fractions might therefore be difficult to measure observationally as long as the widths of the absorption components remain unresolved, which is expected even for the Athena X-IFU \citep[][fig.~4]{wijers_schaye_etal_2019}.     


\begin{table}
\caption{Threshold column densities ($\log_{10} \, \mathrm{N} \; [\txn{cm}^{-2}]$) used for covering fractions for the different ions we show in Figs.~\ref{fig:radprof_fcov_break} and~\ref{fig:radprof_fcov_obs}. The EW cited in the top line of the table is an observer-frame (redshifted) value. It was converted into column densities using the lines from Table~\ref{tab:lines} at $z=0.1$, and the best-fitting $b$ parameters from Table~\ref{tab:bfit} (blue, dot-dashed lines in Fig.~\ref{fig:N-EW}). The sources for the data are described in \S\ref{sec:obs}.}
\label{tab:fcov}
\centering
\begin{tabular}{l l l l l l l}
\hline
&
\multicolumn{1}{c}{\ion{O}{vi}} &
\multicolumn{1}{c}{\ion{Ne}{viii}} &
\multicolumn{1}{c}{\ion{O}{vii}} &
\multicolumn{1}{c}{\ion{Ne}{ix}} &
\multicolumn{1}{c}{\ion{O}{viii}} &
\multicolumn{1}{c}{\ion{Fe}{xvii}} \\
\hline	
$\mathrm{EW} = 0.18 \us \mathrm{eV}$ &	
     &      & 15.4 & 15.4 & 15.6 & 14.8 \\
HST-COS &
13.5 & 13.5     &      &      &      &      \\
CDDF break  &		
14.3 & 13.7 & 16.0 & 15.3 & 16.0 & 15.0 \\
\hline
\end{tabular}
\end{table}

In Fig.~\ref{fig:radprof_fcov_break}, we see that the covering fractions above the CDDF break typically peak close to galaxies.
However, the relatively small cross-section of these central regions means that absorption above the break in the CDDF for blind surveys is dominated by regions outside the inner $30 \pkpc$ around galaxies. We determined this from the covering fraction profiles at different $\Mstellar$, and the total CDDFs for the ions. We compared different sets of absorbers. The first set are the absorbers in the central regions. These are absorbers in the same $6.25 \cMpc$ slice of the simulation, with impact parameters $\mathrm{r}_{\perp} < 30 \pkpc$ ($< 30 \pkpc$ absorbers). The second set is similar, but contains absorbers with $\mathrm{r}_{\perp} \lesssim \Rvir$. In each stellar mass bin, we use the median $\Rvir$ of the parent haloes to define this edge. We estimate the number of absorbers above the column density breaks in the two $\mathrm{r}_{\perp}$ ranges from the covering fraction profiles.

The $< 30 \pkpc$ absorbers contain $\lesssim 10$~per~cent of the absorption above the CDDF break in the $\lesssim \Rvir$ sample, at least in $\Mstellar$ bins responsible for $>10$~per~cent of the total absorption above the CDDF breaks. For $\Mstellar$ bins responsible for less of the total absorption, the $< 30 \pkpc$ absorbers make up $\lesssim 33$~per~cent of the $\lesssim \Rvir$ absorbers (with one exception of $42$~per~cent: \ion{Fe}{xvii} around $\Mstellar = 10^{10}$--$10^{10.5} \Msun$ galaxies). Looking back to Fig.~\ref{fig:radprof_obs}, this also means that absorption above the CDDF break is indeed dominated by scatter in column densities around galaxies at larger radii, rather than typical absorption where column densities are highest.


\subsection{Halo gas as a function of radius}
\label{sec:3dprof}

\begin{figure*}
\includegraphics[width=\textwidth]{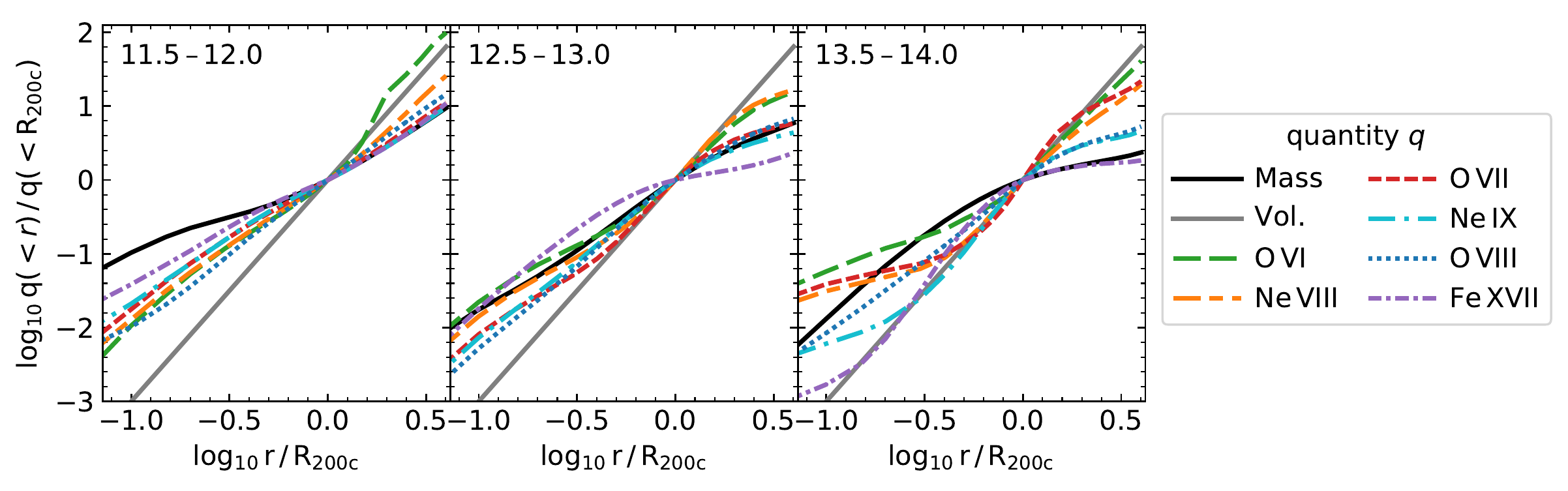}
\caption{Average cumulative volume, gas mass, and ion mass 3D profiles for the different ions in the {\eagle} simulation at $z=0.1$. We show the average enclosed fraction of each quantity, normalized for each halo by the amount enclosed within $\Rvir$. This shows that a large fraction of the ions is near {\Rvir}. The different panels are for different $\Mvir$ ranges, shown in the top left corners in units $\log_{10} \Mvir \, / \Msun$. Since the trend with halo mass is weak, we only show three halo mass bins.}
\label{fig:absloc}
\end{figure*}

In order to better understand the overall contents of haloes, as well as their absorption profiles, we examine the gas and ions in haloes as a function of (3D) radius. 
In Fig.~\ref{fig:absloc}, we show various cumulative 3D profiles for each halo. These profiles come from averaging individual haloes' radial mass distributions, after normalizing those distributions to the amount within $\Rvir$. This means that the combined profiles reflect typical (ion) mass distributions, without weighting by halo mass, baryon fraction, or halo ionization state. 


Most of the ions in these haloes lie in the outer CGM ($r \gtrsim 0.3 \Rvir$). This explains the relatively flat absorption profiles out to $\sim \Rvir$ in Fig.~\ref{fig:radprof_Rvir}. The S-shaped cumulative profiles at large halo masses explain the second peaks around $\Rvir$ in the radial profiles of some of the high-mass haloes: most of the lower energy ions, like \ion{O}{vi}, in these haloes lie in a shell at large radii, which leads to a peak in the 2D-projected column densities. The enclosed ion fractions generally fall between the enclosed mass and volume fractions. Exceptions are lower ions in the inner CGM of high-mass haloes. Also, \ion{Fe}{xvii} is more centrally concentrated than the other ions and gas overall, as Fig.~\ref{fig:radprof_Rvir} also showed. We will discuss this in more detail later. The high spike in \ion{O}{vi} mass at large radii in low-mass haloes is not present in a small, random sample of individual halo \ion{O}{vi} profiles, and is therefore not a typical feature for this halo mass.

\begin{figure*}
	\includegraphics[width=\textwidth]{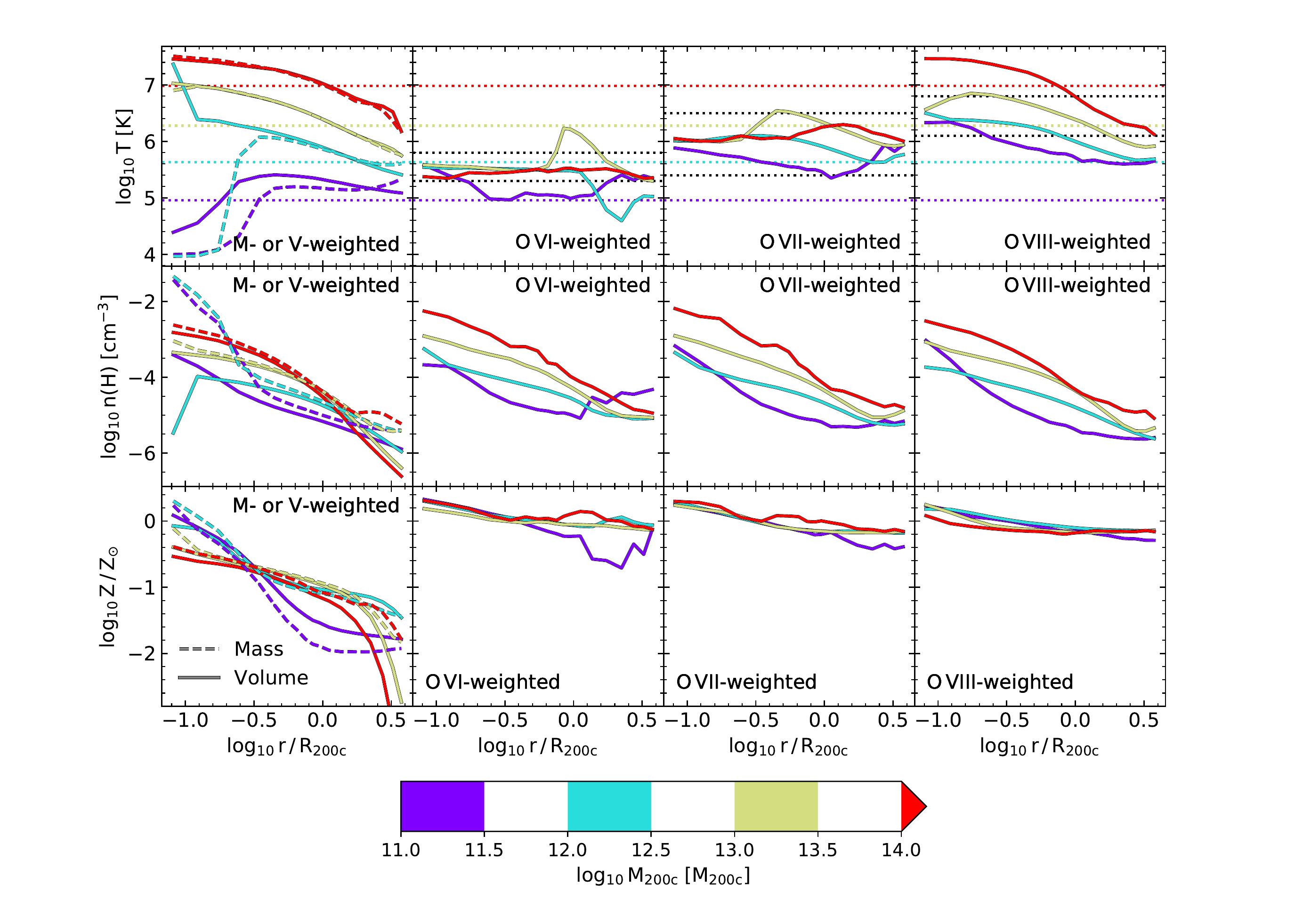}
	\caption{Median temperatures, densities, and metallicities of the halo gas in the {\eagle} simulation at $z=0.1$. The left-hand panels show mass-weighted (dashed) and volume-weighted (solid) medians, while the other columns show medians weighted by the different ions. 
To obtain the medians, we first computed temperature, density, and metallicity histograms of all gas in each radial bin, weighted by mass, volume, or ion mass, for each halo. We then normalized these histograms by the total weighting quantity within $\Rvir$, and averaged the normalized histograms of the haloes in each mass bin. We computed the medians from these stacked histograms. We used the same radial bins in $\Rvir$ units for all haloes.
The dotted lines show $\Tvir$ (eq.~\ref{eq:Tvir}) for the median halo mass in each bin, in colours matching the full range. 
The densities were calculated as mass densities and converted to hydrogen number densities assuming a primordial hydrogen mass fraction of 0.752. The metallicities are oxygen mass fractions (\textsc{SmoothedElementAbundance/Oxygen}), normalized to a 0.00549 solar mass fraction \citep{allendeprieto_lambert_asplund_2001}. The neon and iron mass fractions (not shown) follow similar radial and halo mass trends, though the values differ somewhat. We show a subset of halo masses for legibility. The lowest-radius bin contains all mass/volume/ions within $0.1 \Rvir$. The ion temperatures are mostly set by CIE, while their densities roughly follow the volume-weighted density profile (i.e., the hot gas profile). The ion-weighted metallicities are biased high compared to the mass- and volume-weighted metallicities.}
	\label{fig:3dprof}
\end{figure*}


In Fig.~\ref{fig:3dprof}, we show mass- and volume-weighted median temperature, density, and metallicity profiles (left-hand column). For the temperature profiles, the dotted lines show the simplest prediction: $\Tvir$, as calculated from eq.~\ref{eq:Tvir}. The colours match the median halo mass in each bin.  The profiles show a general rising trend with halo mass, with temperatures at $\Rvir$ matching $\Tvir$ reasonably, and following the $\Tvir$ trend. However, the temperature clearly decreases with radius in most cases. The exceptions are the inner parts of the profiles for low-mass haloes, for which the volume-weighted median temperatures can be much higher than the mass-weighted ones. The haloes are in fact multiphase, with cool gas at $\sim 10^4 \us \mathrm{K}$, some gas at $\sim 10^5 \us \mathrm{K}$, and the hotter volume-filling phase. Sharp transitions in the median profiles occur when the median switches from one phase to another. The multiphase nature is particularly prominent at low mass ($\Mvir \ll 10^{13} \Msun$). 

The gas density also decreases with radius. It is generally higher in higher mass haloes around $\Rvir$, but at larger radii it also drops much faster than in lower mass systems. Volume-weighted densities can be considerably lower than mass-weighted densities, reflecting the multiphase nature of the gas. In the centres of low-mass haloes (especially at smaller radii than shown), median temperatures tend to increase as densities drop. This is likely the result of stellar and/or AGN feedback heating some gas in the halo centres, increasing its temperature and volume. These large volumes for particles centred close to the halo centres may dominate the volume-weighted stacks. Indeed, in this regime, it seems gas at a few discrete temperatures sets these trends (including at $10^{7.5} \us \mathrm{K}$, the heating temperature for stellar feedback), and medians from a few randomly chosen individual galaxies do not show this trend. 

The metallicities also tend to decline with radius, with larger differences in lower mass haloes. Evidently, the metals are better mixed in high-mass haloes, likely because star formation and the accompanying metal enrichment tend to be quenched in these systems, while processes such as mergers and AGN feedback continue to mix the gas. The mass-weighted metallicities are higher than the volume-weighted ones in the inner parts of the halo, while in the outer halo and beyond $\Rvir$, the differences depend on halo mass. For high-mass haloes, the median metallicity in the volume-filling phase drops sharply somewhat outside $\Rvir$. However, the scatter in metallicity at large radii is very large, particularly at high masses.


Fig.~\ref{fig:3dprof} also shows the corresponding ion-mass-weighted temperature,  density and metallicity profiles as a function of halo mass. The coloured, dotted lines in the temperature profiles show $\Tvir$ at the median mass in each bin. The black, dotted lines indicate the CIE temperature range for each ion. Abrupt temperature changes are again a result of the median switching between different peaks in the temperature distribution of multiphase gas.

The ion-weighted temperature mostly follows the CIE temperature range (black dotted lines), rather than the $\Tvir$ range for that set of haloes (coloured dotted lines) within $\Rvir$. Higher ion-weighted temperatures do occur, but in radial regions that contain relatively few ions. 
Ion-weighted temperatures below the CIE range mainly occur at radii $\gtrsim \Rvir$, where ion-weighted hydrogen number densities reach the $\sim 10^{-5} \pcc$ regime where photoionization becomes important and lower temperature gas can become highly ionized (Fig.~\ref{fig:Tvir}). For \ion{O}{vi}, ions at lower temperatures do persist at smaller radii, within the 90~per~cent scatter of the ion-weighted temperature, and especially in  lower mass haloes.  


The ion-weighted densities in the CGM reflect the halo's physical properties: they follow the halo gas density distribution, and in particular, the volume-filling hot phase in cases where the mass- and volume-weighted gas distributions differ. They are however biased to the temperature ranges favoured by CIE and the metallicities are biased high compared to the mass- and volume-weighted values shown in Fig.~\ref{fig:3dprof}.  

These temperature and density effects may explain the `shoulders' around $\Rvir$ in the absorption profiles at some halo masses seen in Fig.~\ref{fig:radprof_Rvir}. This phenomenon occurs at halo masses around or above those for which $\Tvir$ matches the CIE peak for each ion. Since the temperature of the CGM decreases with radius, the ions will preferentially be present at larger radii in higher mass haloes, where they form a `shell', which produces large column densities at projected radii close to the shell radius. This is visible in Fig.~\ref{fig:absloc}, where lower ions in higher mass haloes have S-shaped cumulative ion mass distributions, with relatively little ion mass in the too-hot inner CGM.  

However, around $\Rvir$, as the halo-centric radius increases, the effect of the declining gas temperature is countered by photoionization of the cold phase, which also starts to become important around $\Rvir$. This drives the preferred temperatures of the ions down with radius, along with the gas temperature. The `shoulders' are strong in \ion{Fe}{xvii} and \ion{Ne}{ix} profiles; these ions have the highest ionization energies (Table~\ref{tab:ions}) and are photoionized at lower densities (Fig.~\ref{fig:Tvir}) than the others. 

The sharp drops in the absorption profiles at large radii in Fig.~\ref{fig:radprof_Rvir} may also be explained by these halo properties: the gas density drops outside $\Rvir$, and more sharply for higher halo masses. Similarly, the gas metallicity drops rapidly around these radii in the high-mass haloes. The differences between the ions seem to be consistent with the more easily photoionized ones producing more absorption in the cooler, lower density gas around $\Rvir$. However, the way the CIE temperature range lines up with gas temperatures depends on both the ion and the halo mass, so ion and mass trends are difficult to disentangle.

Though \ion{Fe}{xvii} seems like an outlier in Fig.~\ref{fig:absloc}, in that it is more concentrated in halo centres than the (total) gas mass, this does fit into these trends: the outskirts of most haloes at the masses we consider are simply too cool for this ion. However, in $\Mvir > 10^{14} \Msun$ haloes, which have $\Tvir$ above the preferred range of \ion{Fe}{xvii}, the absorption does extend out to $\Rvir$, albeit at lower column densities.  

We note that the sharp drops in mass- and volume-weighted median metallicity are not in contradiction with the flat ion-mass-weighted metallicities outside $\Rvir$: there is very large scatter in the metallicity at large radii, and metal ions will preferentially exist in whatever metal-enriched gas is present.

\section{Detection prospects}
\label{sec:obs}

To predict what might be observable with different instruments, we first estimate the minimum observable column densities for the different ions. We use column density thresholds which correspond roughly to the detection limits of current, blind (UV) and upcoming (X-ray) surveys. We then use these limits to predict how many absorbers and haloes we should be able to detect per unit redshift, and out to what impact parameters we can expect to find measurable absorption. 

\subsection{Detection limits for different instruments}
\label{sec:detlim}

For the X-ray lines, we estimate the minimum detectable column density from the minimum detectable EW and the $b$-parameters from Table~\ref{tab:bfit}, assuming a single Voigt profile (or a doublet, for \ion{O}{viii}). 
Since these minima depend not just on the instrument, but on the observations (e.g., exposure time, background source flux and spectrum), we take the minimum EWs from the instrument science requirements, which assume a planned observing campaign as well as instrument properties. These are observer-frame minima, which we covert to rest-frame minimum EWs assuming $z=0.1$, the redshift we assume throughout this work.

We focus on what should be detectable with the X-IFU on the planned Athena mission. Here, weak lines around 1~keV should be detectable at $5 \sigma$ significance at observer-frame EWs of 0.18~eV. This is for 50~ks exposure times and a quasar background source with a 2--10~keV flux of $10^{-11} \us \txn{erg}\,\txn{cm}^{-2}\txn{s}^{-1}$ and a photon spectral index $\Gamma=1.8$ \citep{Athena_2017_11}. 
Blind detections of pairs of  \ion{O}{vii} and \ion{O}{viii} absorption lines should be possible at lower EWs than this, at least against bright gamma-ray burst background sources \citep{walsh_mcbreen_etal_2020}. We convert these minimum EWs to minimum column densities using the best-fitting relations shown in Fig.~\ref{fig:N-EW} (blue, dot-dashed lines) and Table~\ref{tab:bfit}. The minimum column densities are shown in Table~\ref{tab:fcov}.

A minimum EW estimate of 0.18~eV is on the rough side for \ion{Fe}{xvii} and \ion{Ne}{ix}, since the oxygen lines have been the main focus of WHIM and hot CGM detection plans. These lines are at different energies, so the energy-dependence of the sensitivity of the instrument and the spectrum of the background source and Galactic absorption, mean that $0.18 \us \mathrm{eV}$ might not be a fully appropriate minimum EW for \ion{Ne}{ix} and \ion{Fe}{xvii}. Besides that, the relation between column density and EW has enough scatter above the minimum observable EW that it does not quite translate into a unique minimum column density, but it is an acceptable approximation in this regime \citep[appendix~B]{wijers_schaye_etal_2019}.


We also make predictions for the proposed Arcus \citep{smith_abraham_etal_2016_arcus, brenneman_smith_etal_2016} mission, and the X-ray Grating Spectrometer (XGS) on the proposed Lynx mission \citep{lynx_2018_08}. For Arcus, we assume a minimum detectable EW of $4 \mA$ (for $5 \sigma$ detections). This is based on bright AGN background sources, which were selected to have a high flux between $0.5$ and $2$~keV \citep{brenneman_smith_etal_2016}, and exposure times $<500$~ks \citep{smith_abraham_etal_2016_arcus}. At least 40 blazars matching the brightness requirements are known \citep{smith_abraham_etal_2016_arcus}. These estimates are based only on \ion{O}{vii} and \ion{O}{viii} (and \ion{C}{vi}), so this minimum EW may not apply to the \ion{Ne}{ix} and \ion{Fe}{xvii} lines at smaller wavelengths. 

Note that Arcus not only aims to find weaker absorption lines than the Athena X-IFU, it is also meant to characterize them in more detail using its higher spectral resolution. Arcus has a $\approx6$--$8\times$ higher spectral resolution than Athena at the wavelengths of \ion{O}{vii} and \ion{O}{viii} at $z=0.1$, which is sufficient \citep[$\approx 120$--$150 \kmps$,][]{smith_abraham_etal_2016_arcus} to determine if absorbers are associated with $\Lstar$ galaxy haloes that have typical virial velocities of $150$--$300 \kmps$, while the Athena X-IFU's resolution \citep[$\approx 900$--$1000 \kmps$,][]{Athena_2018_07} would be insufficient to determine if absorbers belong to individual galactic haloes.  

For the Lynx XGS, the requirement is a detectable EW of $1 \mA$ for \ion{O}{vii} and \ion{O}{viii} \citep{lynx_2018_08}. This applies to 80 bright AGN background sources in a 5~Ms survey, focussed on detecting the CGM of $\sim \Lstar$ galaxies in absorption.

For Arcus and Lynx, we therefore limit our predictions to \ion{O}{vii} and \ion{O}{viii}. The minimum EWs for Arcus translate to column densities of $10^{15.3}$ and $10^{15.6} \pcmsq$ for \ion{O}{vii} and \ion{O}{viii}, respectively. For Lynx, the values are, respectively, $10^{14.6}$ and $10^{14.9} \pcmsq$. 

For the FUV ions, we choose column densities based on what is currently observed with HST-COS. We base estimates on observed column densities and upper limits, and column densities used for covering fractions by observers. We use the data of \citet{tumlinson_thom_etal_2011} and \citet{prochaska_weiner_etal_2011} for \ion{O}{vi}, and of \citet{burchett_tripp_etal_2018} and \citet{meiring_tripp_etal_2013} for \ion{Ne}{viii}.  Note that our limits are for $z=0.1$ for consistency, but the EUV line we discuss for \ion{Ne}{viii} is only observable at higher redshifts. We explore the redshift evolution of the absorption in Appendix~\ref{app:zev}. 

\subsection{Halo-detection rates}
\label{sec:numhaloes}

Based on the CDDFs for gas coming from haloes of different masses, we can estimate how many haloes of different masses should be detectable with the Athena X-IFU over a given total redshift path $\diff z$. Here, it does matter if we use the CDDFs based on the halo-projection method (such as in Fig.~\ref{fig:cddfsplits_abs}), or the pixel-attribution method, where we base the CDDF on column density maps including all gas, but only counting pixels with impact parameter $\mathrm{r}_{\perp} \leq \Rvir$, and that are not closer to another halo in  $\mathrm{r}_{\perp} \, / \Rvir$ units (see \S\ref{sec:radprofmeth}).  


Using estimated minimum column densities of $10^{15.4} \pcmsq$ for \ion{O}{vii} and $10^{15.6} \pcmsq$ for \ion{O}{viii} for the Athena X-IFU, we expect to find, in total, 2.3 \ion{O}{vii} absorbers and 1.0 \ion{O}{viii} absorbers per unit redshift. Of those, 46 and 63~per~cent are within $\Rvir$ of a central galaxy with $\Mvir > 10^{11} \Msun$, respectively. For \ion{O}{vii}, 41~per~cent of all the absorbers is attributed to haloes with $\Mvir = 10^{12}$--$10^{13.5} \Msun$, and for \ion{O}{viii}, 53~per~cent comes from $\Mvir = 10^{12.5}$--$10^{13.5} \Msun$ haloes. (Since the halo-projection CDDFs do not add up to the all-gas CDDF, we do not attempt to derive such fractions from the halo-projection CDDFs.)

We also estimate the total density along lines of sight of observable absorbers coming from haloes of different masses. Here, both the halo-projection CDDFs and the pixel-attribution CDDFs are reasonable starting points.
We expect to find 
0.30 (0.34), 0.39 (0.61), and 0.26 (0.42) \ion{O}{vii} absorbers per unit redshift with the Athena X-IFU blind survey based on the pixel-attribution CDDFs (based on the halo-projection CDDFs), in haloes of $\Mvir = 10^{12}$--$10^{12.5}$, $10^{12.5}$--$10^{13}$, and $10^{13}$--$10^{13.5} \Msun$, respectively.
For \ion{O}{viii}, we similarly expect 
0.21 (0.20) and 0.31 (0.37) absorbers per unit redshift in $\Mvir$ bins of $10^{12.5}$--$10^{13}$ and $10^{13}$--$10^{13.5} \Msun$, respectively. This is assuming most of the redshift path searched is close to $z=0.1$, which is the redshift of the {\eagle} snapshot we extracted the CDDFs from. We compare the absorption to $z=0.5$ in Appendix~\ref{app:zev}.    

Therefore, halo absorbers seem to be somewhat rare, but in a search for \ion{O}{vii} and \ion{O}{viii} absorbers against 100~BLLacs and 100~gamma ray bursts \citep{Athena_2017_11}, we can reasonably expect to find quite a few of these absorbers. There are differences between the different determinations of halo absorption, but they are mostly not too severe. We have also assumed that a single limiting column density is a good estimate for detectability of these absorbers. We can see from Fig.~\ref{fig:N-EW}, and fig.~B1 of \citet{wijers_schaye_etal_2019}, that this should be a reasonable approximation for the limiting-case absorption lines.

\subsection{Extent of detectable absorption}
\label{sec:detext}

\begin{figure*}
	\includegraphics[width=\textwidth]{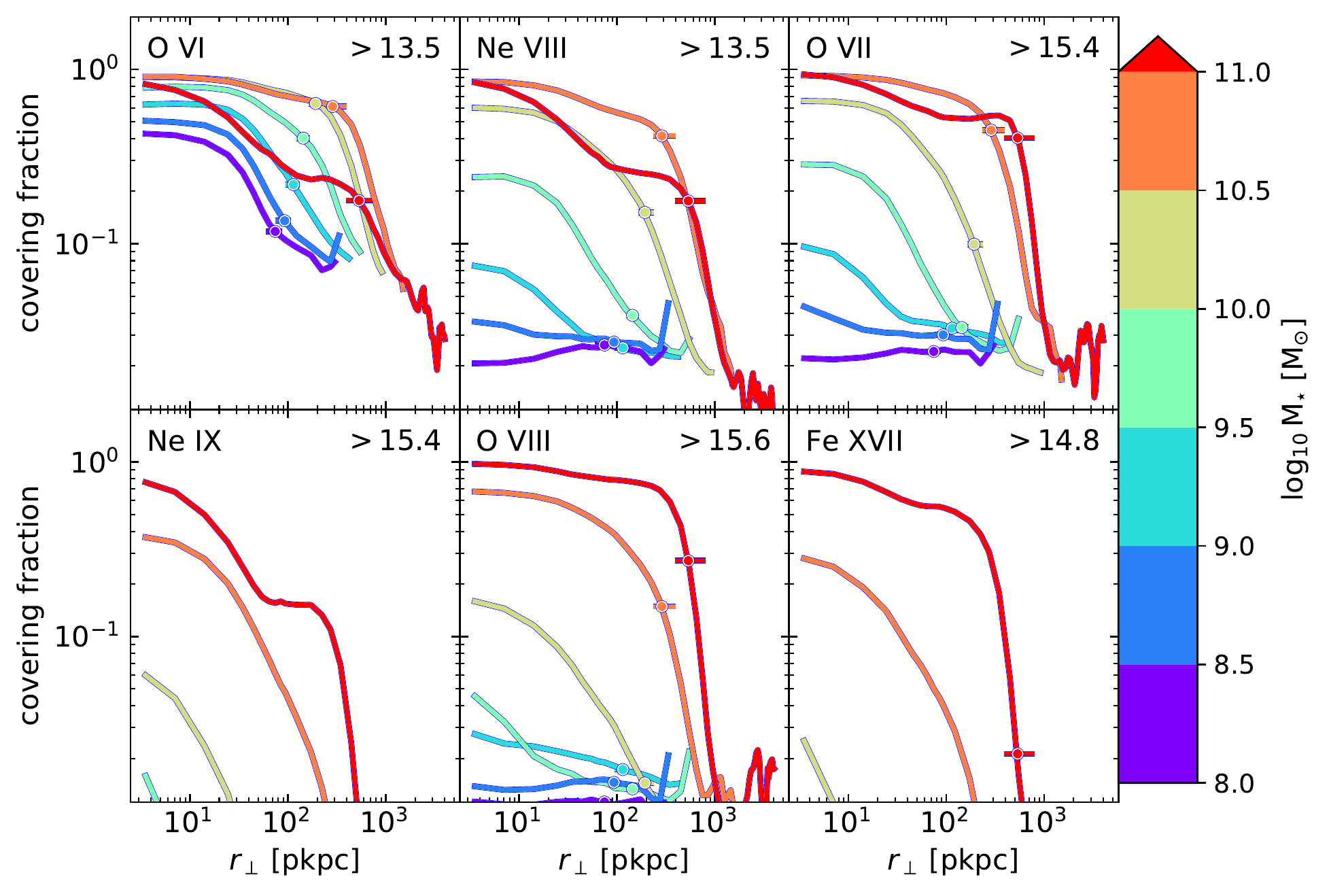}
	\caption{Covering fractions around central galaxies of different stellar masses in {\eagle} as a function of physical impact parameter at $z=0.1$. The covering fraction is the fraction of sightlines with column densities larger than a threshold value at each impact parameter. The threshold column densities, in $\log_{10} \pcmsq$ units, are shown in the panels. Points on each curve mark the median virial radius in each $\Mstellar$ bin, and horizontal lines show the central 80~per~cent of virial radii in those bins. (Some are outside the range of the plot.) These covering fractions are for optimistic HST-COS detection limits (\ion{O}{vi}, \ion{Ne}{viii}) and estimated Athena X-IFU limits (X-ray lines). The Athena limits are for observer-frame EWs of 0.18~eV (0.16~eV rest frame). That is the expected minimum for $5 \sigma$ detections, assuming 50~ks exposure times and a quasar background source with a 2--10~keV flux of $10^{-11} \us \txn{erg}\,\txn{cm}^{-2}\txn{s}^{-1}$ and a photon spectral index $\Gamma=1.8$ \citep{Athena_2017_11}.}
	\label{fig:radprof_fcov_obs}
\end{figure*}

In Fig.~\ref{fig:radprof_fcov_obs}, we investigate how far from halo centres this absorption occurs: we show the fraction of absorbers at different radii above the estimated minimum observable column density for the Athena X-IFU. Those minima are indicated in the panels. For \ion{O}{vi} and \ion{Ne}{viii}, these are more optimistic (but achievable) minimum column densities for HST-COS, reflecting currently possible observations. Many of these have also been done. We discuss some {\eagle} data comparisons in \S\ref{sec:discussion}. For $\sim \Lstar$ galaxies, \ion{O}{vi} should be widely observable according to {\eagle}, and \ion{Ne}{viii} should be widely observed in the $\Lstar$--group mass range. The \citet{burchett_tripp_etal_2018} observations (at $z \approx 0.5$--$1.1$) report \ion{Ne}{viii} absorbers and upper limits around galaxies in a wide mass range ($\Mstellar \sim 10^{9}$--$10^{11} \Msun$), which covers a large range in covering fractions in our predictions and makes a direct comparison to our figures difficult. 

For the X-ray lines, Fig.~\ref{fig:radprof_fcov_obs} shows estimated detection limits for the Athena X-IFU. 
Detection prospects for CGM out to large distances (close to $\Rvir$) look good for \ion{O}{vii} and \ion{O}{viii}, and in a narrower mass range, for \ion{Fe}{xvii}. \ion{Ne}{ix} might, however, prove more difficult to detect at larger impact parameters. However, these limits are for blind detections, and hence conservative for targeted observations. It might thus be possible to find \ion{Ne}{ix} counterparts to absorbers in e.g., \ion{O}{viii}, or to detect weaker lines by searching at known galaxy redshifts. For most of the X-ray ions, the detection thresholds are not very far from the CDDF breaks, so small changes in sensitivity could make large differences for detection prospects, as the difference between Figs.~\ref{fig:radprof_fcov_break} and~\ref{fig:radprof_fcov_obs}  shows.

Using the median column density profiles from Fig.~\ref{fig:radprof_obs}, we find the radii where the covering fractions at the detection limits for the different instruments reach $0.5$. Note that this is the covering fraction in annuli at, not circles of, these radii. These are shown in Table~\ref{tab:detext}. 

\begin{table}
\caption{Impact parameters $r_{\perp} \,/ \pkpc$, rounded to $10 \pkpc$, where the covering fraction $f(>\mathrm{N}) = 0.5$ for different ions, galaxy masses, and column density limits in {\eagle} at $z=0.1$. The column density limits (indicated in the third row in $\log_{10} \pcmsq$) are calculated for the different ions and instruments as explained in the text. We estimate the limits for the Athena X-IFU (Athena), Arcus, and the Lynx XGS (Lynx). A dash (`-') means that the covering fraction is below 0.5 at all radii.} 
\label{tab:detext}
\begin{tabular}{c  | r r r r r r}
\hline
$\mathrm{{M}}_{{\star}}$ & \multicolumn{3}{c}{\ion{O}{vii}} & \multicolumn{3}{c}{\ion{O}{viii}} \\
$\log_{10} \Msun$ & 
\multicolumn{1}{c}{Athena} &  \multicolumn{1}{c}{Arcus} &  \multicolumn{1}{c}{Lynx} &
\multicolumn{1}{c}{Athena} &  \multicolumn{1}{c}{Arcus} &  \multicolumn{1}{c}{Lynx} \\
& 15.4 & 15.3 & 14.6 & 
   15.6 & 15.6 & 14.9 \\
\hline
10.0--10.5 &
30	& 50		& 340 &
- 	& -		& 140 \\ 
10.5--11.0 & 
260	& 310	& 680 &
50	& 50		& 560 \\
11.0--11.7 & 
460	& 540	& 930 &
410	& 410	& 910 \\
\hline
\end{tabular}
\end{table}

This confirms that with the Athena X-IFU, we should be able to find \ion{O}{vii} and \ion{O}{viii} absorption out to close to $\Rvir$ for $\Mstellar > 10^{10.5} \Msun$ galaxies.  Arcus performs similarly; we note that compared to the Athena X-IFU, it does have a much higher spectral resolution for soft X-rays. Lynx should be sensitive to much weaker absorption lines, and should therefore be able to find absorption systems beyond the virial radii of $\Mstellar > 10^{10.5} \Msun$ galaxies, and roughly up to the virial radii of $\Mstellar = 10^{10}$--$10^{10.5} \Msun$ galaxies. Indeed, one of the mission goals is to characterize the CGM of $\sim \Lstar$ galaxies using these absorption lines \citep{lynx_2018_08}.

\section{Discussion and comparison to previous works}
\label{sec:discussion}

All of the predictions reported in the previous sections are based on the {\eagle} simulation. {\eagle} reproduces a number of galaxy and diffuse gas properties, to differing degrees. At the highest halo masses we study ($\mathrm{M}_{\mathrm{500c}} \gtrsim 10^{13.5} \Msun$), there are observations available to which to compare hot gas properties of the ICM. At lower masses, the gas properties are less well constrained. 
\citet{eagle_paper} showed that {\eagle} matches the relation between $\mathrm{M}_{\mathrm{500c}}$ (X-ray) and I-band luminosity ($\sim \Mstellar$) well, but overestimates ICM gas fractions at fixed halo mass (from X-ray measurements) and soft X-ray luminosities at fixed X-ray spectroscopic temperature above $1 \us \mathrm{keV}$. 
Overall, the higher gas fractions mean we might overestimate column densities from these simulations. 



Other authors have studied UV absorption in simulations, particularly CDDFs and absorption around $\Lstar$ to group-mass haloes. 
\citet[{\eagle}]{rahmati_etal_2016} found that the \ion{O}{vi} CDDF is somewhat underpredicted at the high column density end, though uncertainties in the observations and oxygen yields may mean the difference is not severe. They found the cosmic \ion{Ne}{viii} density agreed with that measured by \citet{meiring_tripp_etal_2013}, though the measurements and comparisons are somewhat uncertain. 
   
\citet{oppenheimer_etal_2016} made a comparison of the absorption in $\sim \Lstar$ to group-mass halo zoom-in simulations using the {\eagle} simulation code to COS-Halos \citep{tumlinson_thom_etal_2011} and found similar trends with star formation rates, but overall somewhat too low \ion{O}{vi} column densities in $\sim \Lstar$ haloes. 
\citet{oppenheimer_2018_fossilAGN_cos}, using similar {\eagle}-based zooms, found that those higher \ion{O}{vi} columns could be achieved though ionization by radiation from flickering AGN, in combination with non-equilibrium ionization, but that the CDDF discrepancy is likely not resolvable this way. This radiation and non-equilibrium ionization also affects other ions. The effect on \ion{O}{vii} column densities is, however, small: they decrease by $\lesssim 0.1 \dex$.

\citet{Nelson_etal_2018_hiO} compared their IllustrisTNG data to COS-Halos survey results for \ion{O}{vi} \citep{tumlinson_thom_etal_2011}, complemented by galaxy data, and to the eCGM survey data \citep{johnson_chen_mulchaey_2015}, and found generally good agreement. For their calibrated TNG100-1 volume, the \ion{O}{vi} CDDF might be too large at high column densities.    

For \ion{O}{vi}, there are long-standing difficulties in modelling observed absorbers (in combination with lower ions) due to the uncertain ionization mechanism \citep[e.g.,][]{werk_prochaska_etal_2016}.
We find that, in {\eagle}, the \ion{O}{vi} is mostly collisionally ionized in the inner regions, but photoionized at $\gtrsim \Rvir$, and that photoionized \ion{O}{vi} is present in the inner regions of lower mass haloes.
In IllustrisTNG CIE is most important mechanism in the CGM \citep{Nelson_etal_2018_hiO}, as in {\eagle}, assuming ionization equilibrium. However, the results of \citet{oppenheimer_2018_fossilAGN_cos} demonstrate that, at least for the CGM of $\sim \Lstar$ galaxies, such equilibrium assumptions may underestimate the effect of photoionization. They used non-equilibrium ionization, and a flickering AGN as an additional ionization source, in zoom simulations otherwise using the {\eagle} code and physics. In NIHAO and VELA zoom simulations, \citet{roca-fabrega_dekel_etal_2018} found a roughly equal mix of collisionally and photoionized \ion{O}{vi} in $z=0$, $\Mvir = 10^{11}$--$10^{12.6} \Msun$ haloes. Overall, this supports the picture that observed \ion{O}{vi} does not have one single origin.
 
\citet[IllustrisTNG]{Nelson_etal_2018_hiO} also looked into \ion{O}{vii} and \ion{O}{viii} absorption in the CGM. They found column densities that peak at similar masses as we find, and ion fraction trends with halo mass similar to ours except at the lowest masses we examine, though they measure the fractions in a somewhat different gas selection. \citet{martizzi_vogelsberger_etal_2019}, also studying the TNG100-1 volume, look into the contributions of large-scale structures to the \ion{O}{vii} and \ion{Ne}{ix} CDDFs. Rather than halo contributions, they split the CDDFs into contributions from larger-scale cosmic web structures (knots, filaments, sheets, and voids), and find that absorption at high column densities mainly comes from knots and filaments, and that more collapsed structures contribute more as the column density of absorbers increases.

\section{Conclusions}
\label{sec:conclusions}

Using the {\eagle} simulation, we investigate the contents and properties of the CGM of $\Mvir > 10^{11} \Msun$ haloes, and how they are probed by \ion{O}{vi} (FUV), \ion{Ne}{viii} (EUV) and \ion{O}{vii}, \ion{O}{viii}, \ion{Ne}{ix}, and \ion{Fe}{xvii} (X-ray) line absorption at $z=0.1$. With future X-ray instruments like the Athena X-IFU, Arcus, and the Lynx XGS, we expect that some of these absorption lines can be used to study the hot CGM. The mass of this CGM phase in $\sim \Lstar$ and group-mass haloes is largely unconstrained by current observations, and differs in different cosmological simulations. Determining the mass and metal content of the hot CGM will therefore provide important constraints for our understanding of structure and galaxy formation.

For the baryons, gas, and metals in haloes, we find the following:
\begin{itemize}
\item{The CGM (non-star-forming gas) is the largest baryonic mass component within $\Rvir$ in {\eagle} haloes  at all masses we investigate ($\Mvir > 10^{11} \Msun$), and is particularly dominant at $\Mvir \gtrsim 10^{13} \Msun$ (Fig.~\ref{fig:baryinhalo}).}
\item{Within $\Rvir$, the CGM (non-star-forming gas) also contains more oxygen than the ISM (star-forming gas) for all halo masses, though differences are small at $\Mvir \lesssim 10^{12} \Msun$. However, up to $\Mvir \approx 10^{13} \Msun$, stars contain most of the oxygen ejected by earlier stellar generations that remains within $\Rvir$ (Fig.~\ref{fig:baryinhalo}).}
\item{The ions we study mainly trace CGM gas at $10^{5.5}$--$10^{7} \K$ (Figs.~\ref{fig:Tvir} and~\ref{fig:3dprof}), which constitutes a large fraction of the non-star-forming gas within $\Rvir$ in $\Mvir \approx 10^{12}$--$10^{13.5} \Msun$ haloes (Fig.~\ref{fig:baryinhalo}).}
\item{The mass ranges for which median column densities are highest (Fig.~\ref{fig:radprof_Rvir}) are in line with simple predictions comparing the virial temperature with the temperature where the ion fraction peaks in CIE (Fig.~\ref{fig:Tvir}). This is because these ions mainly trace gas at temperatures around the CIE peak in the volume-filling phase of the CGM (Fig.~\ref{fig:3dprof}). }
\item{In the inner CGM, these ions are all mainly collisionally ionized (although some \ion{O}{vi} is photoionized), but close to $\Rvir$, photoionization becomes relevant (Figs.~\ref{fig:Tvir} and~\ref{fig:3dprof}).  The combination of multiphase gas, a temperature gradient in the volume-filling phase, and different ionization mechanisms means that the haloes (gas at $0.1$--$1 \Rvir$) exhibit a larger diversity of ions than the single-temperature CIE model alone would predict (Fig.~\ref{fig:ionfrac}). }
\end{itemize}

We note an interesting feature in the median column density profiles: the column densities do not always decrease as the halo-centric impact parameter increases. This occurs in haloes with $\Tvir$ above the CIE peak temperature of the absorbing ion, where we find peaks or `shoulders' in the column density profiles (Fig.~\ref{fig:radprof_Rvir}). This occurs because the temperature of the volume-filling phase declines towards the outskirts of the halo (Fig.~\ref{fig:3dprof}), causing the ion fraction to be larger there. Despite the decline of gas density with radius, this leads to a `shell' around the galaxy where most of the metals in a particular ionization state are found (Fig.~\ref{fig:absloc}). In projection, depending on the strength of the shell feature, this leads to a peak or flattening of the column density as a function of halo-centric radius, typically around $\Rvir$.

When we examine absorption as in blind surveys, we find the following:
\begin{itemize}
\item{The CDDFs have shallow slopes at lower column densities and a `tail' with a steep slope at high column densities (Fig.~\ref{fig:cddfs}).}
\item{For the X-ray ions, the high-column-density tail of the CDDF is produced mostly by CGM gas (Fig.~\ref{fig:cddfsplits_abs}): $70$--$80$~per~cent has an impact parameter $\mathrm{r}_{\perp} < \Rvir$ for a halo with $\Mvir > 10^{11} \Msun$. }
\end{itemize}

Finally, we make the following predictions for observational detections:
\begin{itemize}
\item{For most of these ions, column densities remain large out to $\sim \Rvir$ in haloes where $\Tvir$ is around the CIE peak temperature for that ion (Figs.~\ref{fig:radprof_Rvir} and~\ref{fig:Tvir}), and \ion{O}{vii} and \ion{O}{viii} should be detectable with Athena that far out around $\Mstellar > 10^{10.5} \Msun$ galaxies (Fig.~\ref{fig:radprof_fcov_obs}). However, \ion{Fe}{xvii} absorption is more centrally concentrated, and more confined to haloes than the other ions in general (Fig.~\ref{fig:simple_iondist}). }
\item{We expect that the Athena X-IFU can detect \ion{O}{vii} absorption in 77 (59)~per~cent of sightlines passing central galaxies with stellar masses $\Mstellar = 10^{10.5 \txn{--} 11.0} \Msun$ ( $\Mstellar > 10^{11.0} \Msun$) within $100 \pkpc$. For \ion{O}{viii}, this is 46 (82)~per~cent. Hence, the X-IFU will probe covering fractions comparable to those detected with the Cosmic Origins Spectrograph for \ion{O}{vi}.}
\item{\ion{Ne}{ix} and \ion{Fe}{xvii} might prove more difficult to find in the CGM with Athena, because the (roughly estimated) minimum observable column densities for these ions are close to, or in, the high-column-density tail of the CDDF. }
\item{At column densities expected to be detectable with the Athena X-IFU, some of the absorption lines will be saturated. However, the saturation is less severe than thermal line broadening would predict for the temperatures where the CIE ionization fractions peak  (Fig.~\ref{fig:N-EW}). }
\item{Our set of ions is mostly suited to probe haloes of $\Mvir \sim 10^{12}$--$10^{13.5} \Msun$, though \ion{O}{vi}, and \ion{Ne}{viii} at $z=0.5$, also probe lower halo masses, and \ion{Fe}{xvii} also probes somewhat hotter haloes. }
\end{itemize}

\section*{Acknowledgements}

We would like to thank Rob Crain and Fabrizio Nicastro for their involvement in the beginning of the project. NAW and BDO would like to thank Akos Bogdan, Ralph Kraft, Randall Smith, and Alexey Vikhlinin for useful discussions.
We thank Yakov Faerman and the anonymous referee for useful comments. We would additionally like to thank John Helly for programs we used to access {\eagle} data (\textsc{read{\_}eagle}), and Volker Springel for the original version of the code we use to project particles onto a grid (\textsc{HsmlAndProject}). Ali Rahmati helped NAW test the code we use to make projections and helped set up {\specwizard}. We used the \textsc{numpy} \citep{numpy}, \textsc{scipy} \citep{scipy}, \textsc{h5py} \citep{h5py}, and \textsc{matplotlib} \citep{matplotlib} \textsc{python} libraries, and the \textsc{ipython} \citep{ipython} command-line interface.
This paper is supported by the European Union's Horizon 2020 research and innovation programme under grant agreement No 871158, project AHEAD2020. This work used the DiRAC@Durham facility managed by the Institute for Computational Cosmology on behalf of the STFC DiRAC HPC Facility (www.dirac.ac.uk). The equipment was funded by BEIS capital funding via STFC capital grants ST/K00042X/1, ST/P002293/1, ST/R002371/1 and ST/S002502/1, Durham University and STFC operations grant ST/R000832/1. DiRAC is part of the National e-Infrastructure.

\section*{Data availability}

The data in the figures will be shared on a reasonable request to the corresponding author. The {\eagle} data are available at \url{http://icc.dur.ac.uk/Eagle/database.php}; the galaxy and halo catalogues are documented in \citet{mcalpine_helly_etal_2016} and the full simulation data are documented in \citet{eagle-team_2017}. 





\bibliographystyle{mnras}
\bibliography{bibliography_updated}



\appendix

\section{Measuring column densities and EWs}
\label{app:bpar}
In Table~\ref{tab:bfit}, we parametrized the relation between column densities and rest-frame EWs measured in a specific velocity range around the maximum-optical-depth position. Here, we explore how the relation depends on the velocity range over which both are measured. We parametrize this relation with the best-fitting $b$ parameter (eqs.~\ref{eq:bpar}--\ref{eq:logfit}), assuming a single (or doublet, for \ion{O}{viii}) Voigt profile.

\begin{figure}
	\includegraphics[width=\columnwidth]{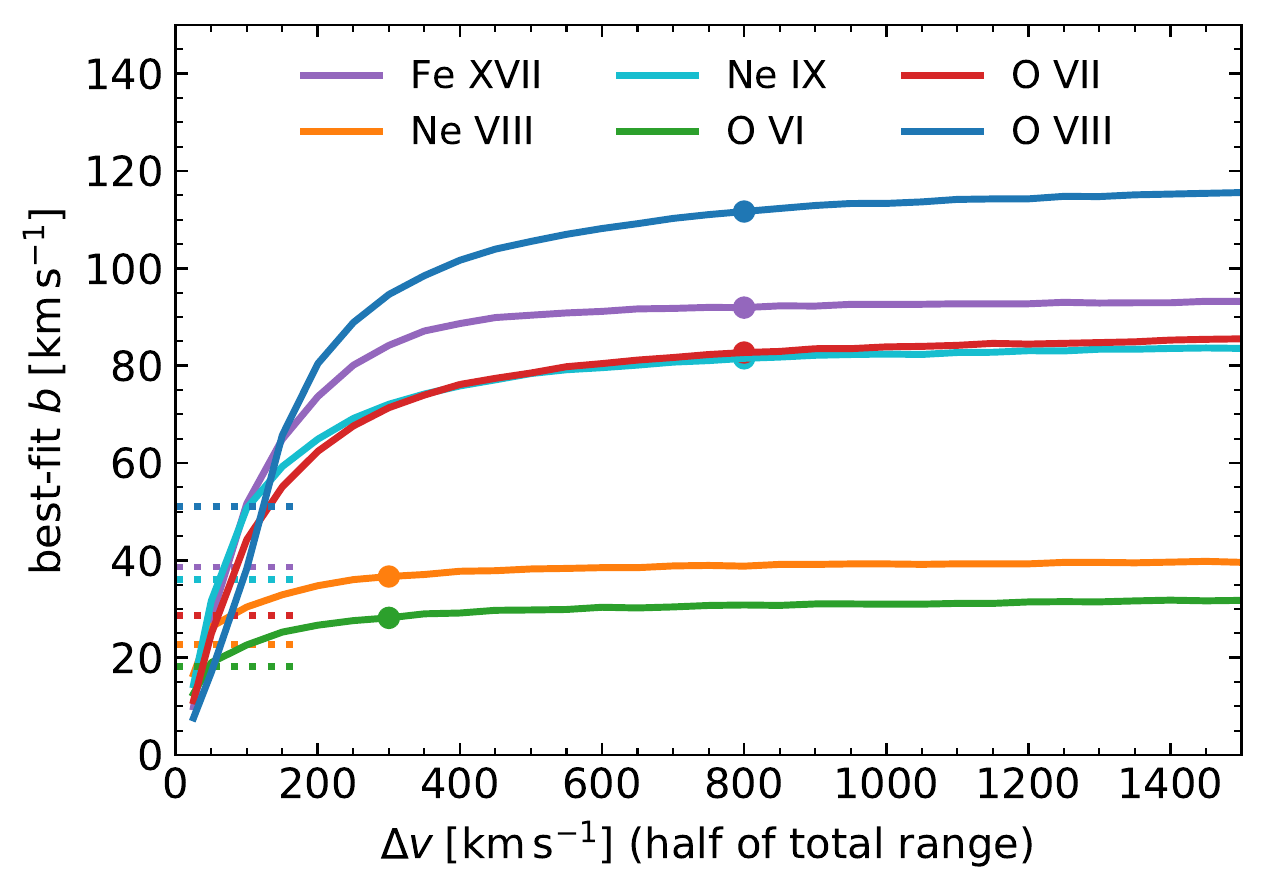}
	\caption{The best-fitting $b$ parameter (eqs.~\ref{eq:bpar}, \ref{eq:cl}, and~\ref{eq:logfit}) for the different ions' absorption lines (Table~\ref{tab:lines}, different colors) in {\eagle} at $z=0.1$ as a function of the line-of-sight velocity range within which the column density and EW are measured. The indicated velocities are the maximum differences relative to the highest optical depth pixel, which is half the full velocity range. The filled points indicate the velocity windows used for each ion in Fig.~\ref{fig:N-EW} and throughout the paper, and the dashed lines indicate the thermal broadening for each ion at the temperature where its CIE ionization fraction is largest.}
	\label{fig:bpar_v}
\end{figure}

Fig.~\ref{fig:bpar_v} shows the dependence of the best-fitting $b$ parameter on the velocity window used. Since structures and correlations on scales approaching half the box size ($\Delta v \approx 3200 \kmps$) cannot be reliably measured in a periodic box, we only show best-fitting values up to $\Delta v = 1500 \kmps$. 


For very small velocity ranges, $b$ parameters are lower than a single absorption component at a typical temperature for high column density absorbers would give. (Low column density absorbers are unsaturated and the fit is therefore not sensitive to their linewidths.) However, the smallest ranges are close to or below the FWHM for such line profiles, so we cannot expect those ranges to be reliable. We interpret the initial rise with velocity range to be due to the velocity window encompassing more absorption from a single absorber or multiple correlated absorbers as the range is increased, and reaching a plateau when all the correlated absorption is included.

For the UV ions (\ion{O}{vi} and \ion{Ne}{viii}), this plateau is reached roughly at the velocity cut $\Delta v = \pm 300 \kmps$ that we used (filled green and orange points in Fig.~\ref{fig:bpar_v}). This cut was motivated by the observations of \citet{tumlinson_thom_etal_2011} and choices by \citet{johnson_chen_mulchaey_2015} and \citet{burchett_tripp_etal_2018}, and is apparently also reasonable for our simulated systems.

For the X-ray ions, we wanted to make sure the velocity range was not too small to probe with the Athena X-IFU, but there was otherwise no clear choice. Based on Fig.~\ref{fig:bpar_v}, we chose a velocity window $\Delta v = \pm 800 \kmps$ for these (filled red, blue, purple, and cyan points in Fig.~\ref{fig:bpar_v}). This is large enough to be in the plateau region for these ions, but stays clear of the half box size.    

Note that \citet{wijers_schaye_etal_2019} found a larger best-fitting $b$ parameter for \ion{O}{viii} using the same set of sightlines, but measuring column densities and EWs over the full sightlines. The difference is not due to the inclusion of damping wings in the spectra in this work, which makes very little difference. Instead, it is driven by a subset of high column density sightlines that contain two strong absorbers, resulting from an alignment of two particular high-mass haloes along the z-axis of the simulation (our line-of-sight direction). This affects best-fitting $b$ parameters at $\Delta v$ approaching half the box size. These haloes also affect the large-$\Delta v$ best-fitting $b$ parameters for \ion{Fe}{xvii} and \ion{Ne}{ix}. 

\section{How to split the CDDFs}
\label{app:techsplit}

When splitting the CDDFs into contributions from absorption by haloes of different masses, we mostly considered only SPH particles in haloes in each chosen mass range (halo-projection method).

For comparison, we also used a method where we attributed a pixel to a halo mass bin by checking whether any central galaxy in the same $6.25 \cMpc$ slice along the line of sight was within $\Rvir$ (projected) for each absorption system, and if so, which one was closest to that system in normalized impact parameter units of $r_{\perp} \, /\Rvir$. However, low-mass haloes show flat absorption profiles and, depending on the ion in question, have virial temperatures well below the temperatures where that ion exists in CIE. Some of the absorption attributed to low-mass haloes in this pixel-attribution method will therefore actually be due to higher mass haloes nearby. We also refer to this as the mask-split method, since the column density maps are split into contributions by different halo masses based on True/False array masks.

\begin{figure*}
 	\includegraphics[width=0.8\textwidth]{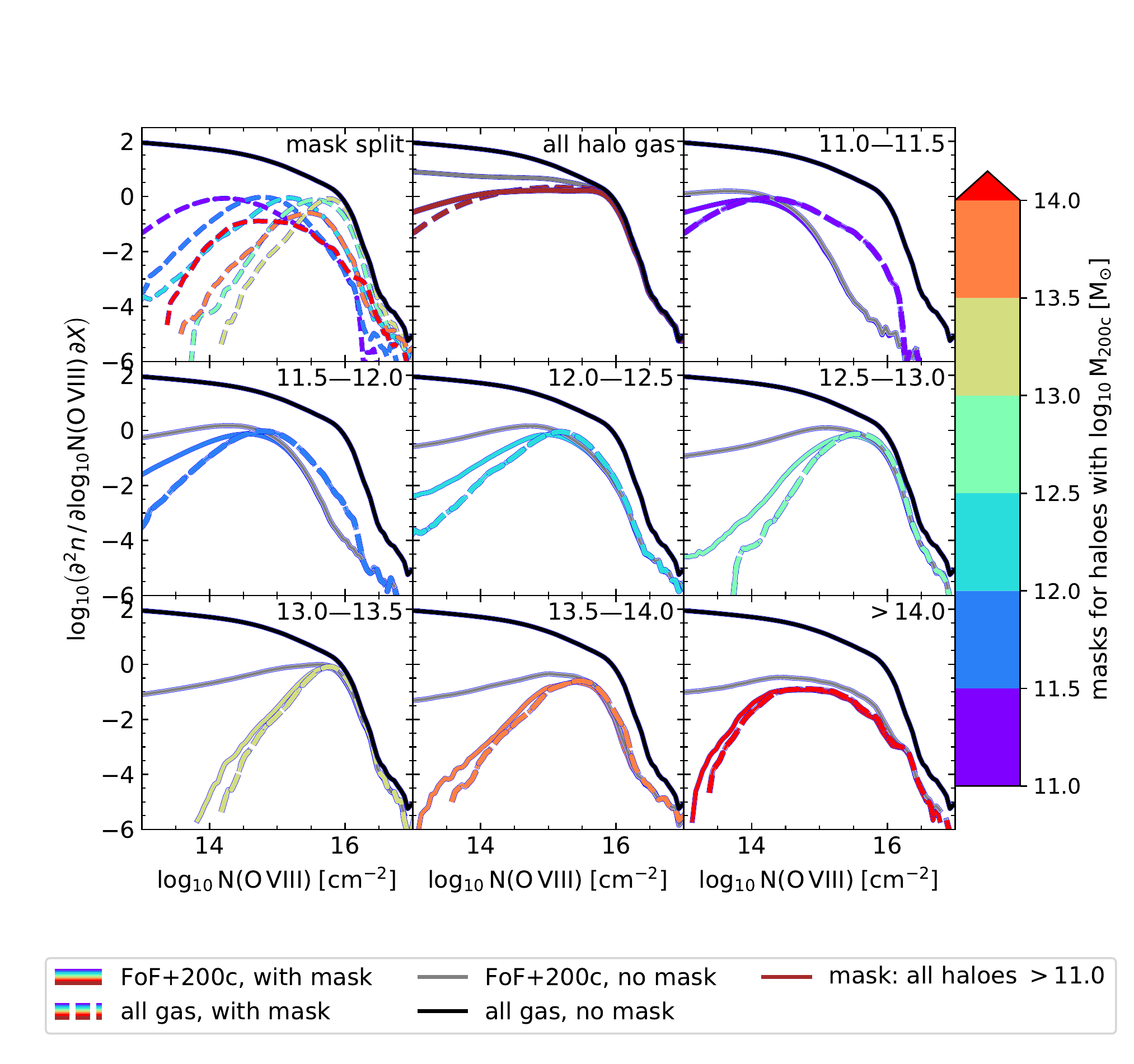}
 	\caption{A comparison between the column density distribution arising from haloes in different mass ranges by two methods. In the first, the CDDF is constructed in the same way as from all gas, but only considering absorption from gas within haloes in each mass range (halo-projection). By within haloes, we mean FoF particles and anything within $\Rvir$ of the centre of potential of each halo. For the second method, the column density distribution attributed to a halo mass bin is determined by, for each pixel in the column density map, checking which halo, if any, matches the line-of-sight slice of the absorber, is within $\Rvir$, and is closest in units of $r_{\perp} \, / \Rvir$ (mask-split or pixel-attribution). This plot shows the distributions for \ion{O}{viii} in {\eagle} at $z=0.1$. The black lines show the total CDDF in all panels. In the top left-hand panel, dashed coloured lines show the part of the total CDDF attributed to the halo mass range corresponding to that colour (colour bar) using the mask-split method. The `all halo gas' panel uses masks for all haloes $\Mvir > 10^{11} \Msun$, cutting out parts attributed to haloes with $10^{9} \Msun < \Mvir < 10^{11} \Msun$. The masks are applied to an absorption map including only gas in haloes (of any mass, including $< 10^{9} \Msun$). The brown lines in this panel come from combining the masks for all the haloes with $\Mvir \geq 10^{11} \Msun$. In the rest of the panels, the range in the top right-hand corner indicates the range of halo masses ($\log_{10} \Mvir \, / \Msun$) for which the gas is included in the column density maps the CDDFs are derived from. The dashed line of the colour matching the mass range shows the contribution according to the mask-split method applied to the column density maps for all gas, and the coloured solid lines show CDDFs from only the gas in the indicated range attributed to the same halo mass range by the mask-split method. The grey lines in the panels for specific halo masses show the halo-gas-only CDDFs used in the main text (Fig.~\ref{fig:cddfsplits_abs}).}
 	\label{fig:cddfsplits_techdep}
\end{figure*}

We show a comparison of which absorption is attributed to which halo using the halo-projection and the pixel-attribution methods in  Fig.~\ref{fig:cddfsplits_techdep}, for \ion{O}{viii}. The top left-hand panel is the same as Fig.~\ref{fig:cddfsplits_abs}, but using the pixel-attribution method. We see that the modal column density increases with halo mass up to the `break' in the CDDF, after which modal column densities decrease again and distributions become flatter.  

In the top, middle panel (`all halo gas'), the grey line shows absorption from haloes at any mass, while the brown lines show CDDFs from all-gas (dashed) and halo-only (solid) column density maps, with only the contributions from pixels within $\Rvir$ of an $\Mvir > 10^{11} \Msun$ halo. The differences between these three methods of measuring how much absorption is due to haloes are relatively small at high column densities, especially between the solid and dashed brown lines, which represent the two main ways to define the CDDF coming from roughly within the virial radii of haloes. This means that the split between halo and extra-halo absorption is relatively robust. (Note that gas outside $\Rvir$ or other virial radius definitions may still be associated with haloes, so where exactly the line between CGM and IGM lies is not generally agreed on.)

In the panels for specific halo masses, we can similarly compare the dashed and solid lines of the colour matching the halo mass in the panel to the grey line to get a sense of how robust the map-split and FoF-only methods are for determining which absorption comes from haloes. The agreement between these methods clearly depends on the halo mass. The higher column density absorption projected within $\Rvir$ of the lowest mass haloes is clearly mostly due to gas outside those haloes, showing that much of the absorption attributed to these haloes using the mask-split method is simply due to these haloes being in roughly the same place as whatever structure is causing the absorption. 

As the halo mass increases, the different methods agree well at column densities at or above the modal column density for that mass. The difference between the grey and coloured solid lines shows that this is largely due to lower column density absorption from gas in the FoF groups of haloes that is outside $\Rvir$ in projection. This will be due, in part, to the FoF groups not agreeing exactly with the overdensity definition of haloes, but also because SPH particles centred inside $\Rvir$ may extend beyond that radius. The very low column densities likely result, at least in part, from `edge effects', where a particular sightline probes (the edge of) one or very few SPH particles at the outer extent of the FoF group.  


Differences between the solid and dashed lines show absorption attributed to haloes (i.e. within $\Rvir$ in the plane of the sky), but not due to gas within the FoF group. This tends to lead to slightly more high-column-density absorbers, but the larger effect is typically at lower column densities, where gas outside the FoF groups is more important (Fig.~\ref{fig:radprof_Rvir}). 

For the other ions, the picture is very similar, except that the mass above which the methods agree at high column densities changes: the largest mass bin with disagreements at most about as bad as $\Mvir = 10^{11.5}$--$10^{12} \Msun$ for \ion{O}{viii} is:
\begin{itemize}
	\item{$\Mvir = 10^{10.5}$--$10^{11} \Msun$ for \ion{O}{vi} (not shown here),}
	\item{$\Mvir = 10^{11}$--$10^{11.5} \Msun$ for \ion{Ne}{viii} and \ion{O}{vii}, and}
	\item{$\Mvir = 10^{11.5}$--$10^{12} \Msun$ for \ion{Ne}{ix}, \ion{Fe}{xvii} (and \ion{O}{viii}).}
\end{itemize}
Above those ranges, we consider the contributions of these ions to the CDDF to be fairly robust. The limiting mass for reliability increases with the CIE temperature range of the ions. 

In summary, the attribution of absorption to haloes shown in Fig.~\ref{fig:cddfsplits_abs} is fairly robust for higher mass haloes ($> 10^{12} \Msun$, and lower for some ions), for column densities above the peak of the CDDF for each $\Mvir$ bin. However, at lower column densities, the CDDF for a set of haloes depends quite strongly on how absorption and haloes are connected. The fraction of absorption beyond the CDDF breaks due to haloes does depend on these choices somewhat, but they do not change the qualitative conclusions.

\section{Redshift evolution}
\label{app:zev}

\begin{figure*}
	\centering
	\includegraphics[width=0.8\textwidth]{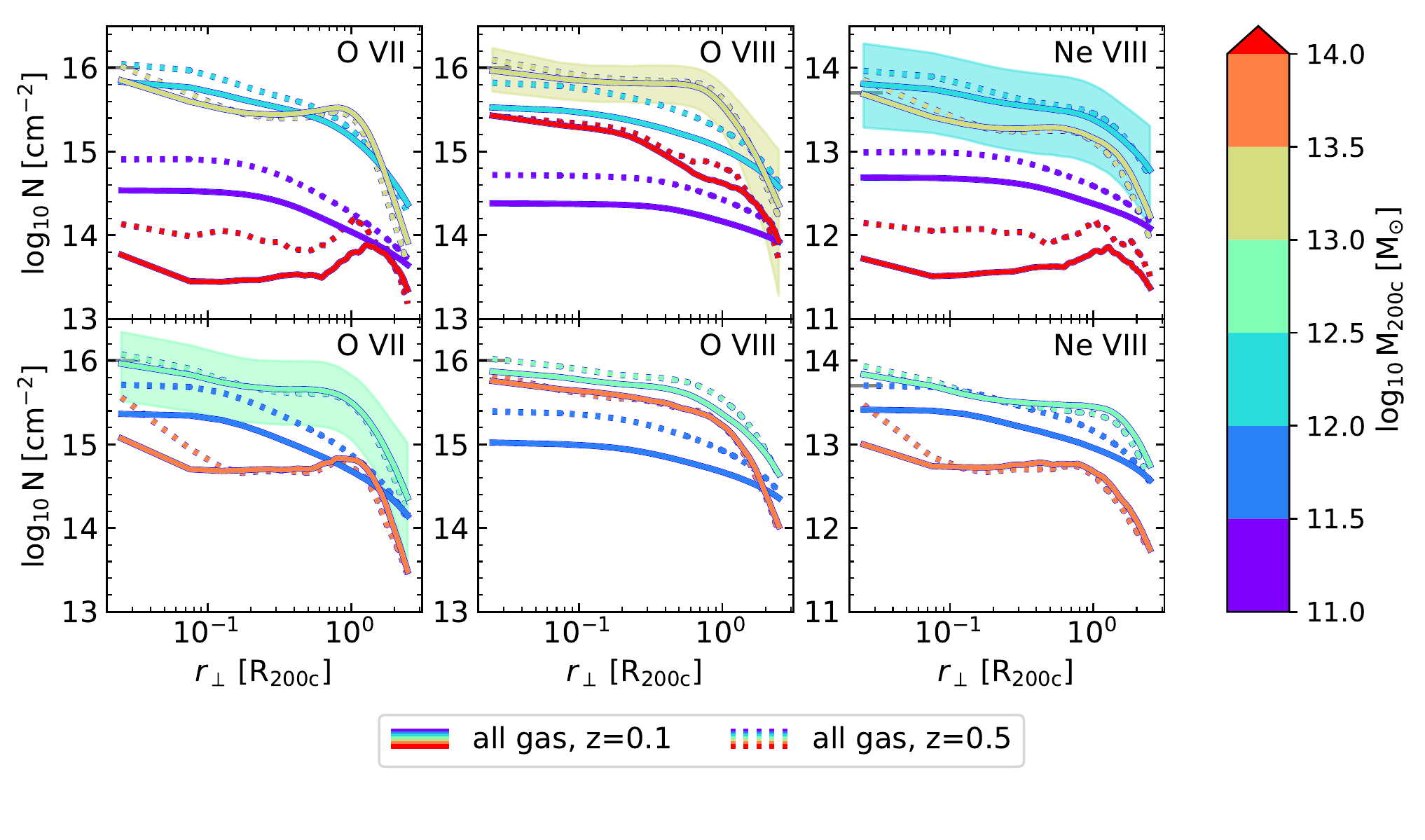}
	\caption{A comparison of the radial profiles for \ion{O}{vii}, \ion{O}{viii}, and \ion{Ne}{viii} in {\eagle} between redshift~0.1, as used throughout this work, and redshift~0.5. The mass bins are split over top and bottom panels for legibility. We see that at halo masses where the column densities peak, the median column density is similar at the two redshifts, but the peak range may be larger at $z=0.5$, and the profiles at higher and lower halo masses change more. The central 80~per~cent scatter is shown for the peak halo mass at $z=0.1$ with the shaded bands.}
	\label{fig:radprof_zvar}
\end{figure*}

In Fig.~\ref{fig:radprof_zvar}, we investigate the redshift evolution of halo radial profiles. This is relevant for all the ions discussed in this work, but we focus on \ion{O}{vii} and \ion{O}{viii}, which should be the most easily observed X-ray lines in this sample, and \ion{Ne}{viii}, which was observed at $z > 0.48$, but is not observable at $z=0.1$ \citep{burchett_tripp_etal_2018}. We compare column densities measured through the same comoving slice thickness ($6.25 \cMpc$) at both redshifts.

The median changes are generally small ($\lesssim 0.2 \dex$) at the halo masses where the column densities are largest, and within the range of the 80~per~cent scatter. The redshift evolution is larger at larger and smaller halo masses, however.   
The changes are in line with how $\Tvir$ evolves: at the same $\Mvir$, $\Tvir$ is somewhat larger at $z=0.5$ than $z=0.1$. Note that for a fixed density profile, $\Mvir$ and $\Rvir$ at the two redshifts will also differ. The differences do mean that any comparisons of absorbers to data should be done at matching redshifts.

\begin{figure*}
	\centering
	\includegraphics[width=0.8\textwidth]{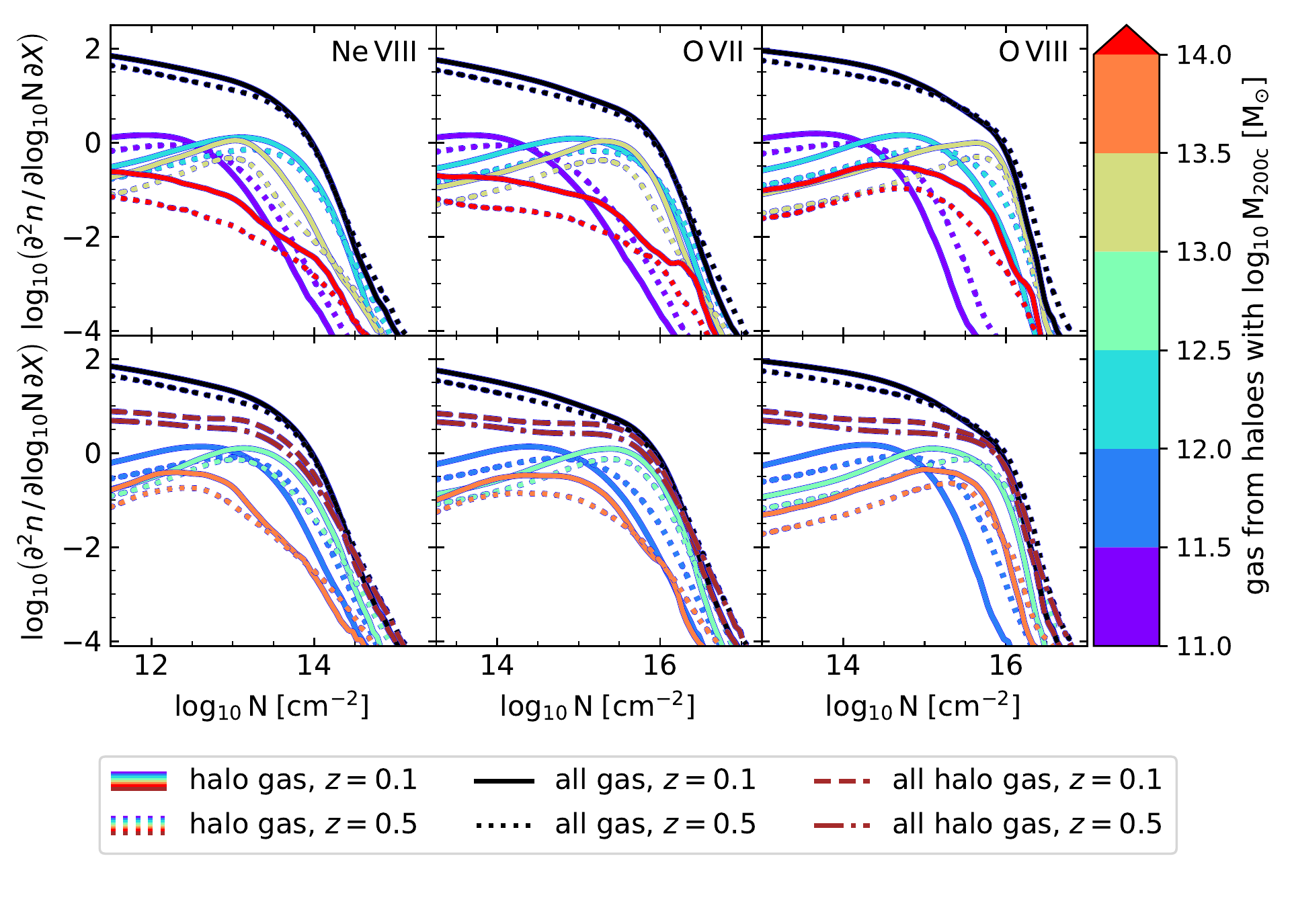}
	\caption{A comparison between the column density distributions of \ion{Ne}{viii}, \ion{O}{vii}, and \ion{O}{viii} in {\eagle} at $z=0.1$ (solid and dashed lines) and $z=0.5$ (dotted and dash-dotted lines). The CDDFs for different halo mass ranges are derived from column density maps using only gas in FoF groups or within $\Rvir$ of haloes of the different masses. The colours match those of Fig.~\ref{fig:cddfsplits_abs}, but the lines for different halo masses are split over top and bottom panels for legibility.}
	\label{fig:cddfsplit_zev}
\end{figure*}

Fig.~\ref{fig:cddfsplit_zev}, we similarly consider the evolution of the CDDFs of these ions from redshift~$0.1$ to $0.5$. We see the CDDFs change in ways consistent with the radial profiles: the distributions look similar, but at $z=0.5$, they look like those for somewhat higher mass $z=0.1$ haloes. Another difference is in the total gas CDDF. All haloes together contribute less to the high-column-density absorption at $z=0.5$ than at $z=0.1$. This might be related to how the FoF groups are defined: by particle separation relative to the average, which means that haloes will be `cut off' at higher densities at the higher redshift. Overall evolution of the {\eagle} \ion{O}{vii} and \ion{O}{viii} CDDFs was discussed in more detail by \citet{wijers_schaye_etal_2019}, and \citet{rahmati_etal_2016} discussed the \ion{Ne}{viii} CDDF evolution, as well as that of \ion{O}{vi}.   

Note that the highest mass bin only has five haloes at $z=0.5$, and 9 at $z=0.1$, so some changes here might be explained by selection effects (lower typical masses at higher redshift) and small sample sizes.

\bsp	
\label{lastpage}
\end{document}